\documentclass[12pt]{article}

\usepackage{arxiv}

\usepackage[utf8]{inputenc} 
\usepackage[T1]{fontenc}    
\usepackage{hyperref}       
\usepackage{url}            
\usepackage{booktabs}       
\usepackage{amsfonts}       
\usepackage{nicefrac}       
\usepackage{microtype}      
\usepackage{graphicx}
\usepackage{doi}
\usepackage{setspace}

\usepackage{soul}
\usepackage{bm}
\usepackage[skip=2pt]{subcaption}
\usepackage{booktabs}
\usepackage{multirow}
\usepackage{lscape}
\usepackage{tikz}
\usepackage{xcolor, colortbl}
\usepackage{xurl}
\usepackage{amsmath,amsfonts,amsthm,amssymb,thmtools}
\usepackage{float}
\usepackage{placeins}
\usepackage[authoryear,round]{natbib}

\newtheorem{assumption}{Assumption}
\newtheorem{orig}{Original Assumption}

\definecolor{spiritblue}{HTML}{6cace4}
\definecolor{maize}{HTML}{FFCB05}
\newcommand{\mathcolorbox}[2]{%
  \begingroup\setlength{\fboxsep}{1pt}%
  \colorbox{#1}{$#2$}%
  \endgroup
}

\newcommand{\mathdisplaycolor}[2]{\colorbox{#1}{$\displaystyle #2$}}

\newcommand{\hlcol}[2]{%
  \begingroup\setlength{\fboxsep}{2pt}%
  \colorbox{#1}{#2}%
  \endgroup
}

\hypersetup{
    colorlinks,
    linkcolor={blue!80!black},
    citecolor={blue!80!black},
    urlcolor={blue!80!black}
}

\newcommand{\V}{\mathbb{V}}
\newcommand{\E}{\mathbb{E}}

\newcommand{\dih}{\hat{d}_i}
\newcommand{\dti}{\tilde{d}_i}
\newcommand{\dtj}{\tilde{d}_j}

\newcommand{\dtm}{\tilde{d}_i^{(M)}}
\newcommand{\dom}{d_{0i}^{(M)}}
\newcommand{\dinfm}{d_{\infty i}}
\newcommand{\dihm}{\hat{d}_i^{(M)}}

\newcommand{\atm}{\tilde{a}_i^{(M)}}
\newcommand{\aom}{a_{0i}^{(M)}}
\newcommand{\ainfm}{a_{\infty i}}
\newcommand{\aihm}{\hat{a}_i^{(M)}}

\newcommand{\btm}{\tilde{b}_i^{(M)}}
\newcommand{\bom}{b_{0i}^{(M)}}
\newcommand{\binfm}{b_{\infty i}}
\newcommand{\bihm}{\hat{b}_i^{(M)}}

\newcommand{\vio}{v^{(1)}_i}
\newcommand{\vit}{v^{(2)}_i}

\newcommand{\viho}{\hat{v}^{(1)}_i}
\newcommand{\viht}{\hat{v}^{(2)}_i}

\newcommand{\thi}{{\hat{\tau}_i}}
\newcommand{\tauh}{{\hat{\tau}}}
\newcommand{\taubo}{\bar{\tau}_{i1}}
\newcommand{\taubt}{\bar{\tau}_{i2}}
\newcommand{\taubk}{\bar{\tau}_{ik}}

\newcommand{\Ybo}{\bar{Y}_{i1}}
\newcommand{\Ybt}{\bar{Y}_{i2}}
\newcommand{\Ybk}{\bar{Y}_{ik}}

\newcommand{\Yjbo}{\bar{Y}_{j1}}
\newcommand{\Yjbt}{\bar{Y}_{j2}}

\newcommand{\Yo}{Y_{i1}}
\newcommand{\Yt}{Y_{i2}}
\newcommand{\Yk}{Y_{ik}}

\newcommand{\no}{n_{i1}}
\newcommand{\nt}{n_{i2}}
\newcommand{\nk}{n_{ik}}

\newcommand{\ybot}{\bar{y}^t_{i1}}
\newcommand{\yboc}{\bar{y}^c_{i1}}
\newcommand{\ybtt}{\bar{y}^t_{i2}}
\newcommand{\ybtc}{\bar{y}^c_{i2}}

\newcommand{\ybkt}{\bar{y}^t_{ik}}
\newcommand{\ybkc}{\bar{y}^c_{ik}}

\newcommand{\ykt}{y^t_{ik}}
\newcommand{\ykc}{y^c_{ik}}

\newcommand{\yot}{y^t_{i1}}
\newcommand{\yoc}{y^c_{i1}}
\newcommand{\ytt}{y^t_{i2}}
\newcommand{\ytc}{y^c_{i2}}

\newcommand{\htYt}{\hat{Y}(t)}
\newcommand{\htYc}{\hat{Y}(c)}
\newcommand{\htNt}{\hat{N}(t)}
\newcommand{\htNc}{\hat{N}(c)}

\newcommand{\Tk}{T_{ik}}

\newcommand{\xo}{\bm x_{i1}}
\newcommand{\xt}{\bm x_{i2}}
\newcommand{\xk}{\bm x_{ik}}

\newcommand{\loomi}{\hat{\tau}[\mbox{LOO MI}, \emptyset]}

\newcommand{\looidpdnx}{\hat{\tau}[\mbox{LOO WLS}, (n,x)]}
\newcommand{\looidpdnxrf}{\hat{\tau}[\mbox{LOO RF}, (n,x)]}

\newcommand{\looidpd}{\hat{\tau}[\mbox{LOO }\cdot,  \cdot]}

\newcommand{\hvt}{\hat{\tau}[0, \emptyset]}
\newcommand{\amw}{\hat{\tau}[\mbox{LOI MI}, \emptyset]}

\newcommand{\wlspdot}{\hat{\tau}^{WLS-P}[\cdot]}
\newcommand{\wlsp}{\hat{\tau}^{WLS-P}[\emptyset]}

\newcommand{\wlspnx}{\hat{\tau}^{WLS-P}[(n,x)]}

\newcommand{\wlsdot}{\hat{\tau}^{WLS}[\cdot]}
\newcommand{\hajek}{\hat{\tau}^{WLS}[\emptyset]}
\newcommand{\wlsn}{\hat{\tau}^{WLS}[n]}
\newcommand{\wlsnx}{\hat{\tau}^{WLS}[(n,x)]}

\title{A General Framework for Design-Based Treatment Effect Estimation in Paired Cluster-Randomized Experiments}

\author{ Charlotte Z.~Mann \\
Statistics Department\\
California Polytechnic State University\\
San Luis Obispo, CA, USA\\
\texttt{czmann@calpoly.edu}
	\And
	Adam C. Sales\\
 Department of Mathematical Sciences\\
	Worcester Polytechnic Institute\\
 Worcester, MA, USA\\
 \texttt{asales@wpi.edu} \\
 \And 
Johann A.~Gagnon-Bartsch\\
Department of Statistics\\
University of Michigan\\
Ann Arbor, MI, USA\\
\texttt{johanngb@umich.edu}
}

\renewcommand{\shorttitle}{pCRT Design-Based Estimation Framework}

\begin{document}
\maketitle

\begin{abstract}
	
Paired cluster-randomized experiments (pCRTs) are common in education program impact evaluation trials. Although common, there is surprisingly no clear consensus regarding how to analyze this randomization design to estimate average treatment effects. Variance estimation is also complicated due to the dependency created through pairing clusters. Therefore, we aim to provide an intuitive and practical comparison between different estimation strategies for pCRTs to inform practitioners' choice of strategy. To this end, we present a general framework for design-based estimation of an average individual effect in pCRTs. This framework offers a novel and intuitive view on the bias-variance trade-off between point estimators and emphasizes the benefits of covariate adjustment for estimation with pCRTs. In addition to providing a general framework for estimation with pCRTs, the point and variance estimators we present support fixed-sample unbiased estimation with similar precision to a common regression model and conservative variance estimation. Through simulation studies based on an educational efficacy trial, we compare the performance of the point and variance estimators reviewed. Our analysis and simulation studies inform the choice of point and variance estimators for analyzing pCRTs in practice.
\end{abstract}

\section{Introduction}

Cluster-randomization is a common experimental design because it can be infeasible to randomize a treatment assignment amongst individuals in a study, for logistical and ethical reasons.  Cluster-randomization is particularly pragmatic when there is natural clustering of individuals -- such as students in classrooms or schools as with many efficacy trials in education. However, clustering reduces the effective sample size (potentially greatly), resulting in a loss of efficiency. One way to overcome this efficiency loss is to pair clusters based on baseline characteristics and randomly assign the treatment within each pair \citep{donner_statistical_1987, dong_statistical_2010,rhodes_pairwise_2014}. Assuming clusters are successfully well paired based on characteristics that are prognostic for the outcome of interest, treatment effect estimates will be more precise. For these reasons, paired cluster-randomized trials (pCRTs) are relatively common across disciplines \citep{donner_statistical_1987, imai_essential_2009, kestler_matched_2013, rhodes_pairwise_2014}.

In practice, researchers commonly use well-known regression estimators to analyze pCRTs \citep{green_analysis_2008, imai_essential_2009, chondros_design_2012, rhodes_pairwise_2014, athey_chapter_2017}. These include weighted cluster-level regression with fixed effects for the pairs, or individual-level hierarchical linear models, with random or fixed effects for the clusters and pairs. The literature suggests cluster-robust standard errors for individual level regression \citep{green_analysis_2008, chondros_design_2012, 
schochet_estimators_2013, athey_chapter_2017, su_model-assisted_2021} and heteroskedasticity-robust standard errors for cluster-level regression \citep{middleton_unbiased_2015, su_model-assisted_2021}. However, inference with these methods, as originally conceived, relies on modeling assumptions which may not be compatible with common causal inference frameworks \citep{freedman_regression_2008, middleton_bias_2008, imai_essential_2009, schochet_is_2010, athey_chapter_2017}. On the other hand, design-based causal inference relies solely on the randomization design of an experiment for inference \citep{imai_rejoinder_2009}, thus alleviating the need for modeling assumptions. The current work focuses on design-based inference for this reason.
 
There is a robust literature addressing the challenges of design-based inference with the two separate randomization elements of pCRTs -- paired \textit{or} clustered randomization. First, there is a well established literature on how to approach design-based estimation with paired experiments \citep[see, e.g.,][]{imai_variance_2008, fogarty_mitigating_2018, fogarty_regression-assisted_2018,  liu_regression-adjusted_2020, pashley_insights_2021}. However, ignoring the weighting that is introduced with clusters when analyzing a pCRT limits the estimands that can be targeted -- namely, it only supports estimation of an average cluster-level effect. On the other hand, the cluster-randomized trial (CRT) literature provides many different approaches to targeting individual-level estimands  \citep[see, e.g.,][]{rosner_use_1999, green_analysis_2008, middleton_bias_2008,hansen_attributing_2009,  schochet_estimators_2013, ding_rank_2018, schochet_analyzing_2020, su_model-assisted_2021,schochet_lasso_2022, wang_model-robust_2023}. See \cite{bugni_inference_2023} and \cite{su_model-assisted_2021} for recent reviews. However, the CRT literature does not provide variance estimators that take paired randomization into account.

 \cite{donner_statistical_1987} provided early insights in the analysis of pCRTs through permutation testing or with a weighted $t$-test.  There has been additional work on randomization inference \citep{small_randomization_2008, zhang_powerful_2012} and design-based parametric estimation  \citep{imai_essential_2009} for pCRTs since. Recent work provides analysis of common regression-based estimators from a design-based perspective \citep{de_chaisemartin_at_2020}.  Other recent work has considered blocked cluster-randomized experiments \citep{middleton_unbiased_2015, schochet_design-based_2021}, which has results that apply to pCRTs as well. A major challenge to treatment effect estimation with pCRTs is that there has been relatively little development of design-based variance estimators for even common point estimators. 

We focus on paired cluster-randomization, addressing the challenges that arise from the combination of both design elements. Our target estimand is the average treatment effect for all individuals in a fixed trial sample (ATE). A somewhat surprising property of the literature on pCRTs is that there is not an agreed upon ``baseline'' estimator for the ATE. By baseline estimator, we mean estimators that only rely on cluster outcomes, potentially weighting by cluster sizes. For example, for most randomization designs, the difference-in-means estimator, or the difference in mean outcomes between the treatment and control groups, is a baseline estimator. Other estimators in the literature may improve upon the precision of the difference-in-means with covariate adjustment or other strategies. In contrast, under paired cluster-randomization, there are many baseline estimators in the literature, which take various weighted differences in mean or total outcomes \citep{imai_essential_2009, middleton_unbiased_2015}. Discussions of the properties of these estimators are spread across different papers with different focuses, making it non-trivial to understand which estimator to use in a given application. 

Therefore, we present a unifying comparison of baseline point and associated variance estimators for the ATE from a fixed-population and design-based perspective. Particularly, we articulate an estimation framework for the ATE in pCRTs. This framework is closely related to those in \cite{middleton_unbiased_2015} and \cite{wu_design-based_2021}, who analyze blocked cluster-randomization and paired randomization, respectively. These works are part of a broader literature for robust design-based covariate adjustment across different study designs \citep{ robins_estimation_1994, robins_analysis_1995, scharfstein_adjusting_1999,rosenbaum:2002a, bang_doubly_2005, laan_targeted_2006, tsiatis_covariate_2008,  moore_covariate_2009, vanderlaan_rose_2011,  aronow_class_2013,  belloni_inference_2014,  middleton_unbiased_2015, wager_high-dimensional_2016, chernozhukov_doubledebiased_2018}.

The estimation framework presented in this paper supports three primary insights for pCRT analysis. First, the framework clarifies that common baseline estimators are equivalent as long as cluster sizes are well matched within each pair, a fact which has been under-discussed in the literature. Second, it provides a novel and intuitive perspective on the bias-variance trade-off between baseline estimators. And finally, the framework emphasizes the importance of covariate adjustment to overcome this bias-variance trade-off. Simulation studies further illustrate the strengths and weaknesses of different point and associated variance estimators with and without covariate adjustment.  We find that, in addition to providing a general framework for estimation in pCRTs, the estimator we present supports fixed-sample unbiased estimation with similar precision to a common regression model.

We focus our comparison and framework to analyzing pCRTs with aggregated cluster-level data rather than individual-level data. There are a number of reasons to take this approach. First, from a design-based perspective, using data aggregated to the cluster level recognizes the experimental design -- that random assignment occurred at the cluster level \citep{imai_essential_2009, rhodes_pairwise_2014}. Second, there may be restrictions on sharing individual-level data, so cluster-level analyses can better support sharing analyses for replication and using publicly available data \citep{schochet_analyzing_2020}. Finally, and perhaps most importantly, there is evidence that individual-level analyses do not see large precision or accuracy gains over cluster-level analyses \citep{green_analysis_2008, athey_chapter_2017, schochet_analyzing_2020,su_model-assisted_2021}. 

We also focus on a specific estimand, from a fixed-population perspective. Previous work has provided unifying frameworks or estimators for a range of estimands of interest that come from viewing the pairs as fixed or sampled from a super-population as well as viewing individuals within a cluster as fixed or sampled from a super-population \citep{rhodes_pairwise_2014, imai_essential_2009}. This is important as there are a number of possible estimands for pCRTs, which may address different policy questions or be more or less difficult to compute \citep{athey_chapter_2017, kahan_estimands_2023}. This work makes a complementary contribution -- articulating a general framework for design-based estimation for a single estimand, explicitly comparing different estimation strategies for the ATE. 

This paper is organized as follows. Section~\ref{sec:sate} establishes our notation, mode of causal inference, and estimand of interest. Section~\ref{sec:review} provides a review of point estimators for the ATE under paired cluster-randomization. Section~\ref{sec:cploop} introduces a point estimator and describes a general framework for design-based estimation of the ATE in pCRTs. Section~\ref{sec:cov} discusses covariate adjustment. Section~\ref{sec:var} reviews design-based variance estimators for the point estimators previously described and defines a variance estimator for the point estimator introduced in Section~\ref{sec:cploop}. Section~\ref{sec:sims} presents the results from simulation studies comparing the point and variance estimators discussed in the paper. Section~\ref{sec:cploopdisc} concludes.

\section{Problem Setup and Inferential Framework}
\label{sec:sate}
Consider a randomized experiment with $M$ pairs of clusters indexed by $i = 1, ..., M$. We will arbitrarily label one cluster in pair $i$ as the ``first'' cluster and the other cluster as the ``second'' cluster, indexed by $k = 1,2$. Denote the number of individuals in the first and second clusters in a pair as $\no$ and $\nt$, respectively. Thus, there are a total of $N = \sum_{i = 1}^{M}\sum_{k = 1}^2 \nk$ individuals in the experiment. We view these clusters and individuals within each cluster as a fixed population -- i.e. they do not arise as a sample from some super population of clusters or individuals.

We assume each individual has two \textit{fixed} potential outcomes \citep{neyman:1935, rubin_estimating_1974, holland_statistics_1986}, one which would be observed if they were assigned to treatment and the other if assigned to control. Denote these $y^t_{ik\ell}$ and $y^c_{ik\ell}$, respectively, for individual $\ell$ in cluster $k$ and pair $i$. Let  $\ybkt$ and $\ybkc$, denote the cluster-average potential outcomes for cluster $k$ in pair $i$. Denote the cluster-level treatment assignment $T_{ik}$. Let $T_i = T_{i1}$ denote the treatment assignment for the first cluster in pair $i$; the assignment for the second cluster in pair $i$ is therefore $T_{i2} = 1-T_i$. The $T_i$ are independent Bernoulli random variables with $P(T_i = 1) = \frac{1}{2}$. Thus, the observed cluster-average outcome in the first cluster in pair $i$ is $\Ybo= T_i\ybot + (1-T_i)\yboc$, and in the second cluster is $\Ybt = (1-T_i)\ybtt + T_i\ybtc$. We denote observed cluster-total outcomes interchangeably as $\nk \bar{Y}_{ik}$ or $Y_{ik}$, for emphasis or notational conciseness. This notation is summarized in Table~\ref{tab:notation} for reference.

Our target estimand is the average treatment effect for all individuals in a given experiment (hereafter ``ATE''), denoted $\bar{\tau}$ and defined:  
\begin{equation}\label{eq:ate}
    \bar{\tau} \equiv \frac{1}{N}\sum_{i = 1}^{M}\sum_{k = 1}^{2}\sum_{\ell = 1}^{\nk} (y^t_{ik\ell} - y^c_{ik\ell}).
\end{equation}
It is additionally useful to consider the ATE in terms of cluster-aggregated potential outcomes:
$$\bar{\tau} = \frac{1}{N}\sum_{i = 1}^{M} \tau_i$$
where $\tau_i = \no(\ybot - \yboc) + \nt(\ybtt - \ybtc)$ is the total causal effect in pair $i$.

\begin{table}[ht]
\centering
\begin{tabular}{ll}
\hline
 $M$ &  Number of pairs\\
 $N$ & Total number of individuals \\
 $i,k, \ell$ & Indices: pair $i$, cluster $k \in (1,2)$, individual $\ell$\\
 $\nk$ & Number of individuals (i.e. cluster size) for cluster $k$ in pair $i$ \\
 $n_i$ & Number of individuals in pair $i$ ($\no + \nt$) \\
 $y^t_{ik\ell}, y^c_{ik\ell}$ & Individual treatment and control potential outcomes\\
 $\ybkt, \ybkc$ & Cluster-mean treatment and control potential outcomes \\
 $T_i$ & Treatment assignment for the ``first'' cluster in pair $i$\\
 $\Ybk$ & Observed cluster-\textit{mean} outcome \\
  $\Yk = \nk\Ybk$ & Observed cluster-\textit{total}  outcome \\
 $\bar{\tau}$ & ATE (\ref{eq:ate})\\
  $\tau_i$ & Total pair causal effect ($\bar{\tau} = \frac{1}{N}\sum_{i = 1}^{M} \tau_i)$\\
 \hline
\end{tabular}
\caption{Summary of notation for paired cluster-randomized experiments.}
\label{tab:notation}
\end{table}

\section{Existing ``Baseline'' Design-Based Estimators}
\label{sec:review}

In this section, we detail four estimators for the ATE under paired cluster-randomization that are commonly discussed in the literature with common notation and discuss their properties. It is notable that these four estimators could all be thought of as ``baseline'' estimators -- i.e. they rely only on weighted cluster-aggregated outcomes and no covariates

\subsection{Horvitz-Thompson (HT) Estimator}

The Horvitz-Thompson estimator \citep{horvitz_generalization_1952} is known for being unbiased for a finite population average treatment effect across any identifiable randomization design \citep{middleton_unified_2018} and is defined as follows:
\begin{equation}\label{eq:ht}
\hat{\tau}^{HT}[\emptyset] = \frac{1}{N}\sum_{i=1}^M 2(2T_i -1) (\no\Ybo - \nt\Ybt)
\end{equation}
We denote the baseline estimators with [$\emptyset$], as above, to indicate that it is not a function of any covariates, i.e., there is no covariate adjustment.  The estimator can be thought of as the sum of unbiased estimates for $\tau_i$, divided by the  (fixed) total number of individuals $N$. Particularly,  $\frac{1}{P(T_i = 1)} = 2$, so (\ref{eq:ht}) can be recognized as an inverse probability weighted (IPW) estimator \citep{robins_estimation_1994,robins_analysis_1995}. \cite{su_model-assisted_2021} note that the Horvitz-Thompson estimator is equivalent to the estimated coefficient for the treatment assignment from an ordinary least squares (OLS) regression of $\frac{M}{N}(\nk\Ybk)$ on the treatment assignment (i.e., $\hat{\beta}$ from the regression model $\frac{M}{N}\nk\Ybk = \alpha + \beta\Tk +\varepsilon_{ik}$ with error term $\varepsilon_{ik}$).   

While the Horvitz-Thompson estimator is unbiased, it is not location invariant, unless the cluster sizes within pairs are equal. Due to this location non-invariance, the Horvitz-Thompson estimator has high variance when the cluster sizes differ greatly between pairs and the outcomes are not zero-centered. Therefore, the Horvitz-Thompson estimator is not typically used in practice. However, it provides a methodologically useful baseline unbiased estimator, as will be discussed in Section~\ref{sec:cploop}.

\subsection{H\'ajek (WLS) Estimator}\label{sec:hajek}

Perhaps the most well known estimator for an average treatment effect is the difference-in-means estimator --- the difference in mean outcomes for the treatment group minus the control group. Under paired cluster-randomization, the difference-in-means estimator takes the form of a H\'ajek estimator \citep{hajek_comment_1971, fredrickson_causal_2018}:
\begin{equation}\label{eq:hj}
\hajek = \frac{\sum_{i=1}^M T_i \no\Ybo + (1-T_i)  \nt\Ybt}{\sum_{i=1}^M T_i \no + (1-T_i)  \nt} - 
\frac{\sum_{i=1}^M (1-T_i)  \no\Ybo + T_i  \nt\Ybt}{\sum_{i=1}^M (1-T_i) \no + T_i  \nt}.
\end{equation}

We use the notation ``WLS'' because the H\'ajek estimator also has a regression equivalent. Namely, the estimated coefficient for the treatment assignment from a weighted least squares (WLS) model of cluster-mean outcomes on the cluster-level treatment assignment, weighted by cluster size (i.e., $\hat{\beta}$ from the regression model $\bar{Y}_{ik} = \alpha + \beta\Tk +\varepsilon_{ik}$ with weights $\nk$).

The H\'ajek estimator gains precision over the Horvitz-Thompson estimator by dividing Horvitz-Thompson estimators for the total treatment and control outcomes by estimates of the sample size based on the treatment and control clusters, rather than by $N$ (see \cite{fredrickson_causal_2018} for a discussion). Specifically, a Horvitz-Thompson estimator of the total treated potential outcomes is of the form: \begin{equation}\label{eq:hty} \hat{Y}(t) = \sum_{i=1}^M2\left\{T_i \no\Ybo + (1-T_i)\nt\Ybt\right\} \end{equation} and a Horvitz-Thompson estimator of the total sample size, based on the treated units is \begin{equation}\label{eq:htn} \hat{N}(t) = \sum_{i=1}^M2\left\{T_i  \no + (1-T_i)\nt\right\}. \end{equation} $\hat{Y}(c)$ and $\hat{N}(c)$ are defined analogously. Then, \begin{equation}\label{eq:hthj} \hajek = \frac{\hat{Y}(t)}{\hat{N}(t)} - \frac{\hat{Y}(c)}{\hat{N}(c)}.\end{equation}
 
This adjustment improves precision over the Horvitz-Thompson estimator and is location invariant. Despite the fact that the difference-in-means estimator is unbiased under many other randomization designs, it is not an unbiased estimator for the ATE under cluster randomization, unless the cluster sizes within each pair are equal. This is because the estimator is ultimately a difference of ratios of random variables since the number of treatment or control \textit{individuals} in a trial is itself a random variable. The bias of the H\'ajek estimator is $-\frac{1}{N}[Cov\{\htYt/\htNt, \htNt\}- Cov\{\htYc/\htNc, \htNc\}]$ \citep{middleton_unbiased_2015}. While biased for the ATE from a fixed-sample perspective, under regularity conditions, the H\'ajek estimator is asymptotically consistent for the ATE, as the number of \textit{pairs} grows to infinity  \citep{middleton_unbiased_2015}. 

\subsection{WLS-P Estimator}

A common estimator in practice is to add a fixed effect for each pair in the WLS model formulation of the H\'ajek estimator. In other words, we may define $\wlsp$ to be the $\hat{\beta}$ from the WLS solution to the regression model 
$ \bar{Y}_{ik} = \alpha + \beta\Tk + \sum_{p=1}^M\tilde{\alpha}_pI_{ik,p} +\varepsilon_{ik}$
with weights $\nk$, where $I_{ik,p}$ is an indicator of whether cluster $ik$ is in pair $p \in 1, ..., M$ and $\varepsilon_{ik}$ is an error term. We will refer to this estimator as the WLS-P estimator (\textbf{w}eighted \textbf{l}east \textbf{s}quares with \textbf{p}air fixed-effects). The estimator can be written in a closed form as a weighted sum of the difference in cluster-mean outcomes within each pair \citep{imai_essential_2009}:  
\begin{equation}\label{eq:fe}
\wlsp = \frac{1}{\sum_{i=1}^M w_i}\sum_{i=1}^M w_i(2T_i -1) (\Ybo - \Ybt)
\end{equation}
where $w_i = \frac{\no\nt}{\no+\nt}$ is proportional to the \textit{harmonic mean} of cluster sizes.

In its regression formulation, this estimator appears similar to the H\'ajek estimator. However, in its closed form, it is clear that these two estimators have distinct properties. Unlike the H\'ajek estimator, the WLS-P estimator does not have a random denominator. However, it is still a biased estimator for the ATE because the individual terms $\frac{w_i}{\sum_{i=1}^M w_i}(2T_i -1) (\Ybo - \Ybt)$ are not unbiased estimates of $\tau_i/N$ unless the cluster sizes within pairs are equal for all pairs in the study \citep{imai_essential_2009}. Importantly, unlike the H\'ajek estimator, this bias does not disappear asymptotically.

\subsection{AMW Estimator}

The WLS-P estimator is motivated by a regression model, but (\ref{eq:fe}) suggests a different and more general form of estimator -- a weighted sum of the difference in cluster-mean outcomes within each pair. \cite{imai_essential_2009} study this form of estimator and propose different weighting schemes. Specifically, they suggest weighting proportionally to the \textit{arithmetic mean} of the cluster sizes ($\tilde{w}_i = \no+\nt$) to estimate the ATE: 
\begin{equation}\label{eq:amw}
    \hat{\tau}^{AMW}[\emptyset] = \frac{1}{N}\sum_{i=1}^M (\no+\nt)(2T_i -1) (\Ybo - \Ybt)
\end{equation}
We therefore call their suggested estimator the ``Arithmetic Mean Weighted'' estimator (AMW). The AMW estimator is also equivalent to the WLS estimated coefficient for the treatment assignment from a linear model of the cluster-mean outcome on the treatment, with each cluster weighted by the \textit{total pair} size, $\no+\nt$ (i.e., $\hat{\beta}$ from the regression model $\bar{Y}_{ik} = \alpha + \beta\Tk +\varepsilon_{ik}$ with weights $\tilde{w}_i$) \citep{de_chaisemartin_at_2020}. 

Let $\taubk = \ybkt - \ybkc$. \cite{imai_essential_2009} show that the finite-sample bias of the AMW estimator is $\frac{1}{2N}\sum_{i=1}^M (\no -\nt)(\taubo - \taubt)$. Therefore, the AMW estimator is in general biased, with two notable exceptions: (1) if the cluster sizes are equal within each pair ($\no = \nt$ for all $i = 1,..., M$), or (2) if the cluster level average treatment effect is equal within each pair ($\taubo = \taubt$ for all $i = 1,..., M$). Moreover, under the general (biased) case, the bias persists asymptotically. This bias arises for the same reason as the WLS-P estimator -- the individual terms $(\no+\nt)(2T_i -1) (\Ybo - \Ybt)$ are not unbiased estimates of $\tau_i$.

The choice of harmonic versus arithmetic mean weights impacts the precision and bias of the WLS-P and AMW estimators.  \cite{imai_essential_2009} note that the harmonic mean weights down-weight pairs that have less well-matched cluster sizes, which can improve precision. However, the WLS-P estimator is still biased even when the cluster level average treatment effect is equal within each pair, as long as the cluster sizes differ, unlike the AMW estimator.
See \cite{imai_essential_2009} for a detailed discussion of the trade-offs between these weights.

\subsection{Additional Approaches}

While the four estimators above are the focus of our comparisons in this paper, it is worth noting additional baseline estimation approaches developed in the literature. \cite{donner_statistical_1987} provided early insights in the analysis of pCRTs through permutation testing or with a weighted $t$-test. Multilevel models using random effects for pairs, rather than fixed-effects, are common in practice. However, this approach does not align with the fixed-sample potential outcomes framework, so we do not include it in our main discussion.

\cite{small_randomization_2008} invert randomization tests using a Wilcoxon Signed-Rank test to calculate a Hodges-Lehman estimator and construct confidence intervals, assuming a constant additive or multiplicative treatment effect. \cite{zhang_powerful_2012} improves upon the power of \cite{small_randomization_2008} by proposing a different test statistic for the same type of permutation testing.

\section{A General Framework for Design-Based Estimation}
\label{sec:cploop}
 The four estimators in the previous section can be thought of as estimates from OLS or WLS regression with different models of the outcome or different weighting schemes. If a researcher was willing to assume one of these outcome models, the choice of an estimator may be clear. However, from a design-based and fixed-population perspective, it is not obvious which of these is preferable.

In this section, we present a general framework for estimation in pCRTs that offers a new, intuitive, view on the distinctions between these baseline estimators. This framework is closely related to those of \cite{middleton_unbiased_2015} and \cite{wu_design-based_2021}, the former focusing on blocked cluster-randomization and the latter on paired randomization. Below, we first articulate the estimation framework (\ref{sec:frame}), then propose a baseline estimator based on this framework (\ref{sec:loomi}), and finally discuss insights gained from the framework for estimation without covariate adjustment (\ref{sec:relation}).

\subsection{Estimation Framework}\label{sec:frame}

Our framework takes the approach of adjusting the Horvitz-Thompson estimator to gain precision. This approach has antecedents in a broad literature for robust design-based covariate adjustment starting with \cite{robins_estimation_1994} (see [removed for blinding] for a review). 

Consider an estimator of the form:
\begin{equation}\label{eq:adj}
    \hat{\tau}^a =  \frac{1}{N}\sum_{i=1}^M(2T_i - 1)\left \{2(\Yo - \Yt)  - a \right \}
\end{equation}
The goal is to find an adjustment $a$ that most improves precision. First, it is helpful to define two quantities:
$$\vio = \no\cdot\ybot - \nt\cdot\ybtc  \mbox{ and } \vit = \nt\cdot\ybtt - \no\cdot\yboc.$$ 
These quantities are the differences in cluster-total treatment and control potential outcomes between clusters one and two. Importantly \textit{either} $\vio$ \textit{or} $\vit$ is observed (if $T_i = 1$ or $T_i = 0$, respectively). Additionally, $\vio + \vit = \tau_i$ (as a reminder $\bar{\tau} = \frac{1}{N}\sum_{i=1}^M\tau_i$). Consider the difference of these total potential differences: $$d_i = \vio - \vit.$$ If we use $d_i$ as the adjustment, $(2T_i - 1)\{2(\Yo - \Yt)  - d_i\}  = \tau_i$ whether $T_i = 1$ or $0$, and thus $\V[\hat{\tau}^{d_i}] = 0$. Therefore, adjusting with $d_i$ minimizes the variance as much as possible. However, $d_i$ is not observed as only one of $\vio$ or $\vit$ is observed for each pair, so it must be imputed. 

Let $\hat{d}_i$ denote some estimate of $d_i$, so we can construct the estimator:
\begin{equation}\label{eq:idpd}
    \hat{\tau} =  \frac{1}{N}\sum_{i=1}^M(2T_i - 1)\{2(\Yo - \Yt)  - \hat{d}_i\}.
\end{equation}
We will refer to this estimator as the ``imputed difference of potential differences estimator'' (IDPD) for the remainder of the paper. This estimator can also be viewed as $\hat{\tau} = \frac{1}{N}\sum_{i=1}^M\thi$, where $\thi \equiv (2T_i - 1)\{2(\Yo - \Yt)  - \hat{d}_i\}$ is an estimator of $\tau_i$.

The IDPD estimator is similar to the augmented inverse propensity weighted (AIPW) estimator developed by \cite{robins_estimation_1994,robins_analysis_1995, scharfstein_adjusting_1999}, but adapted to pCRTs and with a known propensity score due to the randomized design. Additionally, the IDPD estimator is both an extension and special case of the Des Raj difference estimator proposed by \cite{middleton_unbiased_2015}. \cite{middleton_unbiased_2015} suggest fitting specific linear models for choosing the adjustment $a$, the result of which could be thought of as adjusting with a specific estimate of $d_i$. See Supplement A for a detailed explanation of the relationship between the IDPD estimator and the Des Raj difference estimator. 

By articulating this type of estimator as the IDPD estimator in Equation~\ref{eq:idpd} above, flexibility is afforded for how to impute $d_i$. In addition, the IDPD estimator  ultimately reveals a general framework for baseline estimation of the ATE in pCRTs, as described in Section~\ref{sec:relation} below. 

The IDPD estimator is unbiased if $\hat{d}_i \perp T_i$. A way to achieve this independence is sample-splitting \citep{aronow_class_2013, middleton_unbiased_2015,wager_high-dimensional_2016, chernozhukov_doubledebiased_2018, wu_loop_2018, wu_design-based_2021}, for example, estimating $d_i$ in a \textit{leave-one-out} (LOO) manner, using all of the other pairs, excluding pair $i$, to estimate $d_i$. When  $\hat{d}_i \perp T_i$, the variance of the IDPD estimator is $\V[\hat{\tau}]  = \frac{1}{N^2}\left( \sum_{i=1}^M \mbox{MSE}(\hat{d}_i) + \sum_{i \neq j} \gamma_{i,j}   \right),$ 
where $\gamma_{i,j} = \mbox{Cov}(\thi, \tauh_j)$ (See Supplement B). 
The sum of covariance terms are asymptotically negligible compared to the sum of the mean squared errors (MSEs) of the $\hat{d}_i$'s under weak conditions (Supplement B). Thus, essentially, the better $\hat{d}_i$ estimates $d_i$ (in a MSE sense), the smaller the variance of the IDPD estimator. 

Consider how to estimate $\hat{d}_i$ with cluster aggregated outcomes. Recall that $d_i = \vio - \vit = \no\ybot - \nt\ybtc - (\nt\ybtt - \no\yboc)$. Thus, $d_i$ is a function of cluster \textit{total} potential outcomes. Given that cluster totals may vary largely due to different cluster sizes, it is helpful to consider how $d_i$ can be decomposed to separate the cluster sizes from the cluster-mean potential outcomes:
\begin{align}
    d_i 
        =& \underset{\vio}{\underbrace{(\no - \nt)\frac{\ybot + \ybtc}{2} + (\no + \nt)\frac{\ybot - \ybtc}{2}}} \nonumber \\
        &- \underset{\vit}{\underbrace{\left \{ (\nt - \no)\frac{\ybtt + \yboc}{2} + (\no + \nt)\frac{\ybtt - \yboc}{2}\right \}}} \nonumber \\
        &= (\no - \nt)\underset{A_i}{\underbrace{\frac{\ybot + \ybtc +  \ybtt + \yboc}{2}}} + (\no + \nt)\underset{B_i}{\underbrace{\frac{(\ybot  - \ybtc)-(\ybtt - \yboc)}{2}}}  \label{eq:decomp}
\end{align}
Given this decomposition, it is logical to estimate $d_i$ by estimating the average sums $\left ( A_i = \frac{\ybot + \ybtc +  \ybtt + \yboc}{2} \right)$ and differences $\left (B_i = \frac{(\ybot  - \ybtc)-(\ybtt - \yboc)}{2} \right)$ of cluster-mean potential outcomes within a pair. 

Any method or model could be used to impute $d_i$. To denote how imputation is done with the IDPD estimator, we will write $\hat{\tau}[\cdot, \cdot]$ where the first element of the brackets indicates the imputation strategy and the second indicates the covariates used.  In practice, we would likely want to include covariates for this estimation, in particular, to estimate $B_i$. However, we are going to hold that thought for now, because we gain important insights for analyzing pCRTs if we first consider imputing $d_i$ without covariates.

\subsection{Leave-one-out Mean Imputation Estimator}
\label{sec:loomi}  
Consider estimating $d_i$  using leave-one-pair-out mean imputation with the observed outcomes, based on the the decomposition shown in Equation~\ref{eq:decomp}.  Using leave-one-pair-out mean imputation, we estimate $A_i$ with $\frac{1}{M-1} \sum_{j \neq i}  (\Yjbo + \Yjbt)$. Applying this same logic, estimate $B_i$ with $\frac{1}{M-1} \sum_{j \neq i}  \frac{(\Yjbo - \Yjbt) - (\Yjbo - \Yjbt)}{2} = 0$. 

After some algebra, the resulting estimator is an unbiased and scale invariant baseline estimator of the ATE: 
\begin{equation}\label{eq:loo-mi}
    \loomi =  \frac{1}{N}\sum_{i=1}^M(2T_i - 1)\left[2(\Yo - \Yt)  -  \frac{(\no - \nt)}{(M-1)} \sum_{j \neq i}  (\Yjbo + \Yjbt) \right]
\end{equation}
(see Supplement C for details). We will call this estimator the leave-one-out mean imputation (LOO-MI) estimator, and denote it $\loomi$ (representing imputing $d_i$ with \textit{leave-one-out mean imputation} and no covariates, $\emptyset$).

\subsection{LOO-MI Estimator: An Unbiased H\'ajek Estimator}
\label{sec:hajek}  

While most clearly related to the Horvitz-Thompson estimator, the LOO-MI estimator can also be thought of as an unbiased version of the H\'ajek estimator and is more similar to this estimator than may initially appear. As shown in Section~\ref{sec:hajek} and Equations~\ref{eq:hty}-\ref{eq:hthj}, the H\'ajek estimator is the difference between the ratio of Horvitz-Thompson estimators. The LOO-MI estimator can be represented as the difference between the ratio of \textit{adjusted} Horvitz-Thompson estimators. 

First, replace the typical Horvitz-Thompson estimator of the treatment total potential outcomes (Eq.~\ref{eq:hty}) with an adjusted Horvitz-Thompson estimator of the form:
$$\tilde{Y}(t) = \sum_{i=1}^M 2\{T_i\Yo +(1-T_i) \Yt\} + (1-2T_i) \frac{(n_{1, i} - n_{2, i})}{(M-1)} \sum_{j \neq i} \{T_j \Yjbo +(1-T_j) \Yjbt\}.$$
The adjustment term $(1-2T_i) \frac{(n_{1, i} - n_{2, i})}{(M-1)} \sum_{j \neq i} \{T_j \Yjbo +(1-T_j) \Yjbt\}$ can be thought of as an average outcome term $\frac{1}{(M-1)} \sum_{j \neq i} \{T_j \Yjbo +(1-T_j) \Yjbt\}$ weighted by $(n_{1, i} - n_{2, i})$ which vanishes when the cluster sizes are equal.

Then, define $\tilde{N}(t)$ analogously (replacing $\Yo$ with $\no$ and  $\Yt$ with $\nt$), so,
\begin{align*}
    \tilde{N}(t) &= \sum_{i=1}^M  2\left\{(1-T_i)\no + T_i \nt\right\} + (2T_i-1) \frac{(n_{1, i} - n_{2, i})}{(M-1)} \sum_{j \neq i} \left\{(1-T_j) \frac{n_{1j}}{n_{1j}} +T_j \frac{n_{2j}}{n_{2j}}\right\}\\
    &= N
\end{align*}
after some algebra. Define $\tilde{Y}(c)$ and $\tilde{N}(c)$ analogously. Then, the LOO-MI estimator can be written in the form of a H\'{a}jek estimator:
$$\loomi = \frac{\tilde{Y}(t)}{\tilde{N}(t)} - \frac{\tilde{Y}(c)}{\tilde{N}(c)}.$$ Detailed calculations are included in Supplement C. Note that the adjusted Horvitz-Thompson estimators for the total sample sizes  $\tilde{N}(c) = \tilde{N}(t)  = N$ are actually fixed rather than random. 

As is more clearly illustrated in the following section, there is a small trade-off in precision and bias between the H\'ajek and LOO-MI estimators. However, these two estimators are ultimately very similar and in fact asymptotically identical under the regularity conditions discussed in Supplement G.

\subsection{Insights from Estimation Framework} 
\label{sec:relation}
In addition to presenting an unbiased H\'ajek estimator, or a more-precise, scale-invariant, Horvitz-Thompson estimator, the IDPD estimator presents a framework that sheds light on the similarities and distinctions between the baseline point estimators discussed thus far. The primary insight will be that the point estimators discussed in Section~\ref{sec:review} can all be represented as $$\hat{\tau}[\cdot,\emptyset] = \frac{1}{N}\sum_{i=1}^M(2T_i - 1)\{2(\Yo - \Yt)  - \hat{d}_i\},$$ with different imputations $\hat{d}_i$ of $d_i$. As a reminder, the first element of the brackets $[\cdot,\emptyset]$ indicates the imputation strategy and the second indicates the covariates used (which is none in all of these estimators). This presents a general framework for estimation with pCRTs, summarized in Table~\ref{tab:compare}. 

A second insight is that the decomposition of $d_i$ in Equation~\ref{eq:decomp} plays an important role in understanding the relationship between the estimators. $d_i$ can be thought of as containing two elements: the difference in cluster sizes times the sum of cluster-mean potential outcomes in pair $i$ (\hlcol{spiritblue!80}{$A_i$}, highlighted in blue at the top of Table~\ref{tab:compare}), and the sum of cluster sizes times a difference in cluster-mean potential differences in pair $i$ (\hlcol{maize!60}{$B_i$}, highlighted in yellow). The baseline estimators primarily differ in how they estimate the \hlcol{spiritblue!80}{$A_i$}, the sum of cluster-mean potential outcomes. 

\begin{table}[ht]
\centering
\renewcommand{\arraystretch}{1.3}
    \begin{tabular}{lllc}
\multicolumn{4}{c}{$d_i = (\no - \nt)\overset{A_i}{\mathdisplaycolor{spiritblue!80}{\frac{\ybot + \ybtc + \ybtt + \yboc}{2}}} +  (\no + \nt)\overset{B_i}{\mathdisplaycolor{maize!60}{\frac{(\ybot - \ybtc) - (\ybtt - \yboc)}{2}}}$} \\
        \multicolumn{1}{c}{\textbf{Estimator}}& \multicolumn{1}{c}{\textbf{$\hat{\tau}$[$\cdot, \emptyset$]}} &  \multicolumn{1}{c}{$\hat{d}_i$} &  \multicolumn{1}{c}{\textbf{Eq.}} \\
        \hline
        LOO-MI & \textbf{LOO} MI & $(\no - \nt)\mathcolorbox{spiritblue!80}{\frac{1}{(M-1)} \sum_{j \neq i}  (\Yjbo + \Yjbt)}$ & (\ref{eq:loo-mi})\\
         Horvitz-Thompson & 0 &  0 & (\ref{eq:ht})\\
         H\'{a}jek (WLS) &  \textbf{LAI} wMI & $(\no - \nt)\mathcolorbox{spiritblue!80}{\Big(\frac{\hat{Y}(t)}{\hat{N}(t)} + \frac{\hat{Y}(c)}{\hat{N}(c)}\Big)}$ & (\ref{eq:hj})\\
         AMW & \textbf{LOI} MI &  $(\no -\nt)\mathcolorbox{spiritblue!80}{(\Ybo + \Ybt)}$ & (\ref{eq:amw})\\
        WLS-P  & \textbf{LOI} MwDI &  $(\no - \nt)\mathcolorbox{spiritblue!80}{(\Ybo + \Ybt)}$ + & (\ref{eq:fe}) \\
        & & $(\no + \nt)\mathcolorbox{maize!60}{\big\{(\Ybo - \Ybt) -\frac{w_iN}{Wn_i}(\Ybo - \Ybt)\big\}}$ & \\
        \hline
    \end{tabular}
    \centering
    \caption{Summary of baseline estimators for the ATE under the IDPD framework.  In the final row, $n_i = \no + \nt$, $w_i = \frac{\no\nt}{\no+\nt}$, and $W = \sum_{i=1}^Mw_i$.}
    \label{tab:compare}
\end{table}

Table~\ref{tab:compare} summarizes the estimation framework: it  includes the estimate of $d_i$, our notation for the imputation strategy, and a reference to the original equation in this paper for each estimator. See Supplement D for detailed calculations. 

The Horvitz-Thompson estimator is the trivial case of estimating $d_i$ with 0, so we write $\hat{\tau}^{HT}[\emptyset] = \hat{\tau}[0,\emptyset]$. The H\'ajek estimator is an IDPD estimator, estimating the sum of cluster-mean potential outcomes (\hlcol{spiritblue!80}{$A_i$}) in each pair with a weighted average of the observed outcomes across all pairs. Therefore, we can write $\hajek = \hat{\tau}[\mbox{LAI wMI}, \emptyset]$ to indicate leave-\textbf{all-in} (LAI), weighted mean imputation. Notably, this weighted average is a  H\'ajek estimator of the \hlcol{spiritblue!80}{$A_i$}. The AMW estimator uses leave-one-\textbf{in} (LOI) instead of leave-one-\textbf{out} (LOO) mean imputation, estimating the sum of cluster-mean potential outcomes in pair $i$ with the sum of the observed mean outcomes \textit{in} pair $i$. Therefore, we write $\hat{\tau}^{AMW}[\emptyset] = \hat{\tau}[\mbox{LOI MI}, \emptyset]$. The WLS-P estimator incorporates an estimate of the difference in cluster-mean potential differences (\hlcol{maize!60}{$B_i$}) in addition using LOI mean imputation. Therefore, we can write $\wlsp= \hat{\tau}[\mbox{LOI MwDI}, \emptyset]$, where ``wD'' alludes to the weighted estimate for the difference in cluster-mean potential differences. This notation emphasizes the different imputation strategies, however we will continue to refer to the estimators in the same manner as previously in the paper, in the following discussion.

First, this framework makes it clear that if the two cluster sizes within each pair are equal (i.e., $n_{1,i} = n_{2,i}$ for all pairs $i=1, \dots, M$), then all of the point estimators are equivalent (see also Supplement E). In other words, all estimators simplify to the simple difference in means between the treatment and control groups if the cluster sizes within each pair are perfectly matched. \cite{imai_essential_2009} noted that the AMW and WLS-P estimators are unbiased when cluster sizes are equal within each pair, and it is worth noting that this is the setting when weighting is less critical and all estimators simplify to the difference-in-means estimator.

In practice, cluster sizes are not always well matched. From a matching perspective, perfectly matching covariates within each pair may not be feasible nor desirable. If the cluster sizes within pairs are not well matched, this framework can clarify the bias-variance trade-off between these estimators as we discuss below. For the sake of this discussion, let us set aside the difference between the AMW and WLS-P estimators. Then, the estimators are distinguished by using either L\textbf{OI} (AMW and WLS-P), L\textbf{OO} (LOO-MI), or L\textbf{AI} (H\'ajek) mean imputation.

Consider first the variance. The variance of the IDPD estimator is related to the accuracy of the imputation $\hat{d}_i$. More specifically, in the case of LOO and LAI imputation, the variance is a function of the MSE of $\hat{d}_i$ and in the case of LOI imputation the variance is a function of the squared bias of $\hat{d}_i$ (see Supplement B).  Thus, if the variation in cluster-mean potential outcomes between pairs is small and the average treatment effects are homogeneous across pairs, then using LOI, LOO, or LAI imputation should perform similarly. However, if the potential outcomes highly vary between pairs, or if the treatment effect is heterogeneous, LOI imputation will be most accurate.  Therefore, in that setting, the AMW and WLS-P estimators will have precision gains over the H\'ajek and LOO-MI estimators. Additionally, the differences in prediction accuracy for the \hlcol{spiritblue!80}{$A_i$}'s will be magnified if there are large differences in the cluster sizes within each pair.

On the other hand, the bias of estimators in this framework is $-\frac{2}{N}\sum_{i=1}^M \mbox{Cov}(T_i,\hat{d}_i)$ (see Supplement B). 
If there is a heterogeneous treatment effect \textit{within} pairs, then the AMW and WLS-P estimators will be biased, due to the fact that these estimators use LOI imputation and $\hat{d}_i$ is thus highly dependent on $T_i$.  On the other hand, since the H\'ajek estimator relies on the entire sample to estimate $d_i$ (LAI), the dependence between  $\hat{d}_i$ and $T_i$ is relatively small and disappears asymptotically (as does the bias of the estimator). This framework provides more intuition that the bias of the H\'ajek estimator is smaller than the AMW and WLS-P estimators. 

A final observation is that the framework emphasizes the importance of covariate adjustment. \hlcol{spiritblue!80}{$A_i$} can be thought of as an adjustment for there being \textit{different cluster sizes} within pairs, while \hlcol{maize!60}{$B_i$} is what one would normally consider as covariate adjustment. To provide some intuition, note that \hlcol{spiritblue!80}{$A_i$} disappears if the cluster sizes in a pair are equal and \hlcol{maize!60}{$B_i$} is equivalent to the adjustment that is made for paired experiments in \cite{wu_design-based_2021}. Thus, the baseline estimators primarily differ in how they account for different cluster sizes within pairs.  Covariate adjustment can improve precision by estimating \hlcol{maize!60}{$B_i$}, which is difficult to estimate using only observed outcomes, as will be discussed further in the following section.

\section{Covariate Adjustment} \label{sec:cov}
Paired experiments are intended to improve precision and covariate-balance. However, imbalance in prognostic characteristics can remain since there may be few pairs and it is difficult to perfectly align covariates within pairs. Therefore, covariate adjustment can provide precision gains in pCRTs \citep{fogarty_regression-assisted_2018, wang_model-robust_2023, fda_adjusting_2023}.

Further, the estimation framework presented in the previous section provides insight into how covariate adjustment can potentially alleviate the bias-variance trade-off of various estimators of the ATE with pCRTs. It is often likely that an analyst has access to some set of covariates that explain the same variability in experimental outcomes that is mitigated by pair matching, such as the (common) setting where pairs of clusters were formed by matching on baseline covariates. Because these covariates capture the same information that is gained from the paired structure, using them for covariate adjustment with the LAI or LOO estimators (H\'ajek and LOO-MI respectively) could potentially result in precision similar to the LOI estimators (AMW and WLS-P) with much smaller (or no) bias.

For the remainder of this section, we describe methods for covariate adjustment to analyze pCRTs.

\subsection{Covariate Adjusted Horvitz-Thompson Estimator}

\cite{su_model-assisted_2021} propose adjusting the regression formulation for the Horvitz-Thompson estimator using the approach of \cite{lin_agnostic_2013}, although the authors analyze this estimator under simple cluster randomization rather than paired cluster randomization (see Supplement A). The covariate-adjusted LOO IDPD estimator described further below (\ref{sec:covlooidpd}) and the Des Raj difference estimator presented in \cite{middleton_bias_2008} are also both covariate adjusted Horvitz-Thompson estimators. 

\subsection{Covariate Adjusted H\'ajek and WLS-P Estimators}

The most common covariate adjusted estimation strategies for the ATE in pCRTs is to simply add covariates to the regression formulations of the H\'ajek and WLS-P estimators. In other words, for WLS-P, the estimate for the treatment assignment coefficient ($\beta$) in the regression model: 
\begin{equation}\label{eq:per}
    \bar{Y}_{ik} = \alpha_0 + \beta\Tk + \bm \theta ' \bm x_{ik}
 + \sum_{p=1}^M\tilde{\alpha}_pI_{ik,p} +\varepsilon_{ik}
\end{equation} 
weighted by $\nk$, were $\bm x_{ik}$ is a vector of cluster-level covariates. We will denote this estimator with covariates as $\hat{\tau}^{WLS-P}[\bm x]$. Without pair effects ($\tilde{\alpha}_p=0$), the estimated coefficient on the treatment assignment is a covariate-adjusted H\'ajek estimator, which we denote $\hat{\tau}^{WLS}[\bm x]$.

\subsection{Covariate Adjusted Leave-One-Out IDPD Estimator}\label{sec:covlooidpd}


We can straightforwardly incorporate covariate adjustment in the IDPD estimator without introducing bias nor modeling assumptions. As we previously discussed, for the IDPD estimator to be unbiased, the imputation for pair $i$ need only be independent of $T_i$. Therefore, we consider imputing $d_i$ in a leave-one-pair-out manner.  Otherwise, any model or algorithm could be used, including more sophisticated machine learning methods and non-linear models. Importantly, the model need not be correctly specified. We will generally refer to this estimation strategy as the LOO IDPD estimator, denoted, $\looidpd$. So for example, if WLS is used for imputation with covariates $\bm x$, we write $\hat{\tau}[\mbox{LOO WLS},\bm x]$. Practitioners may consider imputing the cluster-mean potential outcomes ($\ybot,\yboc,\ybtt\mbox{, and }\ybtc$) separately, directly imputing $\vio$ and $\vit$, or interpolating between these two strategies to estimate $d_i$. See \cite{wu_design-based_2021} for a detailed discussion of different imputation strategies. 

\subsection{Additional Approaches to Covariate Adjustment with pCRTs}

To the best of our knowledge, there are no direct discussions of how to incorporate covariate adjustment with the AMW estimator currently in the literature. \cite{imai_essential_2009} do not explicitly provide an extension of the AMW estimator to incorporate covariate adjustment. That said, a natural approach could be to simply add covariates to the regression formulation of the AMW estimator.

Outside of adjusted versions of the estimators discussed in this chapter, there are a few additional proposals for analyses of pCRTs with covariate adjustment. \cite{small_randomization_2008} consider a number of covariate adjustment approaches including that of \cite{rosenbaum:2002a}, replacing raw outcomes with model residuals in their randomization-inference based estimators. Finally, \cite{wu_estimation_2014}
propose an estimator that corrects for imbalances in the baseline covariate distributions between the two clusters in each pair by using a specific weighted average outcome within each cluster.  

A final note on covariate adjustment with pCRTs is that, in addition to the typical baseline covariates we may think of, \cite{middleton_unbiased_2015} and \cite{su_model-assisted_2021}, note that the cluster sizes can be a useful covariate in CRT analysis. These authors use the cluster size as a covariate for adjustment (not simply as a weighting factor), noting that the cluster size is a highly predictive covariate for cluster-total outcomes. The cluster size may also be predictive of cluster-mean outcomes, so it could be a worthwhile covariate to include in any adjustment, and can be a first step to improving precision beyond that of the baseline estimators.

\section{Variance Estimation}
\label{sec:var}
Thus far, we have compared point estimators for the ATE under paired cluster-randomization, which presents a number of challenges and trade-offs. Variance estimation under paired cluster-randomization is also challenging, and can distinguish between these different approaches in practice. 

The four baseline estimators discussed in Section~\ref{sec:review} can all be calculated using weighted regression models \citep{imai_essential_2009, su_model-assisted_2021}. Thus, one option for variance estimation is using the associated parametric variance estimator for the appropriate regression coefficient. There is evidence that typical WLS variance estimators can be anti-conservative for the true variance of these point estimators under cluster-randomization (both in the design-based and model-based sense), so some authors have recommended using Huber-White heteroskedasticity robust variance estimators \citep{middleton_unbiased_2015, middleton_unifying_2021, su_model-assisted_2021}. These variance estimators for regression coefficients are based on an outcome model that does not typically align with the fixed-sample potential outcomes causal framework.

On the other hand, the literature on design-based variance estimation for pCRTs is relatively underdeveloped. There has been more focus on variance estimation under blocked cluster-randomized experiments, where there are at least three clusters in each block. In this case, a within-block variance can be estimated, and averaged across blocks \citep{middleton_unbiased_2015, pashley_insights_2021, schochet_design-based_2021}. However, these approaches typically require at least two treatment or control clusters within each block, so do not extend to paired cluster-randomization. Only recently has a design-based variance estimator for the H\'ajek estimator been proposed for paired cluster-randomization \citep{de_chaisemartin_at_2020}, which is telling given the ubiquity of the difference-in-means estimator for treatment effect estimation. 

In this section, we present design-based variance estimators for the point estimators previously discussed, with unified notation. The end of the section discusses these variance estimators.

\subsection{Variance Estimation for the Horvitz-Thompson Estimator}

To the best of our knowledge, the only design-based variance estimator previously proposed for the Horvitz-Thompson estimator under paired cluster-randomization is by \cite{middleton_unbiased_2015}:
\begin{equation}\label{eq:mavar}
\hat{\V}^{MA}\big[\hvt \big]=\frac{16M^2\hat{\sigma}^2(\Yk)}{N^2(2M-1)}
\end{equation}
where $\hat{\sigma}^2(\Yk)$ is an estimate of the variance of the cluster total outcomes. We note that \cite{middleton_unbiased_2015} assumes a sharp null of no treatment effect to derive this variance estimator.

\subsection{Variance Estimation for the H\'ajek Estimator}

 \cite{de_chaisemartin_at_2020} developed a design-based variance estimator by analysing the H\'ajek estimator's regression equivalent, from a design-based perspective. Their main theoretical results assume that the cluster sizes are all equal, although we find in simulation that the estimator performs well when cluster sizes vary as well. Define cluster-level residuals $\tilde{r}_{ik} \equiv Y_{ik} - n_{ik}\big[T_{ik}\frac{\hat{Y}(t)}{\hat{N}(t)}+(1-T_{ik})\frac{\hat{Y}(c)}{\hat{N}(c)}\big]$. In other words, $\tilde{r}_{ik}$ is the residual of the cluster-total outcome from a H\'ajek estimate of the cluster-mean treated or control  outcome (given cluster $ik$'s treatment assignment), times the cluster size. Then,

\begin{equation}\label{eq:crhjvar}
\hat{\V}^{dCR}\big[\hajek \big] = \sum_{i=1}^M\bigg(\frac{T_i\tilde{r}_{i1}+(1-T_i)\tilde{r}_{i2}}{\hat{N}(t)} - \frac{ (1-T_i)\tilde{r}_{i1}+T_i\tilde{r}_{i2}}{\hat{N}(c)}\bigg)^2.
\end{equation}

\subsection{Variance Estimation for the WLS-P and AMW Estimators}

Inference for the WLS-P estimator is the most developed. We will discuss three previously-studied design-based variance estimators below. As a reminder, the WLS-P estimator is a weighted sum of differences in average cluster outcomes within pairs: $\frac{1}{W}\sum_{i=1}^M w_i(2T_i -1) (\Ybo - \Ybt)$, where $w_i = \frac{\no\nt}{\no+\nt}$ and $W = \sum_{i=1}^M w_i$. 

\cite{schochet_design-based_2021} propose the following estimator for the variance of the WLS-P estimator:
\begin{equation}\label{eq:vspmk}
\hat{\V}^{SPMK}\big[\wlsp \big] = \frac{2M}{(M-1)W^2} \sum_{i=1}^M w_i^2(r_{i1}^2 + r_{i2}^2),
\end{equation}
where $r_{ik}$ denotes the residuals from the WLS-P regression formulation (Eq. \ref{eq:per}). 
The authors suggest hypothesis testing and confidence intervals using the $t$-distribution with $M-1$ degrees of freedom. This variance estimator can also be used if covariates are included in the regression model.

\cite{imai_essential_2009} and \cite{de_chaisemartin_at_2020} propose variance estimators that can be used for either the WLS-P or AMW estimators, given that the estimators only differ in pair weights. Let $\hat{\tau}^{w'}$ denote this form of estimator where the weights $w_i'$ are replaced with $w_i = \frac{\no\nt}{\no+\nt}$ or $\tilde{w}_i = \no+\nt$ for the WLS-P or AMW estimator, respectively. \cite{imai_essential_2009} and \cite{de_chaisemartin_at_2020} propose the following variance estimators, respectively for this general form of estimator:
\begin{equation}\label{eq:vikn}
    \hat{\V}^{IKN}[\hat{\tau}^{w'}] = \frac{M}{(M-1)} \sum_{i=1}^M \bigg( \frac{w_i'}{W'} \hat{\bar{\tau}}_i - \frac{1}{M}\hat{\tau}^{w'}\bigg)^2
\end{equation}
\begin{equation}\label{eq:vdcr}
\hat{\V}^{dCR}\big[\hat{\tau}^{w'}\big] = \frac{1}{W^{'2}} \sum_{i=1}^M w_i^{'2}\big( \hat{\bar{\tau}}_i - \hat{\tau}^{w'}\big)^2,
\end{equation}
where, as above, $\hat{\bar{\tau}}_i = (2T_i -1) (\Ybo - \Ybt)$, the difference in mean outcomes within pair $i$. \footnote{$\hat{\V}^{IKN}[\hat{\tau}^{w'}]$ \citep{imai_essential_2009} is implemented in the \texttt{ATEcluster()} function in the \texttt{experiment} package \citep{experiment_2022} and the \texttt{difference\_in\_means()} function of the \texttt{estimatr} package \citep{estimatr_2024} in \texttt{R}.}

\subsection{Variance Estimation for the Leave-One-Out IDPD Estimator}

Let  $V_i = 2(T_i\vio + (1-T_i)\vit)$ and $\hat{V}_i = 2(T_i\viho + (1-T_i)\viht)$. We use the following variance estimator for the LOO IDPD estimator:  
\begin{equation}\label{eq:vloop}
\hat{\V}\big[\hat{\tau}[\mbox{LOO } \cdot, \cdot]\big] = \frac{1}{N^2}\sum_{i=1}^M (V_i - \hat{V}_i)^2,
\end{equation}
based on the analysis in \cite{wu_design-based_2021} for paired trials. Under certain regularity assumptions and assuming that the estimates of $\hat{d}_i$ are independent of the treatment assignment, $T_i$, this is a conservative estimator for the variance of the LOO IDPD estimator (see Supplement F for proofs). To conduct inference with the LOO IDPD estimator, we additionally note that under certain regularity assumptions, it is asymptotically normal (see Supplement G).  

In the case of the LOO-MI estimator, this variance estimator simplifies to:
\begin{equation}\label{eq:vloopmi}
\hat{\V}\big[\loomi\big] = \frac{1}{N^2}\sum_{i=1}^M \Big\{(\no+\nt)(\hat{\bar{\tau}}_i- \bar{\bar{\tau}}_{-i}) +
      (2T_i-1)(\no-\nt)(\bar{S}_i- \bar{\bar{S}}_{-i}) \Big\}^2
\end{equation}
where $\hat{\bar{\tau}}_i = (2T_i - 1)(\Ybo-\Ybt)$,  
$\bar{\bar{\tau}}_{-i} = \frac{1}{(M-1)} \sum_{j \neq i} \hat{\bar{\tau}}_j$,
$\bar{S}_i  = (\Ybo+\Ybt)$, and
$\bar{\bar{S}}_{-i}= \frac{1}{(M-1)} \sum_{j \neq i} \bar{S}_j$. See Supplement H for detailed calculations.

\subsection{Comments on Variance Estimators}

We note the similarity of $\hat{\V}^{IKN}[\hat{\tau}^{w'}]$ (\ref{eq:vikn}) to the well-known variance estimator for the difference-in-means in paired randomized experiments, equivalent to that used for a paired $t$-test \citep{imai_variance_2008, athey_chapter_2017}. Indeed $\hat{\V}^{IKN}[\hat{\tau}^{w'}]$ simplifies to the variance estimator for a paired $t$-test \citep{imai_variance_2008} if all of the cluster sizes are equal (i.e. $\no = \nt$ for all $i = 1, \dots, M$). Thus, $\hat{\tau}^{w'}$ and corresponding variance estimator $\hat{\V}^{IKN}[\hat{\tau}^{w'}]$ have the nice property of generalizing to the difference-in-means estimator and the typical corresponding variance estimator when the cluster sizes are all equal. 

The variance estimator $\hat{\V}^{dCR}\big[\hat{\tau}^{w'}\big]$ \citep{de_chaisemartin_at_2020} (\ref{eq:vdcr}) is similar to $\hat{\V}^{IKN}[\hat{\tau}^{w'}]$ but with a different weighting scheme. The variance estimator we use for the LOO-MI point estimator (\ref{eq:vloopmi}) is also of a similar form, but includes a term that takes into account the variation in the sum of observed outcomes between pair $i$ and all of the other pairs, if the cluster sizes vary. When the cluster sizes within each pair are the same, that term drops out. In fact, when cluster sizes within each pair are the same, the variance estimator we use for the LOO-MI estimator, the variance estimator of \cite{imai_essential_2009}, and the variance estimators of \cite{de_chaisemartin_at_2020} are similar, only differing by some weighting factors (see Supplement I for details).

\section{Simulation Studies}
\label{sec:sims}
We conduct simulation studies to evaluate the reviewed estimators, designed to mimic an educational experiment. The general design is based on an educational trial evaluating the impact of an algebra tutoring intervention on future test performance, where random assignment is at the school-level \citep{pane_effectiveness_2014}.  We designed the simulation studies with simple to more realistic settings for estimation that are still reasonable for a typical pCRT in education.  We identify three major factors that impact the efficiency and bias of the point estimators: (1) how well-matched cluster sizes are \textit{within} pairs, (2) whether there are heterogeneous outcomes \textit{between} pairs,  and (3) whether there are heterogeneous treatment effects \textit{within} pairs, namely when the treatment effect is a function of the cluster size. The treatment effect being correlated with the cluster size provides a very difficult setting for these estimators and is an issue discussed in the CRT literature \citep{bugni_inference_2023}.  In this section, we first provide details of the simulation design, followed by results.

\subsection{Simulation Design}

We generate data to vary these three factors as follows. First, given a number of pairs, M, we generate  cluster (school) sizes as $\nk = n_i + m_{ik}$ where $n_i \sim U\{a,b\}$ and $m_{ik}\sim U\{c,d\}$ ($U\{\}$ representing the discrete uniform distribution). This allows the cluster size to be the same or similar between two clusters in a pair, or to vary completely among clusters, simulating whether cluster sizes are well matched. 

We assume that a single covariate is observed at the cluster-level, which can vary between and within pairs. Thus, we define a cluster-level covariate $x_{ik} = \alpha_i + z_{ik}$ where $\alpha_i\sim N(0, \sigma^2_{\alpha})$ allows for between-pair variability and $z_{ik} \sim N(0, \sigma^2_z)$  allows for within-pair variability that is explained by this covariate ($\alpha_i \perp z_{ik}$). We will refer to the presence of heterogeneous outcomes between pairs as a ``pair effect'' since the pairs themselves can explain some variability in the potential outcomes. We then generate $\nk$ \textit{individual} (student) potential outcomes for each cluster as:
$$y_{ik\ell}^c = \alpha_0 + x_{ik} + \gamma_{ik} + \varepsilon_{ik\ell} \hspace{2cm} y_{ik\ell}^t = y_{ik\ell}^c + \tau_{ik}$$
where $\varepsilon_{ik\ell} \sim N(0,\sigma^2_{\varepsilon})$ introduces individual variability and $\gamma_{ik}\sim N(0,\sigma^2_{\gamma})$ introduces independent, cluster-level variability, neither of which can be explained by the covariate $x_{ik}$. All random variables are independent. Finally, we generate a  cluster-level treatment effect as:
$$\tau_{ik} = \tau_0 + f(n_{ik})$$
where  $f(n_{ik}) = \phi\cdot(n_{ik}-\E[n_{ik}])$. 

This design allows us to control the second and third factors listed above. In defining the covariate, if $\sigma^2_{\alpha} \neq 0$, there is a pair effect while if $\sigma^2_{\alpha} = 0$, no information about the cluster outcomes is gained from the pairing. Additionally, if  $\sigma^2_z \neq 0$, some variance in the outcomes is explained by the cluster-level covariate, in addition to any information gained from the pair. From the generation of the treatment effect, if $\phi = 0$, there is a constant treatment effect, and if $\phi \neq 0$ the cluster size and treatment effect are correlated (but the expected value of the cluster size is always $\tau_0$). 

We choose the simulation parameters to emulate a realistic experiment with an educational intervention and outcomes. A cluster represents one school, where an intervention is applied to students in a certain grade. The average cluster size is 150, but range from 75 to 225. The outcome of interest is a test score. We let the covariate explain 85\% of the variance in the cluster-mean control potential outcome, with the pair element explaining 80\%. This is reasonable if we think about the covariate as a pretest score for the outcome, which is a common adjustment for educational experiments. The treatment effect explains around 15\% of the observed mean outcome. See Supplement J for full simulation details.

In the main text, we will discuss the simulation results for four different settings, listed below. Full results can be found in Supplement L. We choose these settings for the main text to illustrate how the estimators perform in increasingly complex or ``difficult'' settings for estimation, and note that only one of the the three factors changes between each setting (underlined): 
\begin{itemize}
    \item [(\textbf{S1})] There is no pair effect, cluster sizes are well matched, and the treatment effect is constant
    \item [(\textbf{S2})]  There \underline{is a pair effect}, cluster sizes are well matched, and the treatment effect is constant
    \item [(\textbf{S3})]  There is a pair effect, cluster sizes are \underline{not well matched}, and the treatment effect is constant
    \item [(\textbf{S4})]  There is a pair effect, cluster sizes are not well matched, and the treatment effect is \underline{correlated with cluster size}.
\end{itemize}

The simulations are conducted as follows for one data generation and setting. We generate one set of potential outcomes and covariates and 500 treatment assignment vectors. For each treatment assignment vector, we calculate the point (treatment effect) estimates, associated variance estimates, and 95\% confidence intervals. We estimate the finite-population expectation and variance for each point and variance estimator by taking the empirical mean and variances of the estimators across the 500 treatment assignment vectors within each data generation. Finite-population coverage is the proportion of times within these 500 treatment assignments that the true ATE is contained within the confidence interval. We evaluate the variance estimators in terms of what we will call the ``relative bias'' of each variance estimator to the point estimators true variance. This measure is calculated as the empirical mean of the 500 variance estimates, divided by the empirical variance of the 500 point estimates.

We repeat this procedure 400 times and average these 400 finite-population measures for the main results. Standard errors for the simulation results are calculated as the standard deviation of the 400 finite-population simulation estimates divided by $\sqrt{400}$. 

In the main text, we compare the H\'{a}jek ($\hajek$), Horvitz-Thompson ($\hvt$), AMW ($\amw$), WLS-P ($\wlsp$), and LOO-MI ($\loomi$) estimators, for baseline inference with no covariate adjustment. We additionally include $\hat{\tau}^{WLS}[\cdot]$, $\hat{\tau}^{WLS-P}[\cdot]$, and $\hat{\tau}[\mbox{LOO WLS}, \cdot]$ adjusted by the cluster size ($n$) and the cluster size and the covariate ($n, x$). We will use this notation for the estimators for the remainder of the paper. To impute $d_i$ for the LOO IDPD estimator with covariate adjustment, we use leave-one-pair-out WLS with the optimal interpolation between imputing the mean potential outcomes versus directly imputing $\vio$ and $\vio$.\footnote{Options \texttt{(pred = p\_ols\_interp, weighted\_imp = T)} in the \texttt{p\_loop} function of the \texttt{dRCT} package.}

\begin{figure}
  \centering
  \subfloat[]{%
  \includegraphics[width = .9\textwidth]{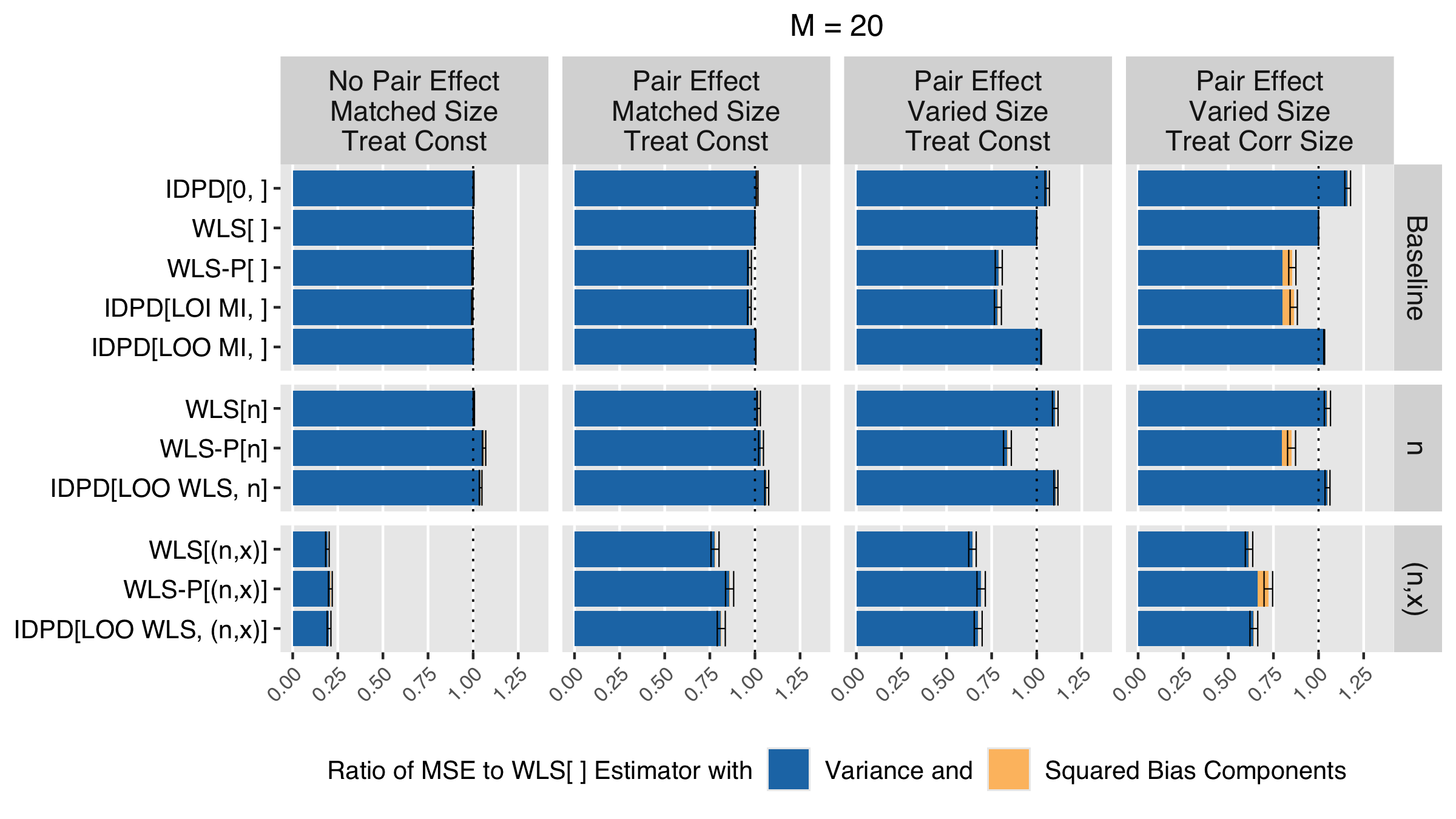}\label{fig:mse-main-m20}}%
  \\
  \subfloat[]{%
  \includegraphics[width = .9\textwidth]{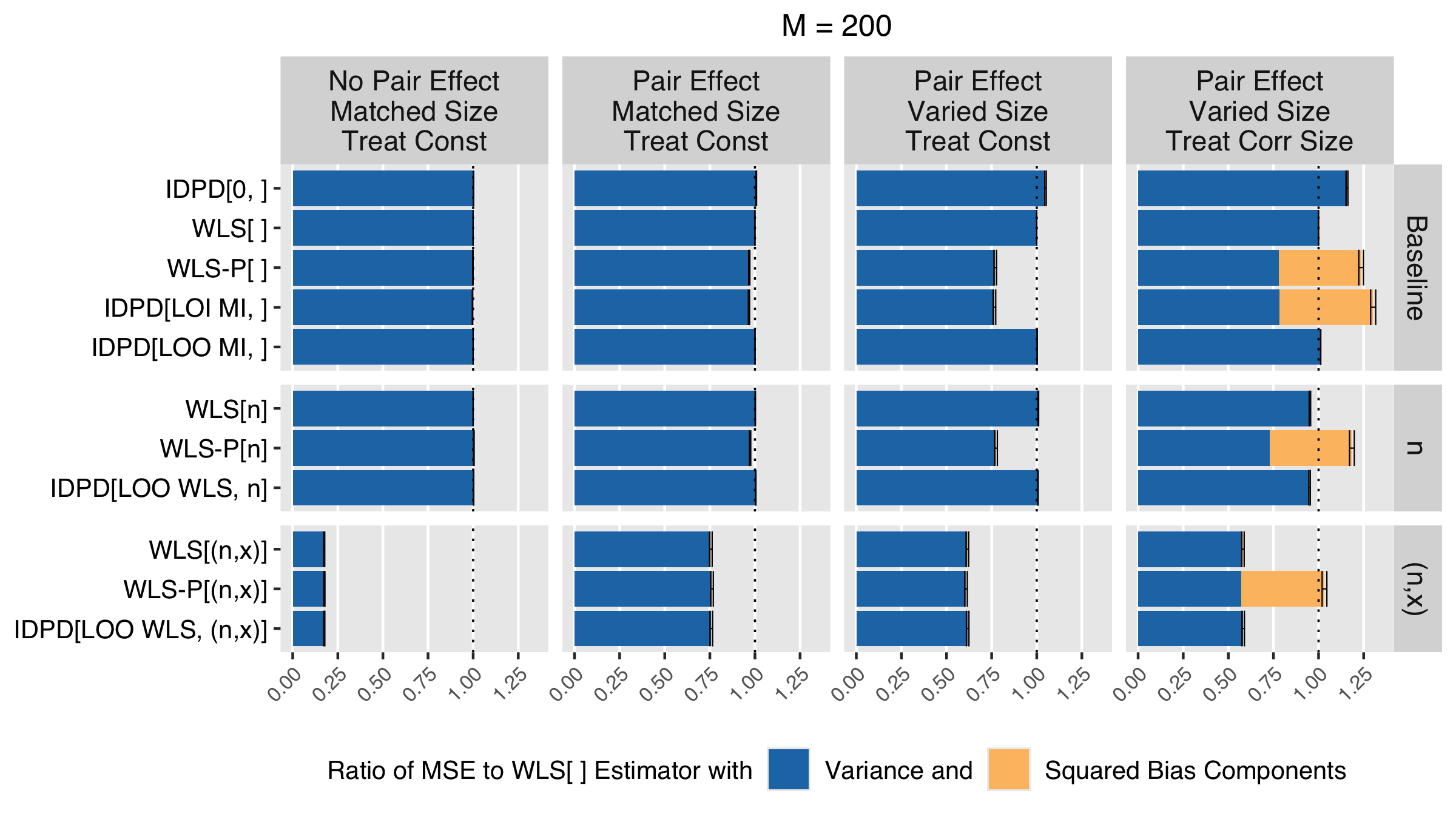} \label{fig:mse-main-m200}%
  }
  \caption{Ratio of the MSE of treatment effect estimators, compared with $\hajek$ (the H\'ajek estimator), when there are (a) $M = 20$ or (b) M = 200 pairs. Facet columns indicate data generation settings and rows indicate the type of covariate adjustment -- none (``Baseline''), with cluster size only (``n''), or with cluster size and the covariate (``(n,x)''). The MSE ratio is partitioned into the variance (left, dark blue) and squared-bias (right, orange) components of the MSE. Error bars represent two simulation standard errors.} \label{fig:AB}
\end{figure}

As discussed in Section~\ref{sec:review}, there are multiple possible variance estimators for certain point estimators, however we show the results for only one in the main text for simplicity (as indicated in the result figure caption). We implemented each of these variance estimators in \texttt{R}, except for the Huber-White robust variance estimator, for which we use the \texttt{sandwich} package \citep{sandwich_2023} with ``HC1'' variance structure.\footnote{For robust variance estimation for the WLS estimators with pCRTs, there is no clear recommendation for which covariance structure to specify. \cite{su_model-assisted_2021} use ``HC0'' with CRTs. There is evidence recommending ``HC3'' for small sample sizes \citep{long_using_2000}. We use ``HC1'' because it is the default parameter in \texttt{STATA}, which is the context in which many practitioners may be using this method.} For 95\% confidence interval estimation, we use the recommended degrees of freedom for a $t$ distribution when available in the literature, and a normal distribution for the LOO IDPD estimators and Horvitz-Thompson estimator. Full simulation results with additional point and variance estimators 
can be found in Supplement L.

\subsection{Simulation Results}

Figures~\ref{fig:mse-main-m20} and~\ref{fig:mse-main-m200} show the ratios of the simulation estimated mean squared error (MSE) of each point estimator as compared to the MSE of $\hajek$. A MSE ratio greater than 1 (to the right of the dotted line) indicates that the estimator has a larger MSE than $\hajek$, while an MSE ratio less than 1 (left of the dotted line) indicates that the estimator has a smaller MSE. The ratio is divided into the variance (dark blue, left) and squared bias (orange, right), with error bars representing two simulation standard errors.

First, consider the baseline estimators when there are a small number of pairs (Figure~\ref{fig:mse-main-m20}, $M = 20$). The results illustrate the first two insights from the estimation framework. First, the baseline estimators perform essentially the same in \textbf{S1}, when there is a constant treatment effect, cluster sizes are well matched, and there is no pair effect. Even when there is a pair effect, (\textbf{S2}), as long as the cluster sizes are essentially the same within pairs, the estimators perform similarly because the estimators are equivalent if the cluster sizes are equal within each pair. Second, allowing the cluster sizes to vary among clusters when there is a pair effect (\textbf{S3}), $\amw$ and $\wlsp$ are the most efficient. When there is a pair effect, the outcomes within each pair vary much less than between pairs. Since $\amw$ and $\wlsp$ use the data within each pair for imputation, the estimate of $d_i$ is more accurate than the other baseline estimators, which use LOO or LAI mean imputation. This difference between using LOI versus LOO/LAI imputation becomes more pronounced between \textbf{S2} and \textbf{S3} because the differences between the cluster sizes within each pair is larger. However, when the treatment is a function of the cluster size, $\amw$ and $\wlsp$ are biased (\textbf{S4}). With a small number of pairs, the bias is small compared to the variance, and therefore $\amw$ and $\wlsp$ are still the most efficient. Across all settings, $\loomi$ and $\hajek$ tend to perform similarly, given their only slight difference in estimation strategy, as discussed in Section~\ref{sec:relation}.

The results with $M = 200$ pairs illustrate the asymptotic behaviors of the estimators (Figure~\ref{fig:mse-main-m200}). The baseline estimators compare similarly to the small sample setting, except for \textbf{S4}, when the treatment effect is a function of the cluster size. The bias of $\amw$ and $\wlsp$ does not disappear asymptotically, and thus the bias outweighs the gains in precision from LOI estimation when there is a large number of pairs. Therefore, when the treatment effect is correlated with cluster size, even with no covariate adjustment, $\hajek$ and $\loomi$ have the smallest MSEs.

When the control potential outcomes are not correlated with the cluster size, and there are a small number of pairs, adjusting with the cluster size actually slightly hurts precision. However, including the cluster size as a covariate does not hurt or slightly improves precision asymptotically ($M= 200$). This aligns with \cite{su_model-assisted_2021}'s analysis of CRTs. Supplement M includes results when the control potential outcomes are correlated with the cluster size, in which case adjusting with the cluster size improves precision.

The estimators that adjust with the cluster size and a prognostic covariate are more precise than the baseline estimators. Figure~\ref{fig:mse-main-m20} also illustrates the point in Section~\ref{sec:cov} -- if a covariate that explains the same effect on the the outcome as the pair matching is used for adjustment, even with a small sample size, $\wlsnx$ and $\looidpdnx$ show comparable or slightly smaller variance than $\wlspnx$, and without any bias. Asymptotically, $\wlspdot$ remains biased, and therefore $\wlsnx$ and $\looidpdnx$ have considerably smaller MSE than  $\wlspdot$.

\begin{figure}
  \centering
  \subfloat[Relative Bias]{%
  \includegraphics[width = \textwidth]{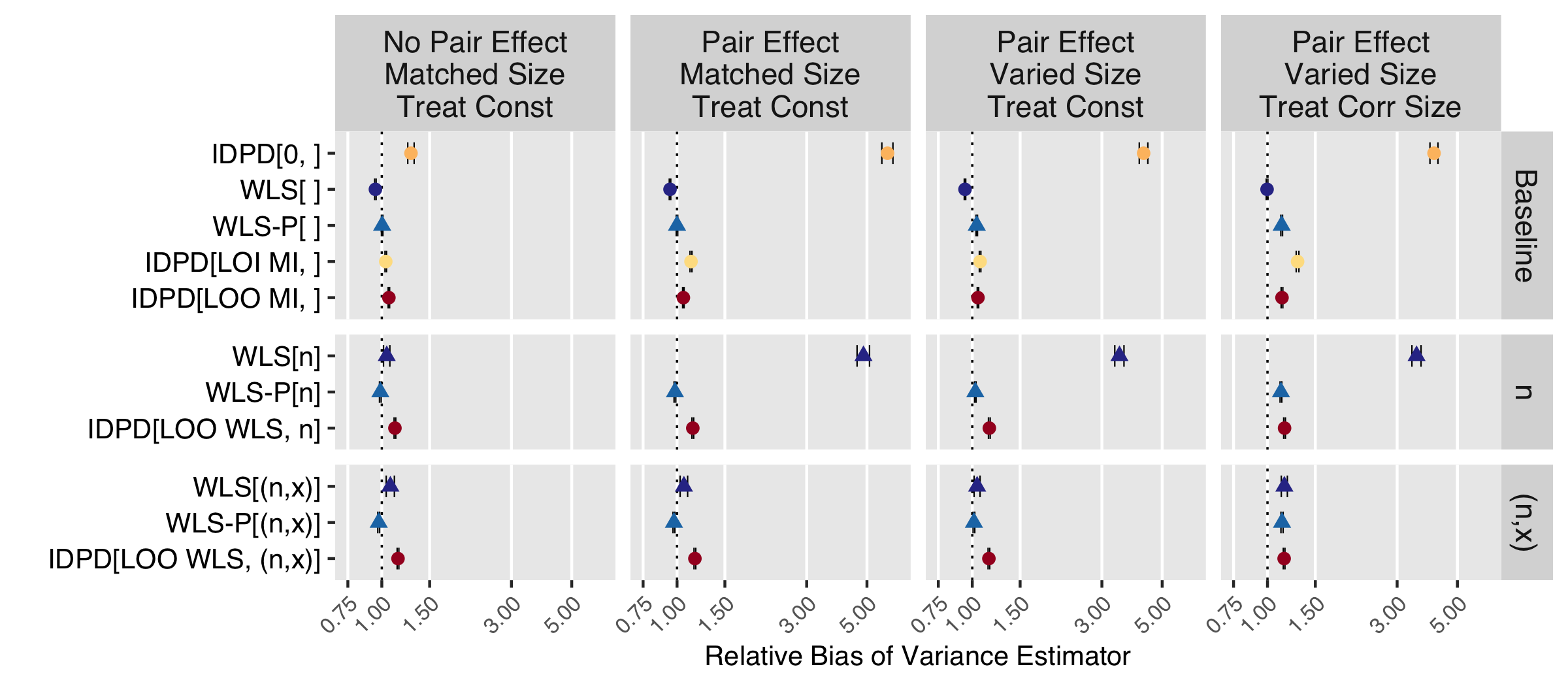}\label{fig:var}}%
  \\
  \subfloat[Coverage]{%
  \includegraphics[width = \textwidth]{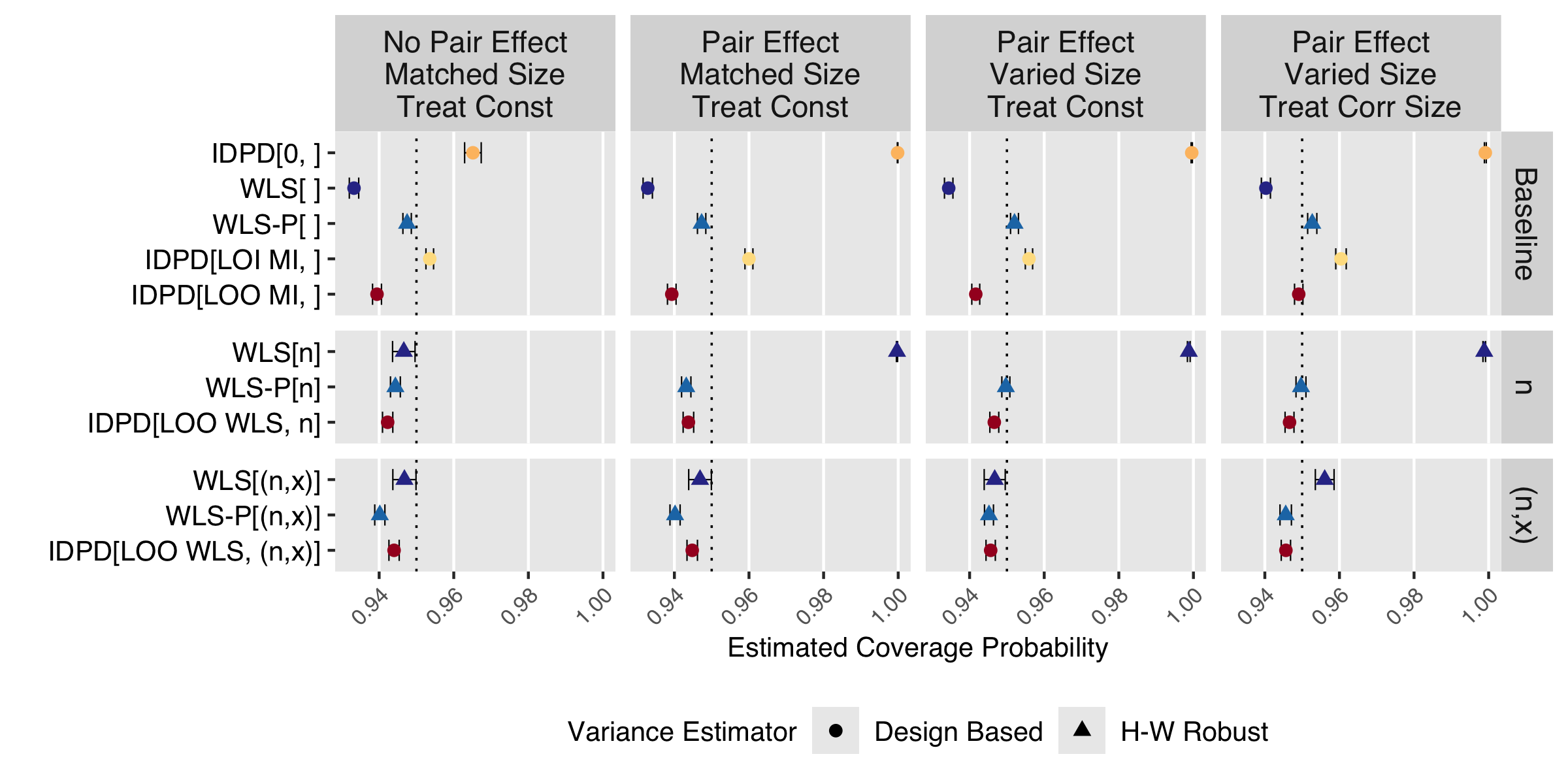} \label{fig:covg}%
  }
  \caption{Simulated relative bias of variance estimators and coverage probability associated with each point estimator when there are $M = 20$ pairs, averaged over 400 data generations. Relative bias is calculated as the empirical mean of the variance estimator divided by the empirical variance of the point estimator across treatment permutations. ``H-W Robust'' refers to the Huber-White heteroskedasticity robust variance estimator with HC1 variance structure. Design-based variance estimators include \citep{middleton_unbiased_2015} (Eq.~\ref{eq:mavar}) for $\hvt$, \citep{de_chaisemartin_at_2020} (Eq.~\ref{eq:vdcr}) for $\hajek$, \citep{imai_essential_2009} (Eq.~\ref{eq:vikn}) for $\amw$, and the estimator presented Eq. \ref{eq:vloop} for the LOO IDPD estimators.}
  \label{fig:var-covg}
\end{figure}

In addition to point estimation, a remaining challenge for inference with pCRTs is variance estimation. Figures~\ref{fig:var} and~\ref{fig:covg} summarize the relative bias of variance estimators and the associated coverage probability for $M = 20$ pairs (results for $M = 200$ can be found in Supplement L). A relative bias greater than 1 (right of the dotted line) indicates that the variance estimator is conservative while a relative bias between 0 and 1 (left of the dotted line) indicates that the variance estimator is anti-conservative. We observe some large differences in the variance estimators, even in cases where the point estimators behave similarly. The design-based variance estimator proposed in \cite{middleton_unbiased_2015} is highly conservative for $\hvt$ except in \textbf{S1}.  Variance estimation for $\hajek$ is slightly anti-conservative when the cluster sizes are well matched and the coverage probabilities show under-coverage when there are a small number of pairs.\footnote{We note that the regression-based cluster-robust variance estimate for $\hajek$ is highly conservative (see Supplement~\ref{chpt:appn_pcrt9}), so we display results for the better-performing variance estimator here.} The \cite{imai_essential_2009} variance estimator for $\amw$ performs variably, depending on the setting. The Huber-White robust variance estimator is also highly conservative for $\wlsn$ when there is a pair effect, although it is only slightly conservative for $\wlsnx$. The Huber-White robust variance estimator for $\wlspdot$ is relatively accurate when cluster sizes are well-matched, and becomes more conservative when cluster sizes vary. As expected, the variance estimator we use for $\looidpd$ is slightly conservative across all settings, most conservative, relative to other settings, when there is covariate adjustment. However, the IDPD variance estimator shows the most consistent results across the settings, while the coverage and relative bias of other estimators depend more heavily on the data setting. 

In addition to these simulation studies, we conducted simulation studies using real data from an educational efficacy trial \cite{pane_effectiveness_2014}. The trail paired schools, and assessed the efficacy of an online algebra tutoring program.  These results are presented in Supplement K. The real-data simulation study results are largely similar to those presented above. The exception is that the Huber-White robust variance estimator is very anti-conservative for the true variance of $\wlsnx$, resulting in under-coverage, when adjusting with a pre-test covariate. On the other hand, the variance estimator for the LOO IDPD estimators remains conservative.

Finally, as noted above, full simulation results, which include additional point and variance estimators are included in Supplement L. The estimators proposed in \cite{middleton_unbiased_2015} and \cite{su_model-assisted_2021} were not included in the main paper here because the variance estimation performed generally worse than other methods across different data generations. Multilevel regression including a random effect for pairs shows a smaller bias than the fixed-effects estimator, but variance estimation tended to be anti-conservative and thus lead to under-coverage.


\section{Discussion}
\label{sec:cploopdisc}
We articulate a general framework for design-based estimation of the individual average treatment effect in paired cluster-randomized experiments. This framework supports a novel viewpoint on the relationships between baseline estimators -- clarifying that they are equivalent when cluster sizes within pairs are equal and providing insight into the bias-variance trade-off in estimation with no covariate adjustment. This framework also emphasises the benefit of incorporating covariate adjustment, even if the only covariates available are the same as those used for constructing pairs; using covariate adjustment in $\looidpd$ or $\wlsdot$ overcomes the original precision loss from LOO/LAI mean imputation, while remaining unbiased. 

Another contribution of this paper is to review estimators for the ATE in pCRTs, with unifying notation. Particularly, design-based variance estimators have not been previously discussed in one source, and few have associated software implementations. By presenting options for variance estimation here, we aim to support comparison and use of these estimators in practice.

These results support a number of practical insights. First, matching clusters closely on size in the design of a pCRT simplifies the choice of estimator in the analysis stage. \cite{imai_essential_2009} strongly encourage matching clusters closely on cluster size, so that $\amw$ (AMW) is unbiased. We echo this suggestion for a broader reason. When cluster sizes are well matched, the baseline estimators are essentially equivalent and thus perform very similarly, regardless of whether there is a pair effect or the treatment effect is a function of cluster size. This is true once adjusting with covariates as well. 

Second, researchers should be aware of the potential drawbacks of analyzing pCRTs with the most common estimator, $\wlspdot$, or a linear regression model with pair fixed effects. Asymptotically, the bias of $\wlspdot$ contributes to a large portion of the estimator's MSE. Thus, if practitioners choose to use linear models for effect estimation, we have shown that you should typically avoid including pair fixed effects in the model.

Relatedly, if clusters were paired using baseline covariates and these covariates are available to the analyst, we suggest that researchers adjust with the covariates that were used to form pairs. While $\wlspdot$ can be the most precise, as long as variability between pairs can be modeled with available covariates, $\looidpd$ or $\wlsdot$ achieve the same precision in addition to being unbiased or less biased. 

Finally, regardless of whether cluster sizes are well matched, $\looidpd$ and $\wlsdot$ perform similarly across data settings. If researchers are comfortable with small fixed-sample bias, $\wlsdot$ is a good option. However, we find that Huber-White robust standard errors are highly conservative with no covariate adjustment (H\'ajek) or adjusting by only the cluster size. If unbiased estimation is a priority for researchers, then they should use $\looidpd$. An additional benefit of $\looidpd$ is that it can accommodate non-linear machine learning models for covariate adjustment. The associated variance estimator is  conservative with a small number of pairs, and only slightly conservative when there are a large number of pairs. 

\section{Software and Supporting Code}

Code supporting results in the main text and appendix can be found at \url{https://github.com/manncz/pcrt-design}. The LOO IDPD estimator and associated variance estimator is implemented in the \texttt{dRCT} package found at \url{https://github.com/manncz/dRCT} \citep{drct_2023}.

\section{Acknowledgements}
The research reported here was supported by the Institute of Education
Sciences, U.S. Department of Education, through Grant R305D210031 to the
University of Michigan. The opinions expressed are those of the authors and do not represent views of the Institute or the U.S. Department of Education nor other funders. Charlotte Z. Mann was additionally supported by the National Science Foundation RTG grant DMS-1646108. 


\bibliographystyle{plainnat}
\bibliography{bibliography}

\newpage
\appendix

\clearpage
\setcounter{table}{0}
\setcounter{figure}{0}

\renewcommand{\thetable}{\thesection\arabic{table}}
\renewcommand{\thefigure}{\thesection\arabic{figure}}
\section{Details of Other Approaches to Adjust the Horvitz-Thompson Estimator}
\label{chpt:appn_pcrt1}

\subsection{Middleton \& Aronow (2015)}

\cite{middleton_unbiased_2015} propose an adjusted Horvitz-Thompson estimator for unbiased estimation of the ATE in blocked cluster randomized experiments. Let $\xk$ denote a vector of cluster total covariates (including the cluster size) for cluster $ik$  and $\bar{\bm x} = \frac{1}{2M}\sum_{i=1}^M \sum_{k=1}^2\xk$ denote a vector of the overall mean total covariates across all clusters in the trial.

\cite{middleton_unbiased_2015} propose a ``Des Raj difference'' estimator \citep{raj_method_1965}, shown below specifically for paired (rather than blocked) cluster randomization:

$$\hat{\tau}^{DR} = \frac{1}{N}\sum_{i=1}^M2(2T_i -1)[\{\Yo - (\xo-\bar{\bm x}) \bm\theta\}-\{\Yt- (\xt-\bar{\bm x}) \bm \theta\}]$$
where $\bm \theta$ is some parameter vector (denoted ($k$, $k_1$, ...) in the original paper). To find the optimal value of $\bm \theta$ to minimize the variance of the estimator, the authors consider minimizing the variance of $u^t_{ik} = \ykt- (\xk-\bar{\bm x})\bm \theta$ and $u^c_{ik} = \ykc- (\xk-\bar{\bm x})\bm \theta$. Their solution for the optimal parameter for treatment potential outcomes ($\bm \theta^t_{optim}$) is the vector of coefficients on an OLS model of the treatment potential outcomes on the cluster total covariates (excluding an intercept). $\bm \theta^c_{optim}$ is found analogously for control cluster total outcomes. These optimal values come from viewing the potential outcomes and covariates as having underlying variances. \cite{middleton_unbiased_2015} argue that in practice one would want one optimal value of the parameter, and prove that $\bm \theta^*=\frac{1}{2}\bm \theta^t_{optim} + \frac{1}{2}\bm \theta^c_{optim}$ is the optimal value for pCRTs based on their analysis. 

Consider estimating  the optimal value $\bm \theta^*_i$ with  $\hat{\bm\theta}^*_i=\frac{1}{2}\hat{\bm\theta}^t_i + \frac{1}{2}\hat{\bm\theta}^c_i$, where $\hat{\bm\theta}^t_i$ and $\hat{\bm\theta}^c_i$ are the estimated coefficient vectors from a leave-one-pair-out OLS model fit on the observed treated and control clusters, respectively (dropping pair $i$). Plugging in these estimates into the Des Raj difference estimator:

\begin{align*}
    \hat{\tau}^{DR} &= \frac{1}{N}\sum_{i=1}^M2(2T_i -1)[\{\Yo - (\xo-\bar{\bm x}) \hat{\bm\theta}^*_i\}-\{\Yt- (\xt-\bar{\bm x}) \hat{\bm\theta}^*_i\}]\\
    &= \frac{1}{N}\sum_{i=1}^M(2T_i -1)[2(\Yo-\Yt)-2(\xo - \xt)\hat{\bm\theta}^*_i]\\
    &= \frac{1}{N}\sum_{i=1}^M(2T_i -1)[2(\Yo-\Yt)-2(\xo(\frac{1}{2}\hat{\bm\theta}^t_i + \frac{1}{2}\hat{\bm\theta}^c_i) - \xt(\frac{1}{2}\hat{\bm\theta}^t_i + \frac{1}{2}\hat{\bm\theta}^c_i)]\\
    &= \frac{1}{N}\sum_{i=1}^M(2T_i -1)[2(\Yo-\Yt)-\{(\hat{y}_{i1}^t + \hat{y}_{i1}^c) -(\hat{y}_{i2}^t + \hat{y}_{i2}^c)\}]\\
    &= \frac{1}{N}\sum_{i=1}^M(2T_i -1)[2(\Yo-\Yt)-\hat{d}_i]
\end{align*}
Where the last line is because another way to write $d_i$ is $d_i = (\yot + \yoc)-(\ytt + \ytc)$. Thus, the Des Raj difference estimator, estimating $\bm \theta$ in this manner, and the IDPD estimator where the cluster average potential outcomes are imputed independently with LOO WLS regression are equivalent. 

In \cite{middleton_unbiased_2015}, the authors estimate $\bm \theta_i$ with a leave-one-out regression model of all observed total outcomes on the total covariates (not distinguishing between treatment and control clusters). In the simulation studies with generated data, we found that this approach to estimate $\bm \theta_i$ did not perform as well as what we described above.

\cite{middleton_unbiased_2015} propose the following variance estimator for the Des Raj difference estimator. Let $U_{ik} = \Yk - (\xk-\bar{\bm x})\hat{\bm\theta}^*_i$ and $\bar{U}$ denote the average of the $U_{ik}$ values across all clusters, and $\hat{\sigma}^2(U_{ik}) = \frac{1}{2M-1} \sum_{i=1}^M\sum_{k=1}^2(U_{ik} - \bar{U})^2$. They propose:

\begin{align*}
    \hat{\V}[\hat{\tau}^{RD}]&=\frac{16M^2\hat{\sigma}^2(U_{ik})}{N^2(2M-1)}\\
    &= \frac{16M^2}{N^2(2M-1)^2}\sum_{i=1}^{M}(U_{i1} - \bar{U})^2 + (U_{i2} - \bar{U})^2
\end{align*}
This variance estimator was derived assuming a null treatment effect. It is essentially the sum of variances for the prediction error $U_{ik}$ for the observed total cluster outcomes. This estimator does not take the covariance between estimates for $\Yo$ and $\Yt$ in the same pair into account, treating them as independent. We found that this variance estimator was highly conservative in practice.

\subsection{Su \& Ding (2021)}

\cite{su_model-assisted_2021} present a unifying framework for regression-based estimation of the ATE in cluster-randomized experiments. Although their design-based analysis of standard error estimation is based on complete randomization, rather than paired randomization, their suggestions for treatment effect point estimation could still be useful for pCRTs. Like \cite{middleton_unbiased_2015}, they suggest adjusting a Horvitz-Thompson estimator by the cluster size, however through a regression model formulation. 

Namely, \cite{su_model-assisted_2021} note that the estimated coefficient for the treatment assignment from an OLS regression of $\frac{M}{N}\Yk$ on the treatment assignment (with an intercept) is equivalent to the Horvitz-Thompson estimator. They propose including the centered cluster size (and other covariates) and the interaction between the centered cluster size (and other covariates) and the treatment assignment as additional covariates in this model, following \cite{lin_agnostic_2013}.\footnote{It has been established that the typical OLS assumptions do not align with the Neyman-Rubin causal model. \cite{lin_agnostic_2013} shows that centering the covariates and including an interaction between the treatment assignment and centered covariates aligns OLS with the Neyman-Rubin causal model. \cite{schochet_design-based_2021} extends this idea to block cluster randomized experiments.} In other words $\hat{\beta}$ from:
$$\frac{M}{N}\Yk = \alpha + \beta T_{ik} + \bm \gamma_1' \tilde{\bm x}_{ik} + \delta_1\tilde{n}_{ik} + \bm \gamma_2 '  T_{ik} \tilde{\bm x}_{ik} + \delta_2T_{ik}\tilde{n}_{ik} + \varepsilon_{ik}$$
where $\tilde{\bm x}_{ik} = (\bm x_{ik} - \bar{\bm x})$ is a centered covariate vector and $\tilde{n}_{ik}  = \nk - \frac{N}{M}$ is the centered cluster size.

\section{Bias and Variance of IDPD Estimator}\label{chpt:appn_pcrt4}

\subsection{Bias of IDPD Estimator}

\begin{align}
    \E[\bar{\tau} - \hat{\tau}] &= \E\Big[\bar{\tau} - \frac{1}{N}\sum_{i=1}^M(2T_i - 1)[2(\Yo - \Yt)  - \hat{d}_i]\Big]\\ 
    &= \E\Big[\bar{\tau} - \frac{1}{N}\sum_{i=1}^M(2T_i - 1)[2(\Yo - \Yt)]\Big]- \E\Big[\frac{1}{N}\sum_{i=1}^M(2T_i - 1)\hat{d}_i\Big]\\
    &= -\frac{1}{N}\sum_{i=1}^M\E[2T_i\hat{d}_i] - \E[\hat{d}_i]\\
    &= -\frac{2}{N}\sum_{i=1}^M\E[T_i\hat{d}_i] - \E[T_i]\E[\hat{d}_i]\\
    &= -\frac{2}{N}\sum_{i=1}^M\mbox{Cov}(T_i,\hat{d}_i)
\end{align}
Where (M.3) is because $\E\Big[\bar{\tau} - \frac{1}{N}\sum_{i=1}^M(2T_i - 1)[2(\Yo - \Yt)]\Big]= 0$.

\subsection{Variance of LOO IDPD Estimator}\label{sec:loo_var}

Define $\thi = (2T_i - 1)[2(\Yo - \Yt) - \hat{d}_i]$ and $\gamma_{i,j} = \mbox{Cov}(\thi, \tauh_j)$. Assume that $\hat{d}_i \perp T_i$. Then,

\begin{align*}
    \V[\looidpd] &= \frac{1}{N^2}\V\Big[\sum_{i=1}^M \thi \Big] \\
    &=  \frac{1}{N^2}\Big( \sum_{i=1}^M \V[\thi] + \sum_{i \neq j} \gamma_{i,j}   \Big)\\
    &=  \frac{1}{N^2}\Big( \sum_{i=1}^M \mbox{MSE}(\dih) + \sum_{i \neq j} \gamma_{i,j}   \Big)
\end{align*}
Since, 

\begin{align*}
    \V[\thi] &= \E\big[\V[\thi | \dih]\big] + \V\big[\E[\thi | \dih]\big] \\
    &= \E\big[\V[(V_i - \dih)T_i + (V_i + \dih)(1-T_i) | \dih]\big] + \V\big[\tau_i\big] \\
    &= \E\big[\V[ 2T_i\vio + 2(1-T_i)\vit + (1-2T_i)\dih | \dih]\big] \\
    &= \E\big[ \V[ 2(\vio -\vit -\dih)T_i + 2\vit + \dih    | \dih]\big] \\
    &= \E\big[4(d_i - \dih)^2 \V[T_i | \dih]\Big] \\
    &= \E[(d_i - \dih)^2]\\
    &= \mbox{MSE}(\dih)
\end{align*}

\subsection{Variance of LOI IDPD Estimator}\label{sec:loi_var}

Consider estimates of $d_i$ like those in the AMW or WLS-P estimator (See Supplement~\ref{chpt:appn_pcrt3b}). We can write them as $\hat{d}_i = T_if_i + g_i$, where $f_i$ and $g_i$ are fixed values. For example, for the AMW estimator where $\hat{d}_i = (\no-\nt)(\Ybo + \Ybt)$, $g_i = (\no-\nt)(\ybot + \yboc - \ybtt - \ybtc)$ and $f_i = (\no-\nt)(\ybtt + \yboc)$.  

Then,
\begin{align}
    \V[\thi] &= \V[2(\vio -\vit -\dih)T_i + 2\vit + \dih ] \\
    &= \V[2\{d_i - f_iT_i - g_i\}T_i  + f_iT_i + g_i ] \\
    &= \V[(2(d_i + \frac{1}{2}f_i - g_i)T_i - 2f_iT_i^2]\\
    &= \V[2(d_i - \frac{1}{2}f_i - g_i)T_i]\\ 
    &= 4\{d_i - (\frac{1}{2}f_i + g_i)\}^2\V[T_i]\\
    &= \{d_i - (\frac{1}{2}f_i + g_i)\}^2\\
    &= \E[d_i - \dih]^2
\end{align}
Where (J.9) is because $T_i = T_i^2$ and (J.12) is due to our definition of $\hat{d}_i$.

\subsection{Variance of LAI IDPD Estimator}\label{sec:lai_var}
Consider estimates of $d_i$ which use outcomes and cluster sizes from all of the pairs, for example using $\hat{d}_i = (\no-\nt)\frac{1}{M}\sum_{j=1}^M(\bar{Y}_{j1} + \bar{Y}_{j2})$. For estimators like this, we  will write $\hat{d}_i = \hat{d}_{-i} +  T_if_i + g_i$,where $\hat{d}_{-i}$ is the part of the estimator that depends on the vector of treatment assignments excluding $T_i$: $\bm T_{-i}$. Thus, $\hat{d}_{-i} \perp T_i$ and $\bm T_{-i} \perp T_i$. $f_i$ and $g_i$ are fixed values as in the previous example. Then,

\begin{align*}
    \V[\thi] &= \V[2(d_i -\dih)T_i + \dih] \\
    &= \V[2\{d_i - \hat{d}_{-i} - f_iT_i - g_i\}T_i + \hat{d}_{-i} + f_iT_i + g_i] \\
    &= \V[2\{d_i - \hat{d}_{-i} - \frac{1}{2}f_i - g_i\}T_i + \hat{d}_{-i} + g_i] \\
    &= \V[\E[2\{d_i - \hat{d}_{-i} - \frac{1}{2}f_i - g_i\}T_i + \hat{d}_{-i} + g_i|\hat{d}_{-i}]]\\
    &\hspace{2cm} +\E[\V[2\{d_i - \hat{d}_{-i} - \frac{1}{2}f_i - g_i\}T_i + \hat{d}_{-i} + g_i|\hat{d}_{-i}]]\\
    &=\V[d_i -\frac{1}{2}f_i] + \E[4\{d_i - \hat{d}_{-i} - \frac{1}{2}f_i - g_i\}^2\V[T_i|\hat{d}_{-i}]]\\
    &= \E[\{d_i - (\hat{d}_{-i} + \frac{1}{2}f_i + g_i)\}^2]\\
    &= \E\Big[\big\{d_i - \E[\hat{d_i}| \bm T_{-i}]\big\}^2\Big]
\end{align*}

\section{Details of LOO-MI Estimator}\label{chpt:appn_pcrt2}

As a reminder, the IDPD estimator is defined as:

$$\hat{\tau} =  \frac{1}{N}\sum_{i=1}^M(2T_i - 1)\{2(\Yo - \Yt)  - \hat{d}_i\}$$
In this supplement, we detail the calculations underlying leave-one-out mean imputation for $\hat{d}_i$ and the formulation of the LOO-MI estimator ($\loomi$) as a  H\'ajek estimator.

\subsection{Leave-One-Out Mean Imputation of $d_i$}

As a reminder, we define $d_i$ as:
\begin{align*}
    d_i =& v^{(1)}_i - v^{(2)}_i \\
        =& (\no - \nt)\underset{(A_i)}{\frac{\ybot + \ybtc + \ybtt + \yboc}{2}} +  (\no + \nt)\underset{(B_i)}{\frac{(\ybot - \ybtc) - (\ybtt - \yboc)}{2}}
\end{align*}
For the LOO-MI estimator, we estimate $A_i$ and $B_i$ with leave-one-out mean imputation. Thus, $A_i$ is estimated as $$\hat{A}_i = \frac{1}{(M-1)} \sum_{j \neq i}  \left\{(\Yjbo + \Yjbt)+ (\Yjbo + \Yjbt)\right\}/2.$$  If one follows the same logic to estimate $B_i$, $$\hat{B}_i = \frac{1}{(M-1)} \sum_{j \neq i}  \left\{(2T_i -1)(\Yjbo - \Yjbt) - (2T_i -1)(\Yjbo - \Yjbt)\right\}/2 = 0.$$ Therefore,
\begin{align*}
    \hat{d}_i^{MI} &= (\no - \nt) \frac{1}{(M-1)} \sum_{j \neq i}  \left\{(\Yjbo + \Yjbt)+ (\Yjbo + \Yjbt)\right\}/2 + (\no + \nt)\cdot0 \\
    &= \frac{(n_{1, i} - n_{2, i})}{(M-1)} \sum_{j \neq i}  (\Yjbo + \Yjbt),
\end{align*}
and 
$$\loomi = \frac{1}{N}\sum_{i=1}^M(2T_i - 1)\left\{2(\Yo - \Yt)  -  \frac{(\no - \nt)}{(M-1)} \sum_{j \neq i}  (\Yjbo + \Yjbt) \right\}$$

\subsection{LOO-MI H\'ajek Estimator}

In Section~\ref{sec:hajek}, we show the well-known result that the H\'ajek estimator can be written as a ratio of Horvitz-Thompson estimators:
$$\hajek = \frac{\hat{Y}(t)}{\hat{N}(t)} - \frac{\hat{Y}(c)}{\hat{N}(c)},$$
where
$$\hat{Y}(t) = \sum_{i=1}^M2\left\{T_i \Yo + (1-T_i)\Yt\right\},$$
$$\hat{N}(t) = \sum_{i=1}^M2\left\{T_i  \no + (1-T_i)\nt\right\},$$
and  $\hat{Y}(c)$ and $\hat{N}(c)$ are defined analogously. Here we will show that the LOO-MI estimator can be written as a ratio of adjusted Horvitz-Thompson estimators ($\loomi = \frac{\tilde{Y}(t)}{\tilde{N}(t)} - \frac{\tilde{Y}(c)}{\tilde{N}(c)}.$), where
$$\tilde{Y}(t) = \sum_{i=1}^M \Big[ 2 \left \{T_i\Yo +(1-T_i) \Yt \right \} + (1-2T_i) \frac{(n_{1, i} - n_{2, i})}{(M-1)} \sum_{j \neq i}\left \{T_j \Yjbo +(1-T_j) \Yjbt \right \} \Big],$$ and $\tilde{Y}(c)$, $\tilde{N}(t)$, and $\tilde{N}(c)$ are defined analogously.

First, note that:
\begin{align}
    \loomi = &  \frac{1}{N}\sum_{i=1}^M(2T_i - 1)\left\{2(\Yo - \Yt)  -  \frac{(\no - \nt)}{(M-1)} \sum_{j \neq i}  (\Yjbo + \Yjbt) \right\} \nonumber \\
    =&  \frac{1}{N}\sum_{i=1}^M \Big[ 2 \left \{T_i\Yo +(1-T_i) \Yt \right \} + \nonumber \\ 
    &  \hspace{1.5cm}  (1-2T_i) \frac{(n_{1, i} - n_{2, i})}{(M-1)} \sum_{j \neq i}\left \{T_j \Yjbo +(1-T_j) \Yjbt \right \} \Big] \label{eq:treatloo} \\ 
    &  - \frac{1}{N} \sum_{i=1}^M \Big[ 2\{(1-T_i)\Yo +T_i \Yt\} + \nonumber\\ 
    & \hspace{1.5cm}  (2T_i-1) \frac{(n_{1, i} - n_{2, i})}{(M-1)} \sum_{j \neq i} \{(1-T_j) \Yjbo +T_j \Yjbt\} \Big] \label{eq:controlloo} \\
    =& \frac{1}{N} \tilde{Y}(t) -  \frac{1}{N} \tilde{Y}(c) \nonumber
\end{align}
(\ref{eq:treatloo}) only includes treated potential outcomes and (\ref{eq:controlloo}) only includes control potential outcomes. Thus,  $\tilde{Y}(t) = \sum_{i=1}^M \Big[ 2 \left \{T_i\Yo +(1-T_i) \Yt \right \} + (1-2T_i) \frac{(n_{1, i} - n_{2, i})}{(M-1)} \sum_{j \neq i}\left \{T_j \Yjbo +(1-T_j) \Yjbt \right \} \Big]$ can be seen as an adjusted Horvitz-Thompson estimator for the total treatment potential outcomes (and similarly for $\tilde{Y}(c)$). 

We replace the total cluster outcomes with the cluster size for estimates of the total treatment and control sample sizes.  We can think of $\Ybo = \Yo / \no$, so the replacement for cluster size for \textit{average} cluster outcomes would be $\no/\no = 1$. Therefore, with this estimator, we can estimate $\tilde{N}(t)$ and $\tilde{N}(c)$ as:

\begin{align*}
    \tilde{N}(t) &=  \sum_{i=1}^M 2\{T_i\no +(1-T_i) \nt \} + (1-2T_i) \frac{(n_{1, i} - n_{2, i})}{(M-1)} \sum_{j \neq i} \{T_j + 1 - T_j\}\\
    &= \sum_{i=1}^M 2T_i\no +2(1-T_i) \nt + (1-2T_i)(n_{1, i} - n_{2, i})\\
    &= \sum_{i=1}^M 2T_i\no + 2\nt - 2T_i\nt + \no - 2T_i\no -\nt + 2T_i\nt\\
    &= \sum_{i=1}^M \no + \nt = N
\end{align*}

\begin{align*}
    \tilde{N}(c) &=  \sum_{i=1}^M 2\{(1-T_i)\no +T_i \nt \} + (2T_i-1) \frac{(n_{1, i} - n_{2, i})}{(M-1)} \sum_{j \neq i} \{1 - T_j + T_j\} \\
    &= \sum_{i=1}^M 2\no - 2T_i\no +2 T_i\nt + (2T_i-1)(n_{1, i} - n_{2, i})\\
    &= \sum_{i=1}^M\no+\nt = N
\end{align*}

Therefore, we can write the LOO-MI estimator as a  H\'ajek estimator, or in other words, as a difference in ratios of Horvitz Thompson estimators:
$$\loomi = \frac{\tilde{Y}(t)}{\tilde{N}(t)} - \frac{\tilde{Y}(c)}{\tilde{N}(c)}.$$
\section{Detailed Calculations for IDPD Estimation Framework}\label{chpt:appn_pcrt3b}

Below, we show that the H\'ajek, AMW, and WLS-P estimators can be expressed as IDPD estimators:
$$\hat{\tau}[\cdot, \emptyset] =  \frac{1}{N}\sum_{i=1}^M(2T_i - 1)\{2(\Yo - \Yt)  - \hat{d}_i\}$$
given different imputations of the difference of potential differences $d_i$.

\subsection{H\'ajek Esitmator}

If $\hat{d}_i = (\no -\nt)\left\{\frac{\htYt}{\htNt} + \frac{\htYc}{\htNc}\right\}$, $\hat{\tau}^{IDPD}[\mbox{LAI wMI}, \emptyset] = \hat{\tau}^{WLS}[\emptyset]$, where ``LAI wMI'' indicates ``leave-all-in weighted mean imputation.'' Let $\htYt = 2\sum_{i=1}^M T_i \Yo + (1-T_i)\Yt$, $\htYc = 2\sum_{i=1}^M (1-T_i) \Yo + T_i\Yt$, $\htNt = 2\sum_{i=1}^M T_i \no + (1-T_i)\nt$, $\htNc = 2\sum_{i=1}^M (1-T_i) \no + T_i\nt$. Note that $\htNc + \htNt = 2N$.

\begin{align*}
     \hat{\tau}^{IDPD}[\mbox{LAI wMI}, \emptyset] &= \frac{1}{N}\sum_{i=1}^M(2T_i - 1)\left\{2(\Yo - \Yt)  - (\no - \nt)\left(\frac{\htYt}{\htNt} + \frac{\htYc}{\htNc}\right)\right\}\\
     &= \frac{1}{N}\htYt - \frac{1}{N}\htYc -  \frac{1}{N}\sum_{i=1}^M(2T_i - 1) (\no - \nt)\left(\frac{\htYt}{\htNt} + \frac{\htYc}{\htNc}\right)\\
     &= \frac{1}{N}\htYt - \frac{1}{N}\htYc - \frac{\htNt -\htNc }{2N}\left(\frac{\htYt}{\htNt} + \frac{\htYc}{\htNc}\right)\\
    &= \frac{2\htNt -\htNt + \htNc}{2\htNt N}\htYt - \frac{2\htNc-\htNc+N_t}{2\htNc N}\htYc\\
    &= \frac{\htYt}{\htNt} - \frac{\htYc}{\htNc} = \hat{\tau}^{WLS}
\end{align*}

\subsection{Arithmetic Mean Weighted Estimator}

If $\hat{d}_i = (\no -\nt)(\Ybo + \Ybt)$, $\hat{\tau}[\mbox{LOI MI}, \emptyset] = \hat{\tau}^{AMW}[\emptyset]$, where ``LOI MI'' indicates ``leave-one-in mean imputation.''

\begin{align*}
    \amw &= \frac{1}{N}\sum_{i=1}^M(2T_i - 1)\{2(\Yo - \Yt)  -\hat{d}_i\}\\
     &= \frac{1}{N}\sum_{i=1}^M(2T_i - 1)\{2(\Yo - \Yt)  - (\no -\nt)(\Ybo + \Ybt)\}\\
     &= \frac{1}{N}\sum_{i=1}^M(2T_i - 1)\{2(\no\Ybo - \nt\Ybt)  - (\no -\nt)(\Ybo + \Ybt)\}\\
     &= \frac{1}{N}\sum_{i=1}^M(2T_i - 1)\{(\no + \nt)\Ybo - (\no+\nt)\Ybt\}\\
     &= \frac{1}{N}\sum_{i=1}^M(2T_i - 1)(\no + \nt)(\Ybo - \Ybt) = \hat{\tau}^{AMW}[\emptyset]
\end{align*}

\subsection{WLS-P Estimator}

We will show that if $\hat{d}_i = (\no - \nt)(\Ybo + \Ybt) + (\no + \nt)\{(\Ybo - \Ybt) - \frac{w_iN}{Wn_i}(\Ybo - \Ybt)\}$, $\hat{\tau}[\mbox{LOI MwDI}, \emptyset] = \hat{\tau}^{WLS-P}[\emptyset]$, where ``LOI MwDI'' indicates ``leave-one-in mean and weighted differences imputation.''

As a reminder, for the fixed effects estimator $w_i = \frac{\no\nt}{\no+\nt}$. Let $W = \sum_{i=1}^M w_i$ and $n_i = \no + \nt$. To find an equivalence between the general adjusted estimator and the fixed-effects estimator, set

$$\frac{1}{N}\sum_{i=1}^M(2T_i - 1)[2(\Yo - \Yt)  -\hat{d}_i] = \frac{1}{W}\sum_{i=1}^M w_i(2T_i -1) (\Ybo - \Ybt)$$
After some algebra, solving for $\hat{d}_i$, you get
\begin{align*}
    \hat{d}_i =& \Big(\frac{2W\no - Nw_i}{W}\Big)\Ybo -\Big(\frac{2W\nt - Nw_i}{W}\Big)\Ybt.
\end{align*}   
Note that 
\begin{align*}
    2W\no - Nw_i &= 2\no\sum_{i=1}^M\frac{\no\nt}{\no+\nt}- \frac{\no\nt}{\no+\nt}\sum_{i=1}^M\no+\nt\\
                 &= (\no -\nt)w_i + 2\no\sum_{j\neq i}w_j - w_i\sum_{j\neq i}n_j\\
    2W\nt - Nw_i &=2\nt\sum_{i=1}^M\frac{\no\nt}{\no+\nt}- \frac{\no\nt}{\no+\nt}\sum_{i=1}^M\no+\nt\\
                 &= (\nt -\no)w_i + 2\nt\sum_{j\neq i}w_j - w_i\sum_{j\neq i}n_j,\\
\end{align*}
so, 
\begin{align*}
    \hat{d}_i =& \frac{(\no - \nt)w_i}{W}(\Ybo + \Ybt)+ 2\frac{\sum_{j\neq i}w_j}{W}(\no\Ybo - \nt\Ybt)- \frac{w_i\sum_{j\neq i}n_j}{W}(\Ybo - \Ybt)\\
    =& \frac{(\no - \nt)w_i}{W}(\Ybo + \Ybt)\\&\hspace{2cm}+ 2\frac{\sum_{j\neq i}w_j}{W}\big [(\no-\nt) (\Ybo+\Ybt)/2 + (\no + \nt)(\Ybo-\Ybt)/2\big]\\&\hspace{2cm}- \frac{w_i\sum_{j\neq i}n_j}{W}(\Ybo - \Ybt)\\
    =& \frac{(\no - \nt)W}{W}(\Ybo + \Ybt) + \frac{(\no +\nt)\sum_{j\neq i}w_j - w_i \sum_{j\neq i}n_j}{W} (\Ybo - \Ybt) \\
    =& (\no - \nt)(\Ybo + \Ybt) + \frac{n_iW - w_iN}{W} (\Ybo - \Ybt) \\
    =& (\no - \nt)(\Ybo + \Ybt) + (\no + \nt)[(\Ybo - \Ybt) - \frac{w_iN}{Wn_i}(\Ybo - \Ybt)]
\end{align*}
\section{Equivalence of Baseline Estimators when Pair Cluster Sizes are Equal}\label{chpt:appn_pcrt3a}

Below, we show that when the cluster sizes within each pair are equal, i.e. $\no = \nt = \frac{n_i}{2}$ for all $i = 1, ..., M$, the Horvitz-Thompson, H\'ajek, AMW, WLS-P, and LOO-MI estimators are equivalent.

\begin{align*}
\hat{\tau}^{HT}[\emptyset] &=  \frac{2}{N}\sum_{i=1}^M (2T_i -1) (\Yo - \Yt) \\
\vspace{1cm}\\
    \loomi &= \frac{1}{N}\sum_{i=1}^M(2T_i - 1)[2(\Yo - \Yt)  -  \frac{(\no - \nt)}{(M-1)} \sum_{j \neq i}  (\bar{Y}_{j1} + \bar{Y}_{j2})]\\
    &= \frac{2}{N}\sum_{i=1}^M(2T_i - 1)(\Yo - \Yt)\\
    \vspace{1cm}\\
    \hat{\tau}^{WLS}[\emptyset]&= \frac{1}{\htNt}\sum_{i=1}^M [T_i \Yo + (1-T_i)  \Yt]
    - \frac{1}{\htNc}\sum_{i=1}^M [(1-T_i) \Yo + T_i \Yt] \\
    &= \frac{2}{N}\sum_{i=1}^M [T_i \Yo + (1-T_i)  \Yt]
    - \frac{2}{N}\sum_{i=1}^M [(1-T_i) \Yo + T_i \Yt]\\
    &= \frac{2}{N}\sum_{i=1}^M(2T_i - 1)(\Yo - \Yt)\\
    \vspace{1cm}
    \end{align*}

\begin{align*}
    \hat{\tau}^{AMW}[\emptyset] &= \frac{1}{\sum_{i=1}^M \no + \nt}\sum_{i=1}^M (\no + \nt)(2T_i -1) (\Ybo - \Ybt)\\
    &= \frac{2}{N}\sum_{i=1}^M \frac{n_i}{2}(2T_i -1) (\Ybo - \Ybt)\\
    &= \frac{2}{N}\sum_{i=1}^M(2T_i - 1)(\Yo - \Yt)\\
    \hat{\tau}^{WLS-P}[\emptyset] &= \frac{1}{\sum_{i=1}^M \frac{\no\nt}{\no + \nt}}\sum_{i=1}^M \frac{\no\nt}{\no + \nt}(2T_i -1) (\Ybo - \Ybt)\\
    &= \frac{1}{\sum_{i=1}^M \frac{n_i^2/4}{n_i}}\sum_{i=1}^M \frac{n_i^2/4}{n_i}(2T_i -1) (\Ybo - \Ybt)\\
    &= \frac{2}{N}\sum_{i=1}^M \frac{n_i}{2}(2T_i -1) (\Ybo - \Ybt)\\
    &= \frac{2}{N}\sum_{i=1}^M(2T_i - 1)(\Yo - \Yt)\\
\end{align*}

\section{Conservative Variance Estimation for LOO IDPD Estimator}\label{chpt:appn_pcrt5a}

As shown in Supplement~\ref{chpt:appn_pcrt4} (similar to the proof in \cite{wu_design-based_2021}),  the true variance of $\looidpd$ is 
$$\V[\looidpd]  = \frac{1}{N^2}\Big( \sum_{i=1}^M \mbox{MSE}(\hat{d}_i) + \sum_{i \neq j} \gamma_{i,j}   \Big)$$ 
where $\gamma_{i,j} = \mbox{Cov}(\thi, \tauh_j)$.  Under the regularity conditions discussed in Supplement~\ref{chpt:appn_pcrt7}, \citep{wu_design-based_2021} show that  $\frac{1}{N^2}\sum_{i \neq j} \gamma_{i,j}$ is negligible compared to $ \frac{1}{N^2}\sum_{i=1}^M \mbox{MSE}(\hat{d}_i)$, so the variance estimation focuses on estimating a bound for $\mbox{MSE}(\hat{d}_i)$.

The results of \cite{wu_design-based_2021} follow for the paired cluster randomized setting, but we repeat their proofs here in the current notation for clarity. Let  $V_i = 2(T_i\vio + (1-T_i)\vit)$ and $\hat{V}_i = 2(T_i\viho + (1-T_i)\viht)$. Note that $V_i$ is observed.
Assuming that $\hat{d}_i \perp T_i$, and therefore $\viho$ and $\viht$ are independent of $T_i$,

\begin{align*}
     MSE[\hat{V}_i] &= \E[(V_i - \hat{V}_i)^2]\\
     &= \E[\E[\{2T_i(\vio - \viho) +  2(1-T_i)(\vit - \viht)\}^2|\viho,\viht]]\\
     &= 2\E[(\vio - \viho)^2 +  (\vit - \viht)^2] \mbox{, and} \\
     &\\
     MSE[\hat{d}_i] &= \E[(d_i - \hat{d}_i)^2]\\
     &= \E[\{(\vio - \vit) - (\viho - \viht)\}^2]\\
     &= \E[\{(\vio - \viho ) - ( \vit- \viht)\}^2]
\end{align*}
It follows that
\begin{align*}
     MSE[\hat{V}_i] - MSE[\hat{d}_i] &= \E[2(\vio - \viho)^2 +  2(\vit - \viht)^2 -\{(\vio - \viho ) - ( \vit- \viht)\}^2]\\
     &= \E[\{(\vio - \viho)  + (\vit - \viht)\}^2]\\
     &= \E[\{(\vio +\vit) -  (\viho +\viht)\}^2] \\
     &= \E[(\tau_i - (\viho +\viht))^2] > 0
\end{align*}
Therefore, $\frac{1}{N^2}\sum_{i=1}^M  MSE[\hat{d}_i] < \frac{1}{N^2}\sum_{i=1}^M  MSE[\hat{V}_i]$. $\frac{1}{N^2}\sum_{i=1}^M  MSE[\hat{V}_i]$ can be unbaisedly estimated as $\frac{1}{N^2}\sum_{i=1}^M (V_i - \hat{V}_i)^2$, so this is a conservative estimator of the variance.
\section{Asymptotic Normality of LOO DPD Estimator}\label{chpt:appn_pcrt7}

Consider an infinite sequence of pairs, $i= 1, 2, 3, ...$ where in the fixed-sample setting, we observe the first $M$ pairs in the sequence. The covariates, potential outcomes and cluster sizes are all considered \textit{fixed} quantities for each pair $i = 1, .., M$ and we analyze the behavior of the estimator as the number of pairs, $M$ grows to infinity. 

$\looidpd$ is equivalent to the estimator of \cite{wu_design-based_2021}, if all clusters are of size 1, but takes the same form even if they are not because the sizes of the clusters are subsumed in the definition of $d_i$. \cite{wu_design-based_2021} show that this estimator is asymptotically normal, given certain regularity assumptions, and these results follow for $\looidpd$, given that the estimators are of the same form, as long as the same assumptions are met. Therefore, we will show, that under certain assumptions, their original assumptions hold, and therefore the estimator $\looidpd$ is asymptotically normal. Below, we first list their original assumptions, followed by assumptions for the paired clustered setting, and finally show that if these new assumptions hold, so do the original ones of \cite{wu_design-based_2021}.
\\

\noindent
\textbf{Assumptions from \cite{wu_design-based_2021}:}

Given the sample with $M$ pairs, let $\dih^{(M)}$ represent the estimate of $d_i$ using the other $M-1$ pairs. Define the quantities 
$\dom = \E\big[ \dihm \big]$ for the finite sample and 
$\dtm = \dihm - \dom $.

\begin{orig}\label{orig1}
There exists some $0 < C < \infty$ and $q >0$ such that for all $i$, $\V[\dih] \leq C/M^q$.
\end{orig}

\begin{orig}\label{orig2}
     Let $\rho_{ij}$ be the correlation of $\dti U_i$ and $\dtj U_j$, $\bar{\rho} = \frac{\sum_{i\neq j}\rho_{ij}}{M(M-1)}$. We assume that $M^{1-q} \bar{\rho} \to 0$.
\end{orig}

\begin{orig}\label{orig3}
    The limit of $\dom$ exists ($\dinfm$) and assume that $$\frac{1}{M}\sum_{i=1}^M (\dom - \dinfm)^2 \to 0.$$
\end{orig}

\begin{orig}\label{orig4}
    Let $V_M = \sum_{i=1}^M (d_i - \dinfm)^2$. There exists a $0 <K < \infty$ such that $\frac{V_M}{M} \to K$ and $$\underset{i = 1, ... , M}{\max} \frac{(d_i - \dinfm)^2}{V_M} \to 0$$
\end{orig}

\noindent
\textbf{Assumptions for estimator under paired cluster randomization:}

Consider the decomposition of $d_i$

\begin{align*}
    d_i &= \frac{\no-\nt}{2}[\ybot + \yboc + \ybtt + \ybtc] + \frac{\no+\nt}{2}[(\ybot -\ybtc) - (\ybtt -\yboc)] \\
    &= \frac{\no-\nt}{2}a_i + \frac{\no+\nt}{2}b_i
\end{align*}

Let $\aihm$ and $\bihm$ denote the estimated $a_i$ and $b_i$ with $M$ pairs. $\aom = \E[\aihm]$ and $\bom = \E[\bihm]$ as the expected values under the possible treatment assignment permutations, and $\atm = \aihm - \aom $ and $\btm = \bihm - \bom$ be the non-random parts of $\aihm$ and $\bihm$. We make the following assumptions

\begin{assumption}\label{astn:n}
$n_{k,i} \leq D$, for some $0 < D < \infty$ for all pairs $i = 1, ..., M$ and clusters $k = 1,2$.
\end{assumption}

\begin{assumption}\label{astn:var}
    There exists some $0 < C' < \infty$ and $q >0$ such that for all $i$, $\mathbb{V}[\aihm] \leq C'/M^q$ and $\mathbb{V}[\bihm] \leq C'/M^q$.
\end{assumption}

\begin{assumption}\label{astn:corr}
     Let $\rho_{ij}$ be the correlation of $\dti U_i$ and $\dtj U_j$, $\bar{\rho} = \frac{\sum_{i\neq j}\rho_{ij}}{M(M-1)}$. We assume that $M^{1-q} \bar{\rho} \to 0$.
\end{assumption}

\begin{assumption}\label{astn:convg}
The limits of $\aom$ and $\bom$ exist ($\ainfm$ and $\binfm$, respectively) and assume that

$$\frac{1}{M}\sum_{i=1}^M (\aom - \ainfm)^2 \to 0,$$
$$\frac{1}{M}\sum_{i=1}^M (\bom - \binfm)^2 \to 0$$
\end{assumption}

\begin{assumption}\label{astn:noover}
Let $V_M = \sum_{i=1}^M (d_i - \dinfm)^2$. There exists a $0 <K < \infty$ such that $\frac{V_M}{M} \to K$ and

 $$\underset{i = 1, ... , M}{\max} \frac{(a_i - \ainfm)^2}{V_M} \to 0$$
 $$\underset{i = 1, ... , M}{\max} \frac{(b_i - \binfm)^2}{V_M} \to 0$$

\end{assumption}

\noindent
\textbf{Proposition: }If Assumptions 1-5 hold, the Original Assumptions of \cite{wu_design-based_2021} hold.

\noindent
\textbf{Proof: }

First, we show that if Assumptions \ref{astn:n} and \ref{astn:var} hold, Original Assumption \ref{orig1} holds. To show this, first note that given random variables $X$ and $Y$, $\V[X+Y] \leq 2(\V[X] + \V[Y])$. This is because

\begin{align}
    2Cov(X,Y) &= 2Corr(X,Y)\sqrt{\V[X]\V[Y]}\\
    &\leq 2\sqrt{\V[X]\V[Y]}\\
    &\leq \V[X]+\V[Y]
\end{align}
where (3) is because for $x, y \in \mathbb{R}$, $x^2+y^2 -2xy = (x-y)^2 \geq 0$, so $x^2+y^2 \geq 2xy$.

Thus,
\begin{align*}
    \V[\hat{d}_i] &= \V \Big [ \frac{(\no-\nt)}{2}\hat{a}_i^{(M)}  +  \frac{(\no+\nt)}{2}\hat{b}_i^{(M)} \Big ] \\
    &\leq \frac{(\no-\nt)^2}{2}\V[\hat{a}_i^{(M)}]+ \frac{(\no+\nt)^2}{2}\V[\hat{b}_i^{(M)}] \\
    &\leq \frac{\no^2 + \nt^2}{2}C'/M^q \\
    &\leq \frac{D^2C'}{M^q} \\
    &\leq \frac{C}{M^q}
\end{align*}

where $C = D^2C'$. $0<D^2C'<\infty$ by Assumption \ref{astn:var} so $0<C<\infty$ and Original Assumption \ref{orig1} holds.

Assumption \ref{astn:corr} is equivalent to Original Assumption \ref{orig2}.

Next, we show that if Assumptions \ref{astn:n} and \ref{astn:convg} hold, Original Assumption \ref{orig3} holds:

\begin{align*}
    0 \leq \frac{1}{M}\sum_{i=1}^M (\dom - \dinfm)^2 \leq & \frac{1}{M}\sum_{i=1}^M \frac{(\no-\nt)^2}{2}(\aom - \ainfm)^2 \\
    &+ \frac{1}{M}\sum_{i=1}^M \frac{(\no+\nt)^2}{2}(\bom -\binfm)^2\\
    \leq& \frac{D^2}{2M}\sum_{i=1}^M(\aom - \ainfm)^2 + \frac{2D^2}{M}\sum_{i=1}^M(\bom - \binfm)^2 \\
    \to& 0
\end{align*}
Where the first inequality is by the fact that for $x, y \in \mathbb{R}$, $(x+y)^2 \leq 2x^2 + 2y^2$ and the second is by Assumption~\ref{astn:n}. Therefore, $\frac{1}{M}\sum_{i=1}^M (\dom - \dinfm)^2 \to 0$ by the Squeeze Theorem.

Finally, we show that if Assumptions \ref{astn:n} and \ref{astn:noover} hold, Original Assumption \ref{orig4} holds: 

\begin{align*}
  0 \leq  \underset{i = 1, ... , M}{\max} \frac{(d_i - \dinfm)^2}{V_M} =&  \underset{i = 1, ... , M}{\max} \frac{(\frac{\no-\nt}{2} (a_i - \ainfm) + \frac{\no+\nt}{2}(b_i - \binfm) )^2}{V_M}\\
     \leq &  \underset{i = 1, ... , M}{\max} \Bigg \{ \frac{(\no-\nt)^2(a_i - \ainfm)^2}{{2V_M}} + \frac{(\no+\nt)^2(b_i - \binfm)^2}{2V_M} \Bigg \}\\
     \leq& \frac{D^2}{2}\underset{i = 1, ... , M}{\max}\frac{(a_i - \ainfm)^2}{V_M} + 2D^2 \underset{i = 1, ... , M}{\max}\frac{(b_i - \binfm)^2}{V_M}\\
    \to& 0
\end{align*}
Where the first inequality is by the fact that for $x, y \in \mathbb{R}$, $(x+y)^2 \leq 2x^2 + 2y^2$ and the second is by Assumption~\ref{astn:n}. Therefore, $\underset{i = 1, ... , M}{\max} \frac{(d_i - \dinfm)^2}{V_M} \to 0$ by the Squeeze Theorem. $\square$

Therefore, under Assumptions 1-5,  $N(\looidpd - \bar{\tau})/V_M$ converges in distribution to a standard normal random variable. See the supplemental materials to \cite{wu_design-based_2021} for a proof.

\section{Variance for LOO-MI Estimator}\label{chpt:appn_pcrt5b}

When $\hat{d}_i$ is estimated using leave-one-out mean imputation, the variance estimator simplifies nicely, as shown below.

While we framed the estimator $\hat{d}_i^{MI}$ in terms of estimates of $A_i$ and $B_i$ in Supplement~\ref{chpt:appn_pcrt2}, we can equivalently think about estimating $\vio$ and $\vit$ using LOO mean imputation. Recall that,
\begin{align*}
    \vio &= (\no - \nt)\frac{\ybot + \ybtc}{2} + (\no + \nt)\frac{\ybot - \ybtc}{2}\\
      \vit  &= (\nt - \no)\frac{\ybtt + \yboc}{2} + (\no + \nt)\frac{\ybtt - \yboc}{2}.
\end{align*}
Therefore, the estimate of $V_i$ using LOO mean imputation is:
\begin{align*}
    \hat{V_i}^{MI} &= 2T_i\hat{v}_i^{(1),MI} + 2(1-T_i)2T_i\hat{v}_i^{(2),MI} \\
    &= 2T_i \big [ \frac{(\no - \nt)}{(M-1)} \sum_{j \neq i}  (\Yjbo + \Yjbt)/2 + \frac{(\no + \nt)}{(M-1)} \sum_{j \neq i} (2T_j-1)(\Yjbo - \Yjbt)/2 \big]\\ 
    & \hspace{1cm}+ 2(1-T_i) \big [\frac{(\nt - \no)}{(M-1)} \sum_{j \neq i}  (\Yjbo + \Yjbt)/2 + \frac{(\no + \nt)}{(M-1)} \sum_{j \neq i} (2T_j-1)(\Yjbo - \Yjbt)/2\big ]\\
    &= 2\big [\frac{(\nt - \no)}{(M-1)} \sum_{j \neq i}  (\Yjbo + \Yjbt)/2 + \frac{(\no + \nt)}{(M-1)} \sum_{j \neq i} (2T_j-1)(\Yjbo - \Yjbt)/2\big ]\\
     & \hspace{1cm}+4T_i \frac{(\no - \nt)}{(M-1)} \sum_{j \neq i}  (\Yjbo + \Yjbt)/2\\
     &= (2T_i - 1)\frac{(\no - \nt)}{(M-1)} \sum_{j \neq i}  (\Yjbo + \Yjbt) +  \frac{(\no + \nt)}{(M-1)} \sum_{j \neq i} (2T_j-1)(\Yjbo - \Yjbt)
\end{align*}

Plugging this into the variance estimator for LOO IDPD estimator, it simplifies to: 
\begin{align*}
     \hat{\V}[\loomi] &= \frac{1}{N^2}\sum_{i=1}^M (V_i - \hat{V_i}^{MI})^2\\
     &= \frac{1}{N^2}\sum_{i=1}^M \Big[(2T_i - 1)\left\{(\no+\nt)(\Ybo-\Ybt) + (\no-\nt)(\Ybo+\Ybt)\right\}\\
    &\hspace{2cm} - (2T_i - 1)\frac{(\no - \nt)}{(M-1)} \sum_{j \neq i}  (\Yjbo + \Yjbt)\\
    &\hspace{2cm} -  \frac{(\no + \nt)}{(M-1)} \sum_{j \neq i} (2T_j-1)(\Yjbo - \Yjbt)\Big ]^2\\
     &= \frac{1}{N^2}\sum_{i=1}^M \Big[(\no+\nt)\{(2T_i - 1)(\Ybo-\Ybt)\\
     &\hspace{2cm} - \frac{1}{(M-1)} \sum_{j \neq i} (2T_j-1)(\Yjbo - \Yjbt)\}\\
     &\hspace{2cm} + (2T_i-1)(\no-\nt)\big\{(\Ybo+\Ybt) - \frac{1}{(M-1)} \sum_{j \neq i} (\Yjbo + \Yjbt)
     \big\} \Big]^2\\
     &= \frac{1}{N^2}\sum_{i=1}^M \left[(\no+\nt)\left\{\hat{\bar{\tau}}_i- \bar{\bar{\tau}}_{-i}\right\} + (2T_i-1)(\no-\nt)\left\{\bar{S}_i- \bar{\bar{S}}_{-i}\right\} \right]^2
\end{align*}

\section{Comparison of Variance Estimators when Pair Cluster Sizes are Equal}\label{chpt:appn_pcrt6}

When $\no = \nt = \frac{n_i}{2}$ for all $i = 1, ..., M$, as a reminder, the H\'ajek, WLS-P, AMW, and LOO-MI estimators are all equivalent. Below we compare different variance estimators for these point estimators. Since the variance estimators were developed with a certain point estimator in mind (when cluster sizes may vary), they are slightly different, even when the point estimators are equivalent. We compare our variance estimator for the LOO-MI estimator ($\hat{\V}[\loomi]$), the variance estimator of \cite{imai_essential_2009} for the AMW and WLS-P estimators ($\hat{\V}^{IKN}[\hat{\tau}^{w'}$]), and the variance estimators of \cite{de_chaisemartin_at_2020} for the H\'ajek and WLS-P/AMW estimators ($\hat{\V}^{dCR}[\hajek]$) and ($\hat{\V}^{dCR}[\hat{\tau}^{w'}]$).

Let
\begin{align*}
    \hat{\bar{\tau}}_i &= (2T_i-1)(\Ybo - \Ybt)\\
    \hat{\bar{\tau}} &= \frac{1}{M} \sum_{i=1}^M(2T_i-1)(\Ybo - \Ybt)\\
    \bar{\bar{\tau}}_{-i} &= \frac{1}{(M-1)} \sum_{j \neq i} (2T_j-1)(\Yjbo - \bar{Y}_{2,j})\\
    \tilde{r}_{k,i} &= Y_{k,i} - n_{k,i} \left\{ T_{k,i}\frac{\hat{Y}(t)}{\hat{N}(t)}+(1-T_{k,i})\frac{\hat{Y}(c)}{\hat{N}(c)} \right\}
\end{align*}

We show below that, when the cluster sizes within each pair are the same, the variance estimators take a similar form, weighted by different quantities, highlighted in \hl{yellow}.

\begin{align*}
    \hat{\V}[\loomi] =& \frac{1}{N^2}\sum_{i=1}^M \left[(\no+\nt)\left\{\hat{\bar{\tau}}_i- \bar{\bar{\tau}}_{-i}\right\} + (2T_i-1)(\no-\nt)\left\{\bar{S}_i- \bar{\bar{S}}_{-i}\right\} \right]^2 \\
      =& \frac{1}{N^2}\sum_{i=1}^M n_i^2 (\hat{\bar{\tau}}_i- \bar{\bar{\tau}}_{-i})^2\\
      =&\frac{1}{N^2}\sum_{i=1}^M n_i^2 (\frac{M}{(M-1)}\hat{\bar{\tau}}_i- \frac{M}{(M-1)}\hat{\bar{\tau}})^2\\
      =&\frac{M^2}{N^2(M-1)^2}\sum_{i=1}^M n_i^2 (\hat{\bar{\tau}}_i- \hat{\bar{\tau}})^2\\
      =& \frac{1}{N^2}\mathdisplaycolor
      {yellow}{\frac{M^2}{(M-1)^2}}\sum_{i=1}^M 
      \left \{ n_i\hat{\bar{\tau}}_i - \frac{1}{M}\sum_{j=1}^M \mathdisplaycolor
      {yellow}{n_i}(2T_j -1)(\Yjbo-\Yjbt) 
      \right \}^2
\end{align*}


\begin{align*}
    \hat{\V}^{IKN}[\hat{\tau}^{w'}] =& \frac{M}{(M-1)} \sum_{i=1}^M \bigg( \frac{w'_i}{W'} \hat{\bar{\tau}}_i - \frac{1}{M}\hat{\tau}^{w'}\bigg)^2\\
    =& \frac{M}{N^2(M-1)}\sum_{i=1}^M (n_i \hat{\bar{\tau}}_i - \frac{N}{M}\hat{\tau}^{w'})^2\\
    &= \frac{1}{N^2}\mathdisplaycolor
      {yellow}{\frac{M}{(M-1)}}\sum_{i=1}^M \Big(n_i\hat{\bar{\tau}}_i - \frac{1}{M}\sum_{j=1}^M \mathdisplaycolor
      {yellow}{n_j}(2T_j -1)(\Yjbo-\Yjbt) \Big)^2
\end{align*}

\begin{align*}
    \hat{\V}^{dCR}[\hat{\tau}^{w'}] =&  \frac{1}{W^2} \sum_{i=1}^M (w_i')^2\big( \hat{\bar{\tau}}_i - \wlsp \big)^2\\
    =&  \frac{1}{N^2}\mathdisplaycolor
      {yellow}{\frac{1}{1}}\sum_{i=1}^M \Big(n_i\hat{\bar{\tau}}_i - \frac{1}{M}\sum_{j=1}^M \mathdisplaycolor
      {yellow}{\frac{M}{4N}n_i n_j}(2T_j -1)(\Yjbo-\Yjbt) \Big)^2
\end{align*}

\begin{align*}
    \hat{\V}^{dCR}[\hajek] =& \sum_{i=1}^M\bigg(\frac{T_i\tilde{r}_{i1}+(1-T_i)\tilde{r}_{i2}}{\hat{N}(t)} - \frac{ (1-T_i)\tilde{r}_{i1}+T_i\tilde{r}_{i2}}{\hat{N}(c)}\bigg)^2\\
    &= \frac{4}{N^2}\sum_{i=1}^M \Big\{ (2T_i-1) n_i \big(\Ybo - \Ybt) - \frac{2n_i}{N}\sum_{j=1}^M(2T_j - 1)(\Yjbo-\Yjbt)\Big\}^2\\
    =&  \frac{4}{N^2}\mathdisplaycolor
      {yellow}{\frac{1}{1}}\sum_{i=1}^M \Big(n_i\hat{\bar{\tau}}_i - \frac{1}{M}\sum_{j=1}^M \mathdisplaycolor
      {yellow}{\frac{M}{4N}n_i n_j}(2T_j -1)(\Yjbo-\Yjbt) \Big)^2
\end{align*}

\section{Simulation Details}\label{chpt:appn_pcrt8}

We generate simulation data with the goal of varying important factors that influence the performance of the treatment effect estimators, while keeping the variance of the outcomes and how much the covariate or treatment effect explains the outcomes consistent.

We generate cluster sizes so that $\nk = n_i + m_{ki}$ where $n_i \sim U\{a,b\}$ and $m_{ik}\sim U\{c,d\}$ ($U\{\}$ representing the discrete uniform distribution). We let $\E[n_{ki}] = 150$, an generate $n_i$ and $m_i$ as follows for our two cluster size settings:

\begin{enumerate}
    \item \textbf{Cluster sizes approximately matched within pairs:} $n_i \sim U\{80,220\}$, $m_{ki}  \sim U\{-27,27\}$
    \item \textbf{Cluster sizes fully vary:} $n_i \sim U\{75,225\}$, $m_{ki}  \sim U\{0,0\}$
\end{enumerate}

Based on these choices, the variance of the cluster sizes is approximately equivalent between the two settings and the expectation is equivalent. 

As a reminder, we generate individual cluster outcomes as:

$$y_{ik\ell}^c = \alpha_0 + x_{ik} + \gamma_{ik} + \epsilon_{ik\ell} \hspace{2cm} y_{ik\ell}^t = y_{ik\ell}^c + \tau_{ik}$$
where $\epsilon_{ik\ell} \sim N(0,\sigma^2_{\epsilon})$ is an individual error and $\gamma_{ik}$ is an independent, cluster-level error. The cluster-level covariate $x_{ik} = \alpha_i + z_{ik}$ where $\alpha_i\sim N(0, \sigma^2_{\alpha})$ and $z_{ik} \sim N(0, \sigma^2_z)$ ($\alpha_i \perp z_{ik}$ and all other errors). These individual outcomes are then averaged to the cluster level. 

To generate data under a simple, but realistic setting, we think about the clusters as schools and the outcome as a centered and standardized test score, so the mean outcome represents the number of standard deviations from the mean score. We let $\sigma^2_{\alpha} + \sigma^2_{z} + \sigma^2_{\gamma} = .25$, representing a variation of about a quarter of a standard deviation from the mean between schools. We let $\sigma^2_{\epsilon} = 1$, which is a reasonable variation between students in test scores. Since the average cluster size is 150, however, the contribution to the variance of cluster mean control potential outcomes is very small, so $\V[\bar{y}_{ik}^c] \approx .25$.  We distribute the variance between the errors as follows to create the pair effect setttings:

\begin{enumerate}
    \item \textbf{No Pair Effect:} $\sigma^2_{\alpha} = 0$,  $\sigma^2_{z} = .2125$,  $\sigma^2_{\gamma} = .0375$
    \item \textbf{Pair Effect:} $\sigma^2_{\alpha} = .2$,  $\sigma^2_{z} = .0125$,  $\sigma^2_{\gamma} = .0375$
\end{enumerate}

We generate a reasonable treatment effect, that could still be detected, by having the treatment effect explain between 10-20\% of the variation in the observed mean outcome. As a reminder, we generate the treatment effect $\tau_{ik} = \tau_0 + \phi\cdot(n_{ik}-\E[n_{ik}])$. Given the way we generate the data, $\V[\bar{Y}_{ik}] = \V[\bar{y}^c_{ik}] + \V[T_i\tau_{ik}] \approx .25 + \frac{1}{2}\V[\tau_{ik}] +\frac{1}{4}\tau_0^2$. Therefore, we let $\tau_0 = .4$. This makes sense as a very meaningful treatment effect, increasing scores by between 1/4 to 1/2 of a standard deviation, and explains a little under 15\% of the mean observed cluster outcomes. We chose the following parameters for the two treatment effect settings:

\begin{enumerate}
    \item \textbf{Constant Treatment Effect:} $\tau_0=.4$, $\phi = 0$
    \item \textbf{Treatment Effect Related to Cluster Size:} $\tau_0=.4$, $\phi=\sqrt{.02/\V[n_{ik}]}$
\end{enumerate}

Given these choices, $\V(T_i\tau_i) = .04$ with the constant treatment effect and $.05$ when the treatment effect is a function of cluster size. 
\section{Real Data Simulations with Cognitive Tutor Algebra I Study}\label{chpt:appn_pcrtcta}

The simulation studies in Section~\ref{sec:sims} of the main paper are designed to illustrate the distinctions between estimators under easily understandable settings. To evaluate the performance on real data, that does not necessarily follow a linear model, we conduct further studies using an efficacy trial. Specifically, we analyze data from a field trial evaluating the Cognitive Tutor Algebra I (CTAI) curriculum, an alternative algebra curriculum \citep{pane_effectiveness_2014}. In this pCRT, schools were pair matched based on prognostic school characteristics. Within each pair, one school was randomly assigned to incorporate the CTAI into their algebra curriculum for two years (the 2007/08 and 2008/9 school years), while the others were to continue with their standard curriculum. We focus on 44 trial schools in Texas, which include 16 pairs of middle schools and six pairs of high schools. 

Following \cite{wu_design-based_2021}, we use publicly available school-level data, published by the Texas Education Agency \citep{tea_aeis}.  Specifically, our outcome of interest is whether a student passed the mathematics section of the Texas Assessment of Knowledge and Skills (TAKS) in 2008 (the first year of the trial). The TAKS was a set of standardized tests administered every year in Texas from 2003-2011 \citep{taks}. For middle schools, we use the math TAKS passing rate for 8th graders, and for high schools, we use the math TAKS passing rate for 9th graders. A common, highly prognostic covariate for education efficacy trials is a pretest score for the outcome assessment. The school's previous year's (2007) math TAKS passing rate is a reasonable pretest score in this case ($x_{ik}$). Finally, we use the reported number of 8th graders in a middle school or 9th graders in a high school as cluster sizes ($\nk$). 

\begin{figure}
 \subfloat[Cluster Size\label{fig:cta-n-bal}]{%
      \includegraphics[width=0.45\textwidth]{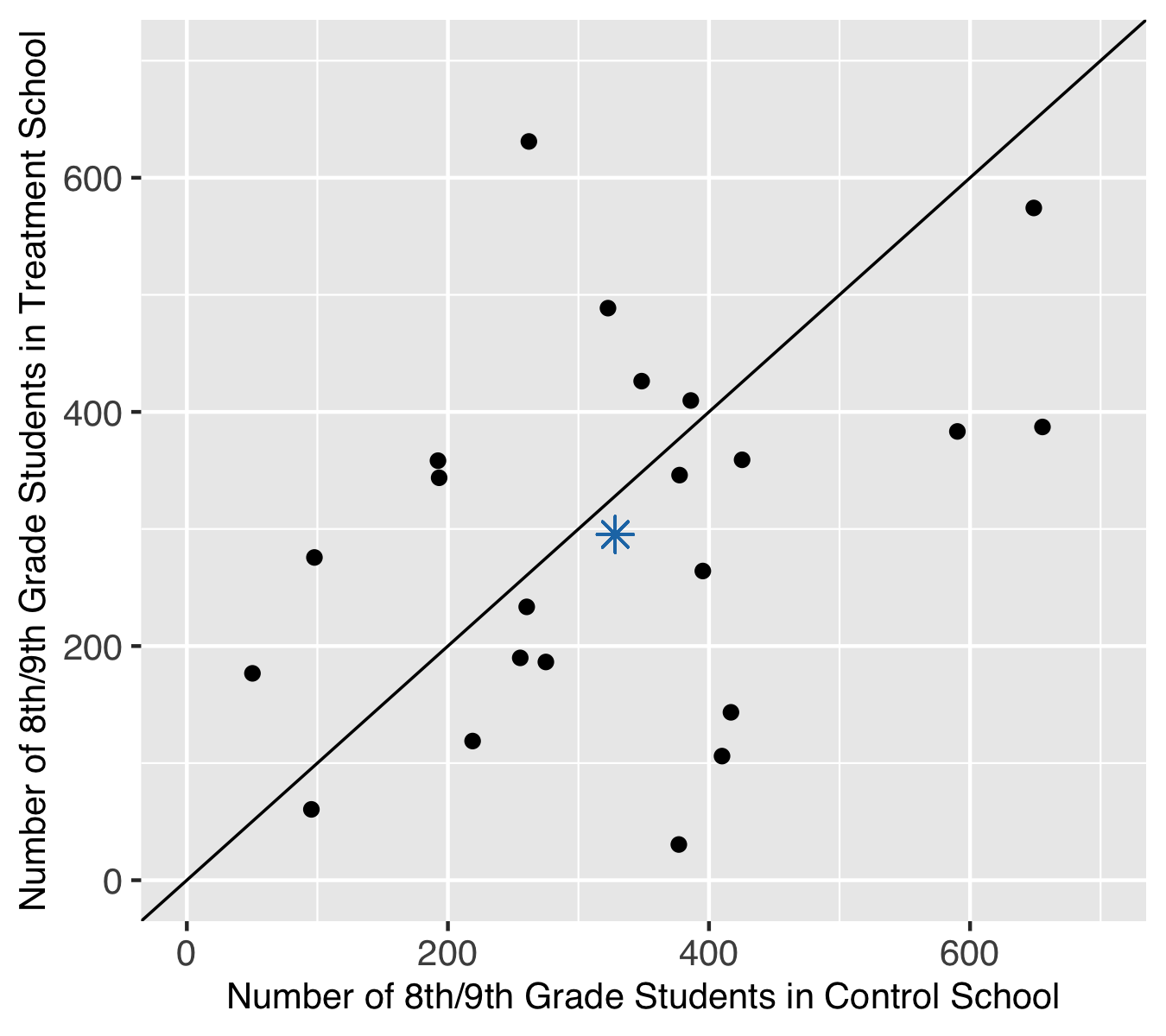}
    }
    \hfill
    \subfloat[Pretest\label{fig:cta-pre-bal}]{%
      \includegraphics[width=0.45\textwidth]{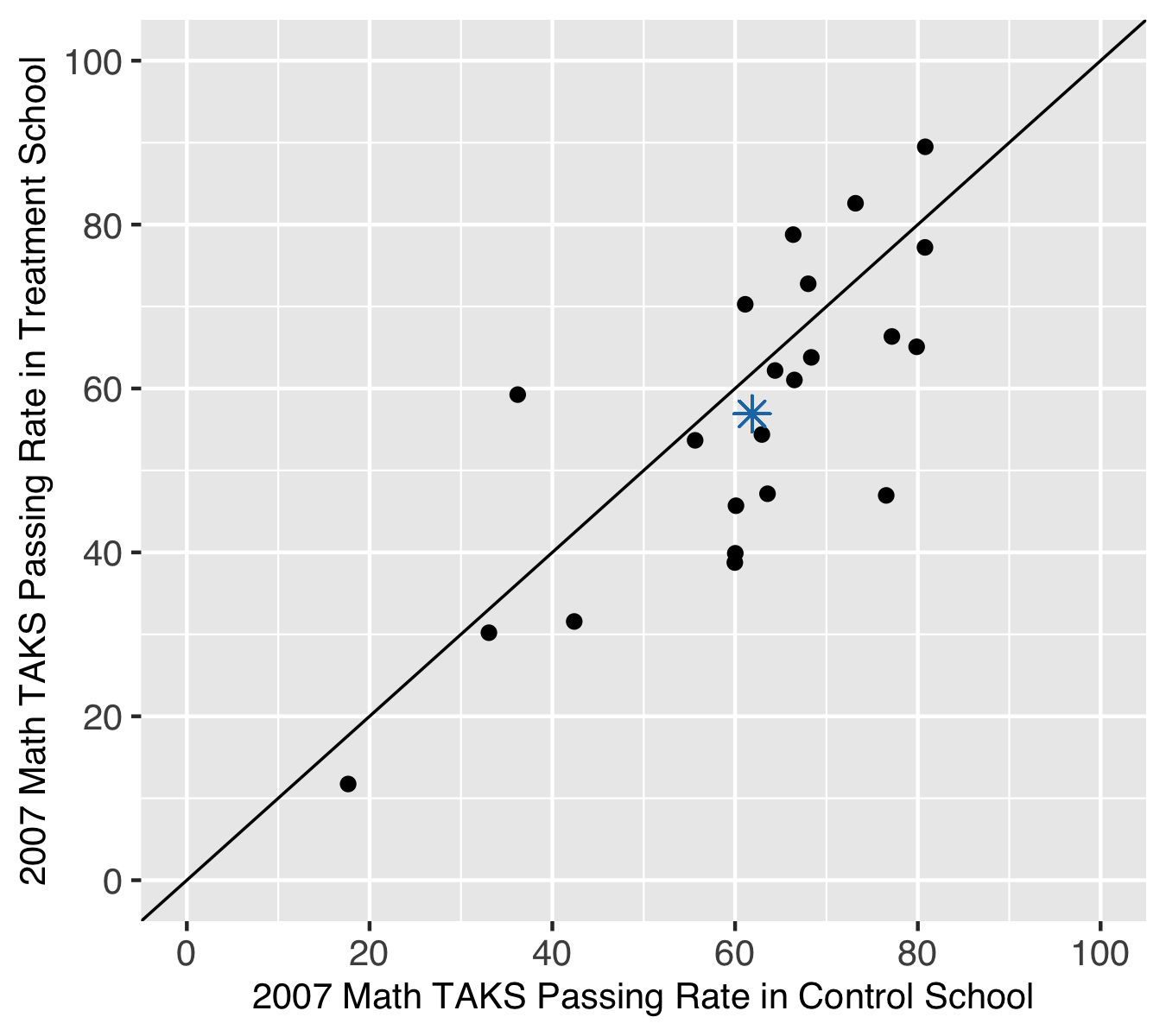}
    }
     \caption{Balance of cluster sizes and pretest scores between treatment and control schools in the CTAI study overall (blue star) and within pairs of schools (black dots). The cluster sizes are the number of students in the appropriate grade in a school (8th for middle schools and 9th for high schools) and the pretest score is the 2007 Math TAKS passing rate for the appropriate grade in a school. To preserve school anonymity, a small amount of random noise is added to the points representing pairs of treatment and control schools.}
\end{figure}

First, the CTAI study provides an example of how it can be difficult to balance baseline characteristics, including the cluster size, in practice. Figures \ref{fig:cta-n-bal} and \ref{fig:cta-pre-bal} compare the distribution of cluster sizes and pretest scores (2007 Math TAKS passing rate) between the treatment and control schools within each of the 22 pairs.  To preserve school anonymity, we add small random noise to the points, but conclusions remain the same. The blue star represents the mean amongst the control group versus the treatment group. If all pairs were perfectly matched on each variable, the dots in Figures \ref{fig:cta-n-bal} and \ref{fig:cta-pre-bal} would fall on the black diagonal line. Figure \ref{fig:cta-n-bal} shows that while \textit{on average} the cluster sizes are reasonably matched between the treatment and control groups in the CTAI study, within each pair of schools, the treatment or control school sizes can greatly differ. Also, there is a wide range in the number of 8th/9th grade students students amongst the 44 schools in the study. Figure \ref{fig:cta-pre-bal} shows that on average the treatment schools performed slightly worse than the control schools on the math TAKS, before the treatment was applied. Additionally, in 16 of the 22 pairs, the treatment school had a lower pretest score than the control school (Figure \ref{fig:cta-pre-bal}). At the same time, as is typical for a pretest score, the 2007 TAKS passing rate is highly predictive of the 2008 TAKS passing rate (explaining 83\% of the variation in the 2008 TAKS passing rate according to a simple regression). Based on a paired $t$-test, the imbalance appears to be due to chance.

To conduct ``real data simulation'' studies, we use the data and paired structure from the original CTAI trial, but imagine that we are applying a new treatment. Thus, we treat the \textit{observed} school-level 2008 TAKS passing rates as \textit{control} cluster-mean potential outcomes ($\bar{y}^c_{ik}$) and impute the set of treatment and control cluster-mean potential outcomes based on a given additive treatment effect ($\bar{y}^t_{ik} = \bar{y}^c_{ik} + \tau_{ik}$). We consider results for a constant treatment effect $\tau_{ik} = 10$ and a treatment effect correlated with cluster size $\tau_{ik} = 10 + \frac{1}{50}(\nk- \frac{N}{2M})$ (which results in an ATE of 11.77). We include an additional point estimator in these simulations - the LOO IDPD estimator using random forests to impute $d_i$ with covariates (IDPD[LOO RF,$\cdot$]) because the outcome is constrained between 0 to 100, which is not well-modeled by a linear model.

Results are based on 10,000 independent treatment assignment permutations. We calculate simulation standard errors for the MSE ratio and relative bias of variance estimators by dividing the 10,000 estimates into 100 groups of 100, estimating the respective measurements within each group, and then dividing the standard deviation of those group estimates by $\sqrt{100}$. We include results for the same estimators as in the Section~\ref{sec:sims} in the main text, and full results can be found in Supplement~\ref{chpt:appn_pcrt11}.

\begin{figure}[ht]
    \centering
    \includegraphics[width = \textwidth]{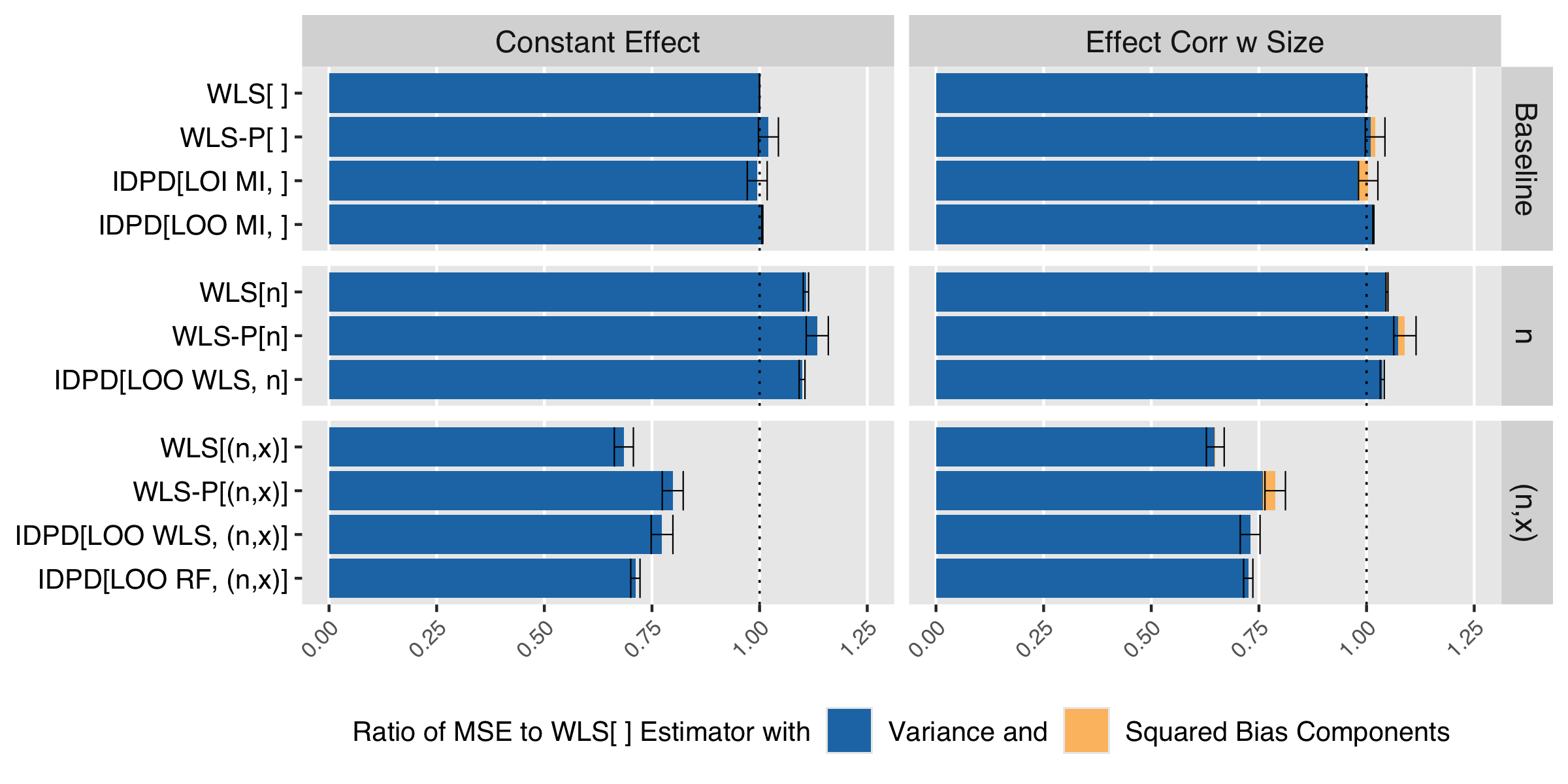}
    \caption{Ratio of the simulated MSE of treatment effect estimators, compared with $\hajek$ (the H\'ajek estimator) for a constant treatment effect or a treatment effect correlated with cluster size with the CTAI study schools. The ratio is partitioned into the variance (left, dark blue) and squared-bias (right, orange) components of the MSE. Error bars represent two simulation standard errors.}
    \label{fig:cta-ri-mse}
\end{figure}

\begin{table}[ht]
\centering
\begingroup\footnotesize
\centering
\resizebox{\ifdim\width>\linewidth\linewidth\else\width\fi}{!}{
\begin{tabular}{llcccc}
\toprule
\multicolumn{2}{c}{\textbf{Estimator}} & \multicolumn{2}{c}{\textbf{Constant Effect}} & \multicolumn{2}{c}{\textbf{Effect Corr w Size}} \\
\cmidrule(l{3pt}r{3pt}){1-2} \cmidrule(l{3pt}r{3pt}){3-4} \cmidrule(l{3pt}r{3pt}){5-6}
\multicolumn{1}{c}{\textbf{Point}} & \multicolumn{1}{c}{\textbf{Variance}} & \multicolumn{1}{c}{\textbf{Var Rel Bias}} & \multicolumn{1}{c}{\textbf{Covg.}} & \multicolumn{1}{c}{\textbf{Var Rel Bias}} & \multicolumn{1}{c}{\textbf{Covg.}}\\
\midrule
\addlinespace[0.3em]
\multicolumn{6}{l}{\textbf{Baseline}}\\
\hspace{1em}IDPD[0, $\emptyset$] & Design (\ref{eq:mavar}) & 1.56 (0.02) & 0.99 & 1.57 (0.02) & 0.99\\
\hspace{1em}WLS[$\emptyset$] & Design (\ref{eq:crhjvar}) & 0.94 (0.01) & 0.94 & 0.98 (0.01) & 0.94\\
\hspace{1em}WLS-P[$\emptyset$] & H-W Robust & 1.01 (0.01) & 0.95 & 1.05 (0.01) & 0.95\\
\hspace{1em}IDPD[LOI MI, $\emptyset$] & Design (\ref{eq:vikn}) & 1.05 (0.01) & 0.95 & 1.21 (0.02) & 0.95\\
\hspace{1em}IDPD[LOO MI, $\emptyset$] & Design (\ref{eq:vloop}) & 1.04 (0.01) & 0.95 & 1.10 (0.02) & 0.95\\
\addlinespace[0.3em]
\multicolumn{6}{l}{\textbf{Cluster Size}}\\
\hspace{1em}WLS[n] & H-W Robust & 2.42 (0.03) & 1.00 & 2.44 (0.03) & 1.00\\
\hspace{1em}WLS-P[n] & H-W Robust & 0.98 (0.01) & 0.94 & 1.02 (0.01) & 0.94\\
\hspace{1em}IDPD[LOO WLS, n] & Design (\ref{eq:vloop}) & 1.12 (0.02) & 0.95 & 1.12 (0.02) & 0.95\\
\addlinespace[0.3em]
\multicolumn{6}{l}{\textbf{Cluster Size and Pretest}}\\
\hspace{1em}WLS[(n,x)] & H-W Robust & 0.67 (0.01) & 0.88 & 0.71 (0.01) & 0.89\\
\hspace{1em}WLS-P[(n,x)] & H-W Robust & 0.92 (0.01) & 0.92 & 0.95 (0.01) & 0.92\\
\hspace{1em}IDPD[LOO WLS, (n,x)] & Design (\ref{eq:vloop}) & 1.22 (0.02) & 0.90 & 1.22 (0.02) & 0.90\\
\hspace{1em}IDPD[LOO RF, (n,x)] & Design (\ref{eq:vloop}) & 1.13 (0.02) & 0.95 & 1.17 (0.02) & 0.96\\
\bottomrule
\end{tabular}}

\endgroup
\caption{Simulated relative bias of variance estimators and coverage associated with each point estimator, using the CTAI study.  Simulation standard errors are in parentheses for the relative bias of variance estimators. The simulation standard errors for coverage probabilities are all less than 0.0035.  Design-based variance estimators include a reference to the relevant equation in this chapter. ``H-W Robust'' refers to the Huber-White heteroskedasticity robust variance estimator with HC1 structure.}
\label{tab:var-cta}
\end{table}

Figure~\ref{fig:cta-ri-mse} shows the estimated ratios of each point estimator's MSE with the MSE of $\hajek$. We exclude $\hvt$ from the figure because its MSE is much larger than the other estimators - resulting in MSE ratios of 8.5 (1.29) and 8.7 (1.39) when there is a constant effect or effect correlated with cluster size, respectively (simulation standard errors in parentheses). Otherwise, the baseline estimators perform similarly for both treatment effects. In this case, including the cluster size as a covariate hurts precision. However, adjusting with the cluster size and the pretest score (n,x) improves precision, as expected, given that the pretest is highly prognostic for the outcome. $\wlsnx$ and $\looidpdnxrf$ have the greatest precision. However, $\wlsnx$ performs better than $\looidpdnxrf$ when the treatment effect is correlated with cluster size. We suspect this is because, as currently implemented, the LOO IDPD estimator fits separate outcome models for treatment and control clusters, equivalent to including the interaction between the treatment assignment and all covariates, while $\wlsnx$ does not include these interactions (see supplementary materials, ``05-cta-apx.html).

Table~\ref{tab:var-cta} shows the estimated relative bias of variance estimators (``Var Rel Bias'') and associated coverage rates (``Covg.''). The results are similar to the simulation studies in Section~\ref{sec:sims} in the main text, except when adjusting by the cluster size and pretest score. In this case, the Huber-White robust variance estimator is very anti-conservative for the true variance of $\wlsnx$, resulting in under-coverage. This phenomenon is less severe for $\wlspnx$. We also find that $\looidpdnx$ shows under-coverage in this case, even though the associated variance estimator is conservative in expectation, indicating that there are some irregular predictions. However, the variance estimator for $\looidpdnxrf$ is conservative and shows almost exact 95\% coverage.

\FloatBarrier

\section{Additional Figures: Full Simulation Results}\label{chpt:appn_pcrt9}

This supplement contains the full simulation results for the eight settings that result from varying the presence of a pair effect, whether the cluster sizes are well matched within pairs, and whether there is a constant treatment effect. We additionally include estimators discussed in the literature that we did not include in the main paper results for brevity and focus.

The additional point estimators include WLS with a pair random effect instead of a pair fixed effect (WLS-R[$\cdot$]), the Des Raj difference estimator (IDPD[LOO DR, $\cdot$]) as implemented in \cite{middleton_unbiased_2015} and described in Supplement~\ref{chpt:appn_pcrt1}, and the weighted regression estimator of \cite{su_model-assisted_2021} described in Supplement~\ref{chpt:appn_pcrt1} (WLS-HT[$\cdot$]). For each point estimator, we also show full results for available, associated variance estimators. For design-based variance estimators, the relevant equation is included. In addition to the Huber-White robust variance estimators, we also show results for the typical parametric variance estimator for OLS or WLS (``Regression'').


\begin{figure}[ht]
    \centering
    \subfloat[M=20]{%
    \includegraphics[width = .9\textwidth]{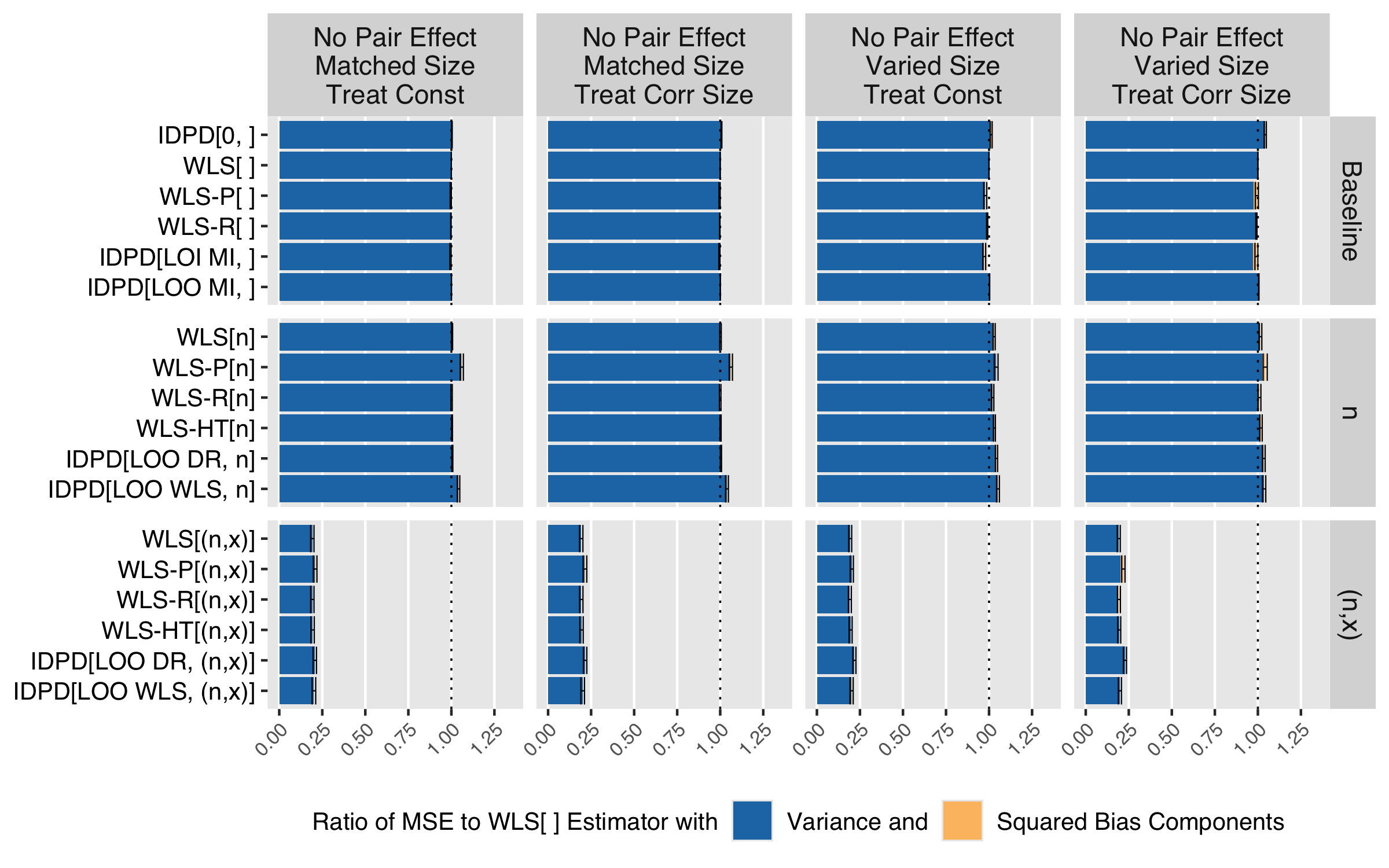}%
    }
    
    \subfloat[M=200]{%
     \includegraphics[width = .9\textwidth]{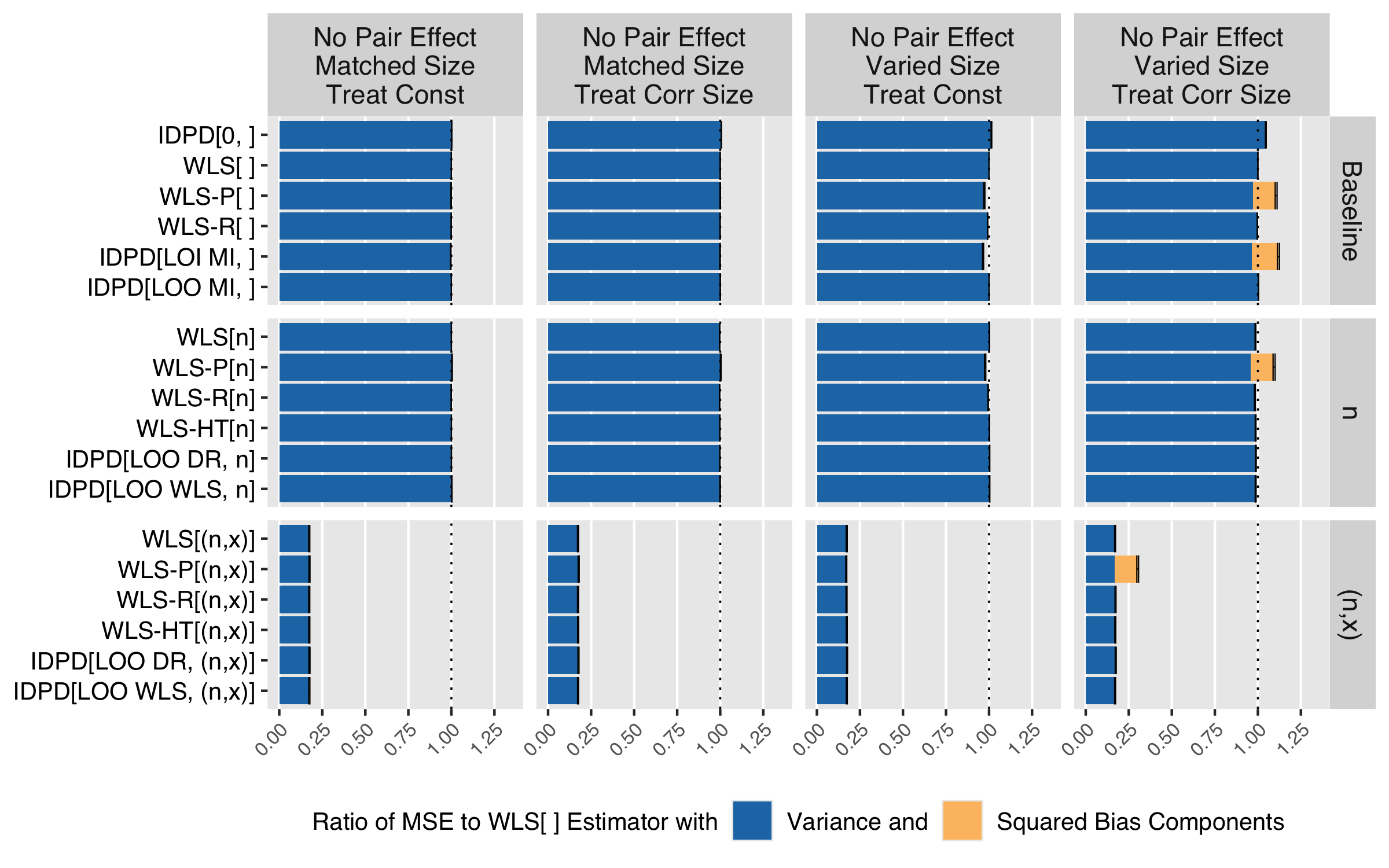}%
    }
    \caption{Ratio of the MSE of treatment effect estimators, compared with the H\'ajek estimator ($\hajek$) when there is no pair effect and there are $M = 20$ or $M= 200$ pairs.}
    \label{fig:mse-200-apdx}
\end{figure}

\begin{figure}[ht]
    \centering
    \subfloat[M=20]{%
    \includegraphics[width = .9\textwidth]{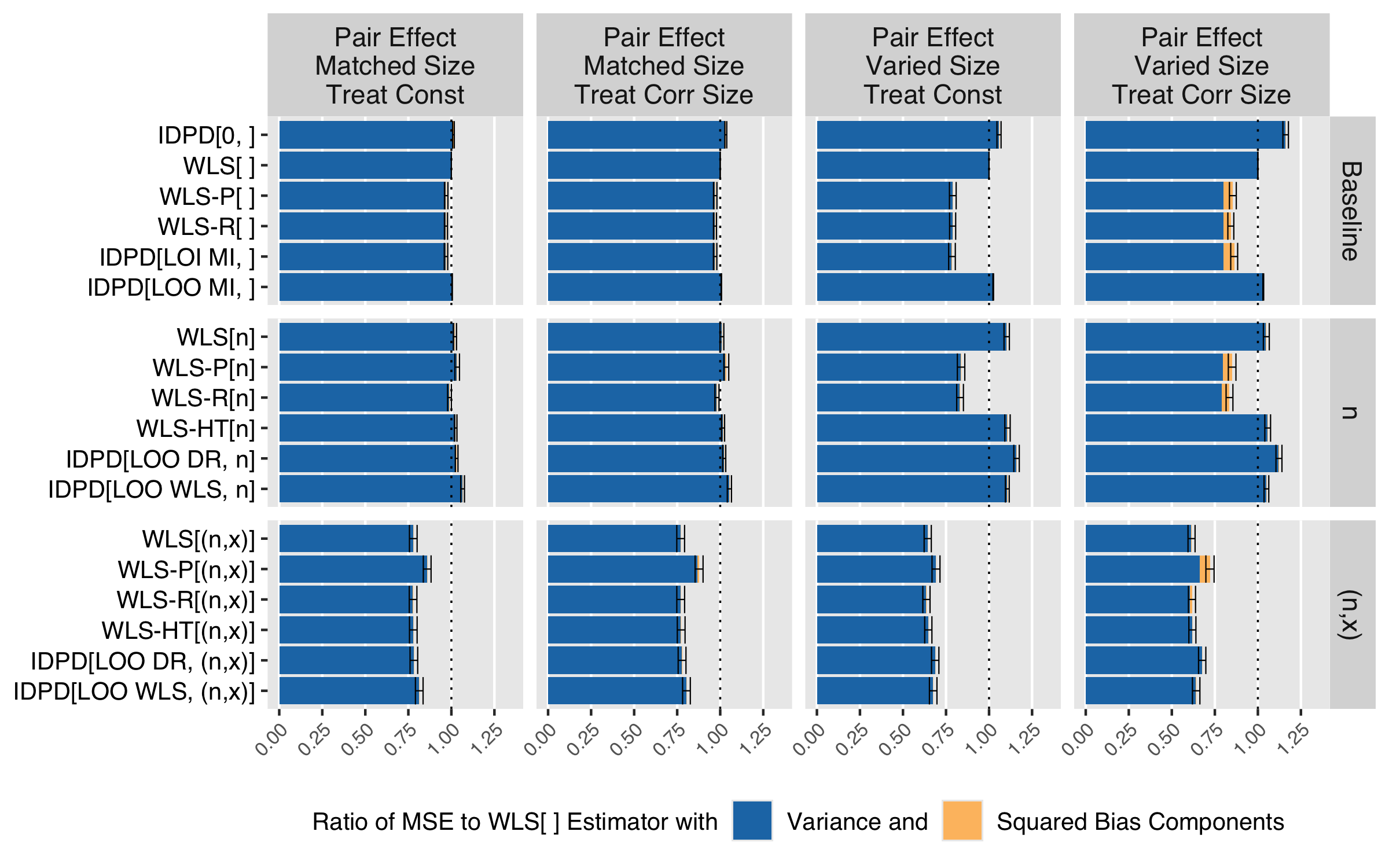}%
    }
    
    \subfloat[M=200]{%
     \includegraphics[width = .9\textwidth]{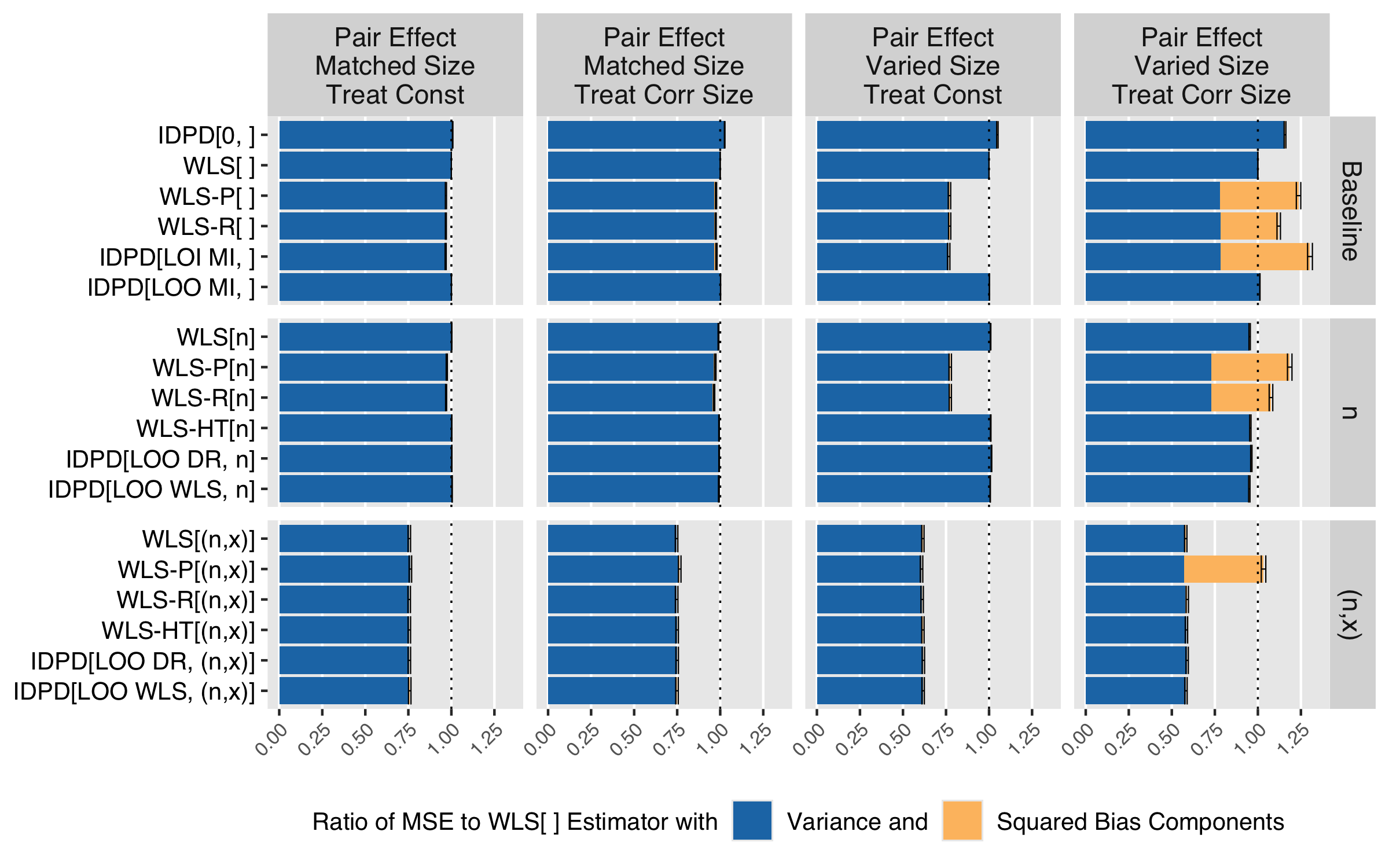}%
    }
    \caption{Ratio of the MSE of treatment effect estimators, compared with the H\'ajek estimator ($\hajek$) when there is a pair effect and there are $M = 20$ or $M= 200$ pairs.}
    \label{fig:mse-200-appdx}
\end{figure}


\begin{table}[ht]
\centering
\begingroup \footnotesize
\centering
\resizebox{\ifdim\width>\linewidth\linewidth\else\width\fi}{!}{
\begin{tabular}{llcccc}
\toprule
\multicolumn{6}{c}{\textbf{Relative Bias of Variance Estimators}} \\
\multicolumn{6}{c}{\textbf{M = 20}} \\
\cmidrule(l{3pt}r{3pt}){1-6}
\multicolumn{2}{c}{\textbf{ }} & \multicolumn{4}{c}{\textbf{No Pair Effect}} \\
\multicolumn{2}{c}{\textbf{ }} & \multicolumn{2}{c}{\textbf{Matched Size}} & \multicolumn{2}{c}{\textbf{Varied Size}} \\
\multicolumn{1}{c}{\textbf{\textbf{Method}}} & \multicolumn{1}{c}{\textbf{Var Estimator}} & \multicolumn{1}{c}{\textbf{Tr Const}} & \multicolumn{1}{c}{\textbf{Tr Cor. Size}} & \multicolumn{1}{c}{\textbf{Tr Const}} & \multicolumn{1}{c}{\textbf{Tr Cor. Size}}\\
\midrule
\addlinespace[0.3em]
\multicolumn{6}{l}{\textbf{Baseline}}\\
\hspace{1em}\cellcolor{gray!10}{IDPD[0, $\emptyset$]} & \cellcolor{gray!10}{H-W Robust} & \cellcolor{gray!10}{1.08 (0.015)} & \cellcolor{gray!10}{1.19 (0.017)} & \cellcolor{gray!10}{1.05 (0.013)} & \cellcolor{gray!10}{1.10 (0.014)}\\
\hspace{1em}\cellcolor{gray!10}{IDPD[0, $\emptyset$]} & \cellcolor{gray!10}{Design (\ref{eq:mavar})} & \cellcolor{gray!10}{1.28 (0.017)} & \cellcolor{gray!10}{1.42 (0.020)} & \cellcolor{gray!10}{1.24 (0.015)} & \cellcolor{gray!10}{1.32 (0.016)}\\
\hspace{1em}WLS[$\emptyset$] & H-W Robust & 1.05 (0.014) & 1.08 (0.015) & 1.03 (0.013) & 1.05 (0.013)\\
\hspace{1em}WLS[$\emptyset$] & Regression & 0.98 (0.014) & 1.02 (0.014) & 0.97 (0.012) & 0.99 (0.013)\\
\hspace{1em}WLS[$\emptyset$] & Design (\ref{eq:crhjvar}) & 0.95 (0.003) & 0.98 (0.003) & 0.95 (0.003) & 0.96 (0.003)\\
\hspace{1em}\cellcolor{gray!10}{WLS-P[$\emptyset$]} & \cellcolor{gray!10}{H-W Robust} & \cellcolor{gray!10}{1.00 (0.003)} & \cellcolor{gray!10}{1.04 (0.003)} & \cellcolor{gray!10}{1.04 (0.003)} & \cellcolor{gray!10}{1.06 (0.003)}\\
\hspace{1em}\cellcolor{gray!10}{WLS-P[$\emptyset$]} & \cellcolor{gray!10}{Regression} & \cellcolor{gray!10}{0.94 (0.005)} & \cellcolor{gray!10}{0.98 (0.005)} & \cellcolor{gray!10}{0.97 (0.005)} & \cellcolor{gray!10}{0.99 (0.005)}\\
\hspace{1em}\cellcolor{gray!10}{WLS-P[$\emptyset$]} & \cellcolor{gray!10}{Design (\ref{eq:crhjvar})} & \cellcolor{gray!10}{0.95 (0.003)} & \cellcolor{gray!10}{0.98 (0.003)} & \cellcolor{gray!10}{0.95 (0.003)} & \cellcolor{gray!10}{0.97 (0.003)}\\
\hspace{1em}\cellcolor{gray!10}{WLS-P[$\emptyset$]} & \cellcolor{gray!10}{Design (\ref{eq:vikn})} & \cellcolor{gray!10}{1.03 (0.003)} & \cellcolor{gray!10}{1.14 (0.005)} & \cellcolor{gray!10}{1.02 (0.003)} & \cellcolor{gray!10}{1.08 (0.004)}\\
\hspace{1em}\cellcolor{gray!10}{WLS-P[$\emptyset$]} & \cellcolor{gray!10}{Design (\ref{eq:vspmk})} & \cellcolor{gray!10}{1.00 (0.003)} & \cellcolor{gray!10}{1.04 (0.003)} & \cellcolor{gray!10}{1.04 (0.003)} & \cellcolor{gray!10}{1.06 (0.003)}\\
\hspace{1em}WLS-R[$\emptyset$] & Regression & 0.85 (0.007) & 0.88 (0.007) & 0.85 (0.006) & 0.87 (0.006)\\
\hspace{1em}\cellcolor{gray!10}{IDPD[LOI MI, $\emptyset$]} & \cellcolor{gray!10}{H-W Robust} & \cellcolor{gray!10}{1.05 (0.015)} & \cellcolor{gray!10}{1.08 (0.015)} & \cellcolor{gray!10}{1.04 (0.014)} & \cellcolor{gray!10}{1.06 (0.014)}\\
\hspace{1em}\cellcolor{gray!10}{IDPD[LOI MI, $\emptyset$]} & \cellcolor{gray!10}{Regression} & \cellcolor{gray!10}{0.99 (0.014)} & \cellcolor{gray!10}{1.02 (0.014)} & \cellcolor{gray!10}{1.01 (0.014)} & \cellcolor{gray!10}{1.03 (0.014)}\\
\hspace{1em}\cellcolor{gray!10}{IDPD[LOI MI, $\emptyset$]} & \cellcolor{gray!10}{Design (\ref{eq:vikn})} & \cellcolor{gray!10}{1.03 (0.003)} & \cellcolor{gray!10}{1.14 (0.005)} & \cellcolor{gray!10}{1.02 (0.003)} & \cellcolor{gray!10}{1.07 (0.004)}\\
\hspace{1em}IDPD[LOO MI, $\emptyset$] & Design (\ref{eq:vloop}) & 1.06 (0.003) & 1.11 (0.004) & 1.06 (0.003) & 1.08 (0.003)\\
\addlinespace[0.3em]
\multicolumn{6}{l}{\textbf{Cluster Size}}\\
\hspace{1em}\cellcolor{gray!10}{WLS[n]} & \cellcolor{gray!10}{H-W Robust} & \cellcolor{gray!10}{1.04 (0.014)} & \cellcolor{gray!10}{1.06 (0.014)} & \cellcolor{gray!10}{1.02 (0.013)} & \cellcolor{gray!10}{1.04 (0.013)}\\
\hspace{1em}WLS-P[n] & H-W Robust & 0.99 (0.003) & 1.02 (0.004) & 1.02 (0.004) & 1.04 (0.004)\\
\hspace{1em}WLS-P[n] & Design (\ref{eq:vspmk}) & 0.89 (0.003) & 0.92 (0.003) & 0.93 (0.003) & 0.95 (0.003)\\
\hspace{1em}\cellcolor{gray!10}{WLS-R[n]} & \cellcolor{gray!10}{Regression} & \cellcolor{gray!10}{0.86 (0.007)} & \cellcolor{gray!10}{0.88 (0.007)} & \cellcolor{gray!10}{0.85 (0.006)} & \cellcolor{gray!10}{0.87 (0.006)}\\
\hspace{1em}WLS-HT[n] & H-W Robust & 1.05 (0.014) & 1.06 (0.014) & 1.03 (0.013) & 1.03 (0.013)\\
\hspace{1em}\cellcolor{gray!10}{IDPD[LOO DR, n]} & \cellcolor{gray!10}{Design (\ref{eq:mavar})} & \cellcolor{gray!10}{1.30 (0.018)} & \cellcolor{gray!10}{1.40 (0.019)} & \cellcolor{gray!10}{1.23 (0.015)} & \cellcolor{gray!10}{1.31 (0.016)}\\
\hspace{1em}IDPD[LOO WLS, n] & Design (\ref{eq:vloop}) & 1.12 (0.004) & 1.12 (0.004) & 1.11 (0.004) & 1.11 (0.004)\\
\addlinespace[0.3em]
\multicolumn{6}{l}{\textbf{Covariate and Cluster Size}}\\
\hspace{1em}\cellcolor{gray!10}{WLS[(n,x)]} & \cellcolor{gray!10}{H-W Robust} & \cellcolor{gray!10}{1.08 (0.019)} & \cellcolor{gray!10}{1.19 (0.021)} & \cellcolor{gray!10}{1.04 (0.014)} & \cellcolor{gray!10}{1.15 (0.015)}\\
\hspace{1em}WLS-P[(n,x)] & H-W Robust & 0.97 (0.004) & 1.17 (0.007) & 1.01 (0.004) & 1.13 (0.005)\\
\hspace{1em}WLS-P[(n,x)] & Design (\ref{eq:vspmk}) & 0.78 (0.003) & 0.93 (0.005) & 0.82 (0.003) & 0.91 (0.004)\\
\hspace{1em}\cellcolor{gray!10}{WLS-R[(n,x)]} & \cellcolor{gray!10}{Regression} & \cellcolor{gray!10}{0.87 (0.007)} & \cellcolor{gray!10}{1.02 (0.011)} & \cellcolor{gray!10}{0.87 (0.006)} & \cellcolor{gray!10}{0.96 (0.008)}\\
\hspace{1em}WLS-HT[(n,x)] & H-W Robust & 1.08 (0.019) & 1.09 (0.019) & 1.04 (0.014) & 1.04 (0.014)\\
\hspace{1em}\cellcolor{gray!10}{IDPD[LOO DR, (n,x)]} & \cellcolor{gray!10}{Design (\ref{eq:mavar})} & \cellcolor{gray!10}{2.12 (0.036)} & \cellcolor{gray!10}{2.54 (0.043)} & \cellcolor{gray!10}{1.91 (0.024)} & \cellcolor{gray!10}{2.24 (0.029)}\\
\hspace{1em}IDPD[LOO WLS, (n,x)] & Design (\ref{eq:vloop}) & 1.15 (0.004) & 1.15 (0.004) & 1.14 (0.004) & 1.14 (0.004)\\
\bottomrule
\end{tabular}}

\endgroup
\caption{Simulated \textbf{relative bias} of variance estimators associated with each point estimator when there are \textbf{20 pairs} and \textbf{no pair effect} with simulation standard errors in parentheses. Relative bias is calculated as the empirical mean of the variance estimator divided by the empirical variance of the point estimator.}
\label{tab:var-20-no}
\end{table}

\begin{table}[ht]
\centering
\begingroup \footnotesize
\centering
\resizebox{\ifdim\width>\linewidth\linewidth\else\width\fi}{!}{
\begin{tabular}{llcccc}
\toprule
\multicolumn{6}{c}{\textbf{Relative Bias of Variance Estimators}} \\
\multicolumn{6}{c}{\textbf{M = 20}} \\
\cmidrule(l{3pt}r{3pt}){1-6}
\multicolumn{2}{c}{\textbf{ }} & \multicolumn{4}{c}{\textbf{Pair Effect}} \\
\multicolumn{2}{c}{\textbf{ }} & \multicolumn{2}{c}{\textbf{Matched Size}} & \multicolumn{2}{c}{\textbf{Varied Size}} \\
\multicolumn{1}{c}{\textbf{\textbf{Method}}} & \multicolumn{1}{c}{\textbf{Var Estimator}} & \multicolumn{1}{c}{\textbf{Tr Const}} & \multicolumn{1}{c}{\textbf{Tr Cor. Size}} & \multicolumn{1}{c}{\textbf{Tr Const}} & \multicolumn{1}{c}{\textbf{Tr Cor. Size}}\\
\midrule
\addlinespace[0.3em]
\multicolumn{6}{l}{\textbf{Baseline}}\\
\hspace{1em}\cellcolor{gray!10}{IDPD[0, $\emptyset$]} & \cellcolor{gray!10}{H-W Robust} & \cellcolor{gray!10}{5.16 (0.126)} & \cellcolor{gray!10}{5.47 (0.130)} & \cellcolor{gray!10}{3.68 (0.070)} & \cellcolor{gray!10}{3.47 (0.061)}\\
\hspace{1em}\cellcolor{gray!10}{IDPD[0, $\emptyset$]} & \cellcolor{gray!10}{Design (\ref{eq:mavar})} & \cellcolor{gray!10}{5.95 (0.140)} & \cellcolor{gray!10}{6.42 (0.146)} & \cellcolor{gray!10}{4.27 (0.077)} & \cellcolor{gray!10}{4.10 (0.070)}\\
\hspace{1em}WLS[$\emptyset$] & H-W Robust & 5.03 (0.127) & 5.14 (0.127) & 3.73 (0.072) & 3.64 (0.068)\\
\hspace{1em}WLS[$\emptyset$] & Regression & 4.74 (0.121) & 4.85 (0.121) & 3.52 (0.070) & 3.43 (0.065)\\
\hspace{1em}WLS[$\emptyset$] & Design (\ref{eq:crhjvar}) & 0.94 (0.003) & 1.09 (0.006) & 0.94 (0.003) & 1.00 (0.003)\\
\hspace{1em}\cellcolor{gray!10}{WLS-P[$\emptyset$]} & \cellcolor{gray!10}{H-W Robust} & \cellcolor{gray!10}{1.00 (0.003)} & \cellcolor{gray!10}{1.16 (0.006)} & \cellcolor{gray!10}{1.04 (0.003)} & \cellcolor{gray!10}{1.13 (0.004)}\\
\hspace{1em}\cellcolor{gray!10}{WLS-P[$\emptyset$]} & \cellcolor{gray!10}{Regression} & \cellcolor{gray!10}{0.95 (0.005)} & \cellcolor{gray!10}{1.11 (0.008)} & \cellcolor{gray!10}{0.97 (0.004)} & \cellcolor{gray!10}{1.06 (0.005)}\\
\hspace{1em}\cellcolor{gray!10}{WLS-P[$\emptyset$]} & \cellcolor{gray!10}{Design (\ref{eq:crhjvar})} & \cellcolor{gray!10}{0.94 (0.003)} & \cellcolor{gray!10}{1.10 (0.006)} & \cellcolor{gray!10}{0.95 (0.003)} & \cellcolor{gray!10}{1.03 (0.004)}\\
\hspace{1em}\cellcolor{gray!10}{WLS-P[$\emptyset$]} & \cellcolor{gray!10}{Design (\ref{eq:vikn})} & \cellcolor{gray!10}{1.13 (0.005)} & \cellcolor{gray!10}{1.62 (0.017)} & \cellcolor{gray!10}{1.08 (0.004)} & \cellcolor{gray!10}{1.32 (0.008)}\\
\hspace{1em}\cellcolor{gray!10}{WLS-P[$\emptyset$]} & \cellcolor{gray!10}{Design (\ref{eq:vspmk})} & \cellcolor{gray!10}{1.00 (0.003)} & \cellcolor{gray!10}{1.16 (0.006)} & \cellcolor{gray!10}{1.04 (0.003)} & \cellcolor{gray!10}{1.13 (0.004)}\\
\hspace{1em}WLS-R[$\emptyset$] & Regression & 0.95 (0.005) & 1.11 (0.008) & 0.96 (0.005) & 1.04 (0.006)\\
\hspace{1em}\cellcolor{gray!10}{IDPD[LOI MI, $\emptyset$]} & \cellcolor{gray!10}{H-W Robust} & \cellcolor{gray!10}{5.24 (0.133)} & \cellcolor{gray!10}{5.35 (0.134)} & \cellcolor{gray!10}{4.86 (0.109)} & \cellcolor{gray!10}{4.63 (0.098)}\\
\hspace{1em}\cellcolor{gray!10}{IDPD[LOI MI, $\emptyset$]} & \cellcolor{gray!10}{Regression} & \cellcolor{gray!10}{4.95 (0.127)} & \cellcolor{gray!10}{5.07 (0.128)} & \cellcolor{gray!10}{4.72 (0.105)} & \cellcolor{gray!10}{4.49 (0.094)}\\
\hspace{1em}\cellcolor{gray!10}{IDPD[LOI MI, $\emptyset$]} & \cellcolor{gray!10}{Design (\ref{eq:vikn})} & \cellcolor{gray!10}{1.13 (0.005)} & \cellcolor{gray!10}{1.61 (0.017)} & \cellcolor{gray!10}{1.07 (0.003)} & \cellcolor{gray!10}{1.29 (0.007)}\\
\hspace{1em}IDPD[LOO MI, $\emptyset$] & Design (\ref{eq:vloop}) & 1.06 (0.003) & 1.27 (0.007) & 1.05 (0.003) & 1.13 (0.004)\\
\addlinespace[0.3em]
\multicolumn{6}{l}{\textbf{Cluster Size}}\\
\hspace{1em}\cellcolor{gray!10}{WLS[n]} & \cellcolor{gray!10}{H-W Robust} & \cellcolor{gray!10}{4.85 (0.128)} & \cellcolor{gray!10}{4.94 (0.130)} & \cellcolor{gray!10}{3.48 (0.068)} & \cellcolor{gray!10}{3.54 (0.068)}\\
\hspace{1em}WLS-P[n] & H-W Robust & 0.98 (0.004) & 1.14 (0.006) & 1.03 (0.003) & 1.12 (0.004)\\
\hspace{1em}WLS-P[n] & Design (\ref{eq:vspmk}) & 0.89 (0.003) & 1.03 (0.005) & 0.93 (0.003) & 1.01 (0.004)\\
\hspace{1em}\cellcolor{gray!10}{WLS-R[n]} & \cellcolor{gray!10}{Regression} & \cellcolor{gray!10}{0.95 (0.005)} & \cellcolor{gray!10}{1.11 (0.008)} & \cellcolor{gray!10}{0.96 (0.005)} & \cellcolor{gray!10}{1.04 (0.006)}\\
\hspace{1em}WLS-HT[n] & H-W Robust & 4.96 (0.127) & 4.96 (0.129) & 3.49 (0.069) & 3.49 (0.070)\\
\hspace{1em}\cellcolor{gray!10}{IDPD[LOO DR, n]} & \cellcolor{gray!10}{Design (\ref{eq:mavar})} & \cellcolor{gray!10}{6.02 (0.147)} & \cellcolor{gray!10}{6.44 (0.157)} & \cellcolor{gray!10}{3.96 (0.069)} & \cellcolor{gray!10}{4.17 (0.072)}\\
\hspace{1em}IDPD[LOO WLS, n] & Design (\ref{eq:vloop}) & 1.14 (0.004) & 1.14 (0.004) & 1.16 (0.004) & 1.16 (0.004)\\
\addlinespace[0.3em]
\multicolumn{6}{l}{\textbf{Covariate and Cluster Size}}\\
\hspace{1em}\cellcolor{gray!10}{WLS[(n,x)]} & \cellcolor{gray!10}{H-W Robust} & \cellcolor{gray!10}{1.06 (0.017)} & \cellcolor{gray!10}{1.18 (0.019)} & \cellcolor{gray!10}{1.04 (0.013)} & \cellcolor{gray!10}{1.15 (0.015)}\\
\hspace{1em}WLS-P[(n,x)] & H-W Robust & 0.97 (0.004) & 1.17 (0.007) & 1.01 (0.004) & 1.13 (0.005)\\
\hspace{1em}WLS-P[(n,x)] & Design (\ref{eq:vspmk}) & 0.78 (0.003) & 0.93 (0.005) & 0.82 (0.003) & 0.91 (0.004)\\
\hspace{1em}\cellcolor{gray!10}{WLS-R[(n,x)]} & \cellcolor{gray!10}{Regression} & \cellcolor{gray!10}{0.87 (0.007)} & \cellcolor{gray!10}{1.02 (0.011)} & \cellcolor{gray!10}{0.87 (0.006)} & \cellcolor{gray!10}{0.97 (0.008)}\\
\hspace{1em}WLS-HT[(n,x)] & H-W Robust & 1.07 (0.018) & 1.08 (0.018) & 1.04 (0.014) & 1.05 (0.014)\\
\hspace{1em}\cellcolor{gray!10}{IDPD[LOO DR, (n,x)]} & \cellcolor{gray!10}{Design (\ref{eq:mavar})} & \cellcolor{gray!10}{2.31 (0.043)} & \cellcolor{gray!10}{2.90 (0.057)} & \cellcolor{gray!10}{2.07 (0.028)} & \cellcolor{gray!10}{2.51 (0.035)}\\
\hspace{1em}IDPD[LOO WLS, (n,x)] & Design (\ref{eq:vloop}) & 1.16 (0.004) & 1.16 (0.004) & 1.15 (0.004) & 1.15 (0.004)\\
\bottomrule
\end{tabular}}

\endgroup
\caption{Simulated \textbf{relative bias} of variance estimators associated with each point estimator when there are \textbf{20 pairs} and a \textbf{pair effect} with simulation standard errors in parentheses.}
\label{tab:var-20-pair}
\end{table}

\begin{table}[ht]
\centering
\begingroup \footnotesize
\centering
\resizebox{\ifdim\width>\linewidth\linewidth\else\width\fi}{!}{
\begin{tabular}{llcccc}
\toprule
\multicolumn{6}{c}{\textbf{Relative Bias of Variance Estimators}} \\
\multicolumn{6}{c}{\textbf{M = 200}} \\
\cmidrule(l{3pt}r{3pt}){1-6}
\multicolumn{2}{c}{\textbf{ }} & \multicolumn{4}{c}{\textbf{No Pair Effect}} \\
\multicolumn{2}{c}{\textbf{ }} & \multicolumn{2}{c}{\textbf{Matched Size}} & \multicolumn{2}{c}{\textbf{Varied Size}} \\
\multicolumn{1}{c}{\textbf{\textbf{Method}}} & \multicolumn{1}{c}{\textbf{Var Estimator}} & \multicolumn{1}{c}{\textbf{Tr Const}} & \multicolumn{1}{c}{\textbf{Tr Cor. Size}} & \multicolumn{1}{c}{\textbf{Tr Const}} & \multicolumn{1}{c}{\textbf{Tr Cor. Size}}\\
\midrule
\addlinespace[0.3em]
\multicolumn{6}{l}{\textbf{Baseline}}\\
\hspace{1em}\cellcolor{gray!10}{IDPD[0, $\emptyset$]} & \cellcolor{gray!10}{H-W Robust} & \cellcolor{gray!10}{1.03 (0.005)} & \cellcolor{gray!10}{1.12 (0.006)} & \cellcolor{gray!10}{1.02 (0.005)} & \cellcolor{gray!10}{1.06 (0.005)}\\
\hspace{1em}\cellcolor{gray!10}{IDPD[0, $\emptyset$]} & \cellcolor{gray!10}{Design (\ref{eq:mavar})} & \cellcolor{gray!10}{1.17 (0.006)} & \cellcolor{gray!10}{1.30 (0.007)} & \cellcolor{gray!10}{1.17 (0.005)} & \cellcolor{gray!10}{1.23 (0.006)}\\
\hspace{1em}WLS[$\emptyset$] & H-W Robust & 1.00 (0.005) & 1.03 (0.005) & 1.01 (0.005) & 1.02 (0.005)\\
\hspace{1em}WLS[$\emptyset$] & Regression & 0.93 (0.005) & 0.96 (0.005) & 0.93 (0.004) & 0.95 (0.004)\\
\hspace{1em}WLS[$\emptyset$] & Design (\ref{eq:crhjvar}) & 1.00 (0.003) & 1.03 (0.003) & 1.00 (0.003) & 1.02 (0.003)\\
\hspace{1em}\cellcolor{gray!10}{WLS-P[$\emptyset$]} & \cellcolor{gray!10}{H-W Robust} & \cellcolor{gray!10}{1.01 (0.003)} & \cellcolor{gray!10}{1.04 (0.003)} & \cellcolor{gray!10}{1.04 (0.003)} & \cellcolor{gray!10}{1.06 (0.003)}\\
\hspace{1em}\cellcolor{gray!10}{WLS-P[$\emptyset$]} & \cellcolor{gray!10}{Regression} & \cellcolor{gray!10}{0.93 (0.003)} & \cellcolor{gray!10}{0.96 (0.003)} & \cellcolor{gray!10}{0.96 (0.003)} & \cellcolor{gray!10}{0.98 (0.003)}\\
\hspace{1em}\cellcolor{gray!10}{WLS-P[$\emptyset$]} & \cellcolor{gray!10}{Design (\ref{eq:crhjvar})} & \cellcolor{gray!10}{1.00 (0.003)} & \cellcolor{gray!10}{1.03 (0.003)} & \cellcolor{gray!10}{1.00 (0.003)} & \cellcolor{gray!10}{1.02 (0.003)}\\
\hspace{1em}\cellcolor{gray!10}{WLS-P[$\emptyset$]} & \cellcolor{gray!10}{Design (\ref{eq:vikn})} & \cellcolor{gray!10}{1.03 (0.003)} & \cellcolor{gray!10}{1.12 (0.003)} & \cellcolor{gray!10}{1.02 (0.003)} & \cellcolor{gray!10}{1.08 (0.003)}\\
\hspace{1em}\cellcolor{gray!10}{WLS-P[$\emptyset$]} & \cellcolor{gray!10}{Design (\ref{eq:vspmk})} & \cellcolor{gray!10}{1.01 (0.003)} & \cellcolor{gray!10}{1.04 (0.003)} & \cellcolor{gray!10}{1.04 (0.003)} & \cellcolor{gray!10}{1.06 (0.003)}\\
\hspace{1em}WLS-R[$\emptyset$] & Regression & 0.87 (0.003) & 0.90 (0.003) & 0.88 (0.003) & 0.89 (0.003)\\
\hspace{1em}\cellcolor{gray!10}{IDPD[LOI MI, $\emptyset$]} & \cellcolor{gray!10}{H-W Robust} & \cellcolor{gray!10}{1.00 (0.005)} & \cellcolor{gray!10}{1.03 (0.005)} & \cellcolor{gray!10}{1.01 (0.005)} & \cellcolor{gray!10}{1.02 (0.005)}\\
\hspace{1em}\cellcolor{gray!10}{IDPD[LOI MI, $\emptyset$]} & \cellcolor{gray!10}{Regression} & \cellcolor{gray!10}{0.93 (0.005)} & \cellcolor{gray!10}{0.96 (0.005)} & \cellcolor{gray!10}{0.97 (0.005)} & \cellcolor{gray!10}{0.99 (0.005)}\\
\hspace{1em}\cellcolor{gray!10}{IDPD[LOI MI, $\emptyset$]} & \cellcolor{gray!10}{Design (\ref{eq:vikn})} & \cellcolor{gray!10}{1.03 (0.003)} & \cellcolor{gray!10}{1.12 (0.003)} & \cellcolor{gray!10}{1.02 (0.003)} & \cellcolor{gray!10}{1.07 (0.003)}\\
\hspace{1em}IDPD[LOO MI, $\emptyset$] & Design (\ref{eq:vloop}) & 1.01 (0.003) & 1.05 (0.003) & 1.01 (0.003) & 1.03 (0.003)\\
\addlinespace[0.3em]
\multicolumn{6}{l}{\textbf{Cluster Size}}\\
\hspace{1em}\cellcolor{gray!10}{WLS[n]} & \cellcolor{gray!10}{H-W Robust} & \cellcolor{gray!10}{1.00 (0.005)} & \cellcolor{gray!10}{1.02 (0.005)} & \cellcolor{gray!10}{1.01 (0.005)} & \cellcolor{gray!10}{1.02 (0.005)}\\
\hspace{1em}WLS-P[n] & H-W Robust & 1.01 (0.003) & 1.04 (0.003) & 1.04 (0.003) & 1.06 (0.003)\\
\hspace{1em}WLS-P[n] & Design (\ref{eq:vspmk}) & 1.00 (0.003) & 1.03 (0.003) & 1.03 (0.003) & 1.05 (0.003)\\
\hspace{1em}\cellcolor{gray!10}{WLS-R[n]} & \cellcolor{gray!10}{Regression} & \cellcolor{gray!10}{0.87 (0.003)} & \cellcolor{gray!10}{0.89 (0.003)} & \cellcolor{gray!10}{0.88 (0.003)} & \cellcolor{gray!10}{0.89 (0.003)}\\
\hspace{1em}WLS-HT[n] & H-W Robust & 1.00 (0.005) & 1.01 (0.005) & 1.01 (0.005) & 1.01 (0.005)\\
\hspace{1em}\cellcolor{gray!10}{IDPD[LOO DR, n]} & \cellcolor{gray!10}{Design (\ref{eq:mavar})} & \cellcolor{gray!10}{1.17 (0.006)} & \cellcolor{gray!10}{1.25 (0.006)} & \cellcolor{gray!10}{1.17 (0.005)} & \cellcolor{gray!10}{1.25 (0.006)}\\
\hspace{1em}IDPD[LOO WLS, n] & Design (\ref{eq:vloop}) & 1.02 (0.003) & 1.02 (0.003) & 1.02 (0.003) & 1.02 (0.003)\\
\addlinespace[0.3em]
\multicolumn{6}{l}{\textbf{Covariate and Cluster Size}}\\
\hspace{1em}\cellcolor{gray!10}{WLS[(n,x)]} & \cellcolor{gray!10}{H-W Robust} & \cellcolor{gray!10}{1.00 (0.005)} & \cellcolor{gray!10}{1.10 (0.005)} & \cellcolor{gray!10}{1.01 (0.005)} & \cellcolor{gray!10}{1.11 (0.005)}\\
\hspace{1em}WLS-P[(n,x)] & H-W Robust & 1.00 (0.003) & 1.18 (0.004) & 1.04 (0.003) & 1.15 (0.004)\\
\hspace{1em}WLS-P[(n,x)] & Design (\ref{eq:vspmk}) & 0.98 (0.003) & 1.16 (0.004) & 1.02 (0.003) & 1.13 (0.004)\\
\hspace{1em}\cellcolor{gray!10}{WLS-R[(n,x)]} & \cellcolor{gray!10}{Regression} & \cellcolor{gray!10}{0.88 (0.003)} & \cellcolor{gray!10}{1.01 (0.004)} & \cellcolor{gray!10}{0.89 (0.003)} & \cellcolor{gray!10}{0.98 (0.004)}\\
\hspace{1em}WLS-HT[(n,x)] & H-W Robust & 1.00 (0.005) & 1.01 (0.005) & 1.01 (0.005) & 1.02 (0.005)\\
\hspace{1em}\cellcolor{gray!10}{IDPD[LOO DR, (n,x)]} & \cellcolor{gray!10}{Design (\ref{eq:mavar})} & \cellcolor{gray!10}{1.91 (0.010)} & \cellcolor{gray!10}{2.37 (0.013)} & \cellcolor{gray!10}{1.93 (0.010)} & \cellcolor{gray!10}{2.39 (0.013)}\\
\hspace{1em}IDPD[LOO WLS, (n,x)] & Design (\ref{eq:vloop}) & 1.02 (0.003) & 1.02 (0.003) & 1.02 (0.003) & 1.02 (0.003)\\
\bottomrule
\end{tabular}}

\endgroup
\caption{Simulated \textbf{relative bias} of variance estimators associated with each point estimator when there are \textbf{200 pairs} and \textbf{no pair effect} with simulation standard errors in parentheses.}
\label{tab:var-200-no}
\end{table}

\begin{table}[ht]
\centering
\begingroup \footnotesize
\centering
\resizebox{\ifdim\width>\linewidth\linewidth\else\width\fi}{!}{
\begin{tabular}{llcccc}
\toprule
\multicolumn{6}{c}{\textbf{Relative Bias of Variance Estimators}} \\
\multicolumn{6}{c}{\textbf{M = 200}} \\
\cmidrule(l{3pt}r{3pt}){1-6}
\multicolumn{2}{c}{\textbf{ }} & \multicolumn{4}{c}{\textbf{Pair Effect}} \\
\multicolumn{2}{c}{\textbf{ }} & \multicolumn{2}{c}{\textbf{Matched Size}} & \multicolumn{2}{c}{\textbf{Varied Size}} \\
\multicolumn{1}{c}{\textbf{\textbf{Method}}} & \multicolumn{1}{c}{\textbf{Var Estimator}} & \multicolumn{1}{c}{\textbf{Tr Const}} & \multicolumn{1}{c}{\textbf{Tr Cor. Size}} & \multicolumn{1}{c}{\textbf{Tr Const}} & \multicolumn{1}{c}{\textbf{Tr Cor. Size}}\\
\midrule
\addlinespace[0.3em]
\multicolumn{6}{l}{\textbf{Baseline}}\\
\hspace{1em}\cellcolor{gray!10}{IDPD[0, $\emptyset$]} & \cellcolor{gray!10}{H-W Robust} & \cellcolor{gray!10}{4.45 (0.032)} & \cellcolor{gray!10}{4.73 (0.033)} & \cellcolor{gray!10}{3.56 (0.022)} & \cellcolor{gray!10}{3.33 (0.020)}\\
\hspace{1em}\cellcolor{gray!10}{IDPD[0, $\emptyset$]} & \cellcolor{gray!10}{Design (\ref{eq:mavar})} & \cellcolor{gray!10}{5.08 (0.035)} & \cellcolor{gray!10}{5.48 (0.037)} & \cellcolor{gray!10}{4.06 (0.024)} & \cellcolor{gray!10}{3.85 (0.023)}\\
\hspace{1em}WLS[$\emptyset$] & H-W Robust & 4.37 (0.031) & 4.46 (0.032) & 3.63 (0.022) & 3.54 (0.022)\\
\hspace{1em}WLS[$\emptyset$] & Regression & 4.04 (0.028) & 4.14 (0.028) & 3.36 (0.020) & 3.28 (0.020)\\
\hspace{1em}WLS[$\emptyset$] & Design (\ref{eq:crhjvar}) & 1.00 (0.003) & 1.13 (0.004) & 1.00 (0.003) & 1.06 (0.003)\\
\hspace{1em}\cellcolor{gray!10}{WLS-P[$\emptyset$]} & \cellcolor{gray!10}{H-W Robust} & \cellcolor{gray!10}{1.01 (0.003)} & \cellcolor{gray!10}{1.15 (0.004)} & \cellcolor{gray!10}{1.04 (0.003)} & \cellcolor{gray!10}{1.13 (0.004)}\\
\hspace{1em}\cellcolor{gray!10}{WLS-P[$\emptyset$]} & \cellcolor{gray!10}{Regression} & \cellcolor{gray!10}{0.93 (0.003)} & \cellcolor{gray!10}{1.07 (0.004)} & \cellcolor{gray!10}{0.97 (0.003)} & \cellcolor{gray!10}{1.05 (0.004)}\\
\hspace{1em}\cellcolor{gray!10}{WLS-P[$\emptyset$]} & \cellcolor{gray!10}{Design (\ref{eq:crhjvar})} & \cellcolor{gray!10}{1.00 (0.003)} & \cellcolor{gray!10}{1.13 (0.004)} & \cellcolor{gray!10}{1.00 (0.003)} & \cellcolor{gray!10}{1.08 (0.004)}\\
\hspace{1em}\cellcolor{gray!10}{WLS-P[$\emptyset$]} & \cellcolor{gray!10}{Design (\ref{eq:vikn})} & \cellcolor{gray!10}{1.11 (0.003)} & \cellcolor{gray!10}{1.52 (0.006)} & \cellcolor{gray!10}{1.08 (0.003)} & \cellcolor{gray!10}{1.31 (0.005)}\\
\hspace{1em}\cellcolor{gray!10}{WLS-P[$\emptyset$]} & \cellcolor{gray!10}{Design (\ref{eq:vspmk})} & \cellcolor{gray!10}{1.01 (0.003)} & \cellcolor{gray!10}{1.15 (0.004)} & \cellcolor{gray!10}{1.04 (0.003)} & \cellcolor{gray!10}{1.13 (0.004)}\\
\hspace{1em}WLS-R[$\emptyset$] & Regression & 0.93 (0.003) & 1.07 (0.004) & 0.96 (0.003) & 1.04 (0.004)\\
\hspace{1em}\cellcolor{gray!10}{IDPD[LOI MI, $\emptyset$]} & \cellcolor{gray!10}{H-W Robust} & \cellcolor{gray!10}{4.50 (0.033)} & \cellcolor{gray!10}{4.60 (0.033)} & \cellcolor{gray!10}{4.61 (0.033)} & \cellcolor{gray!10}{4.40 (0.031)}\\
\hspace{1em}\cellcolor{gray!10}{IDPD[LOI MI, $\emptyset$]} & \cellcolor{gray!10}{Regression} & \cellcolor{gray!10}{4.19 (0.029)} & \cellcolor{gray!10}{4.29 (0.030)} & \cellcolor{gray!10}{4.43 (0.032)} & \cellcolor{gray!10}{4.23 (0.030)}\\
\hspace{1em}\cellcolor{gray!10}{IDPD[LOI MI, $\emptyset$]} & \cellcolor{gray!10}{Design (\ref{eq:vikn})} & \cellcolor{gray!10}{1.10 (0.003)} & \cellcolor{gray!10}{1.52 (0.006)} & \cellcolor{gray!10}{1.07 (0.003)} & \cellcolor{gray!10}{1.28 (0.005)}\\
\hspace{1em}IDPD[LOO MI, $\emptyset$] & Design (\ref{eq:vloop}) & 1.01 (0.003) & 1.18 (0.004) & 1.01 (0.003) & 1.09 (0.004)\\
\addlinespace[0.3em]
\multicolumn{6}{l}{\textbf{Cluster Size}}\\
\hspace{1em}\cellcolor{gray!10}{WLS[n]} & \cellcolor{gray!10}{H-W Robust} & \cellcolor{gray!10}{4.35 (0.031)} & \cellcolor{gray!10}{4.43 (0.032)} & \cellcolor{gray!10}{3.61 (0.022)} & \cellcolor{gray!10}{3.67 (0.022)}\\
\hspace{1em}WLS-P[n] & H-W Robust & 1.00 (0.003) & 1.15 (0.004) & 1.04 (0.003) & 1.13 (0.004)\\
\hspace{1em}WLS-P[n] & Design (\ref{eq:vspmk}) & 0.99 (0.003) & 1.13 (0.004) & 1.03 (0.003) & 1.12 (0.004)\\
\hspace{1em}\cellcolor{gray!10}{WLS-R[n]} & \cellcolor{gray!10}{Regression} & \cellcolor{gray!10}{0.93 (0.003)} & \cellcolor{gray!10}{1.07 (0.004)} & \cellcolor{gray!10}{0.96 (0.003)} & \cellcolor{gray!10}{1.04 (0.004)}\\
\hspace{1em}WLS-HT[n] & H-W Robust & 4.36 (0.031) & 4.36 (0.031) & 3.61 (0.022) & 3.61 (0.022)\\
\hspace{1em}\cellcolor{gray!10}{IDPD[LOO DR, n]} & \cellcolor{gray!10}{Design (\ref{eq:mavar})} & \cellcolor{gray!10}{5.07 (0.035)} & \cellcolor{gray!10}{5.42 (0.036)} & \cellcolor{gray!10}{4.17 (0.024)} & \cellcolor{gray!10}{4.44 (0.025)}\\
\hspace{1em}IDPD[LOO WLS, n] & Design (\ref{eq:vloop}) & 1.02 (0.003) & 1.02 (0.003) & 1.02 (0.003) & 1.02 (0.003)\\
\addlinespace[0.3em]
\multicolumn{6}{l}{\textbf{Covariate and Cluster Size}}\\
\hspace{1em}\cellcolor{gray!10}{WLS[(n,x)]} & \cellcolor{gray!10}{H-W Robust} & \cellcolor{gray!10}{1.00 (0.005)} & \cellcolor{gray!10}{1.10 (0.005)} & \cellcolor{gray!10}{1.01 (0.005)} & \cellcolor{gray!10}{1.11 (0.005)}\\
\hspace{1em}WLS-P[(n,x)] & H-W Robust & 1.00 (0.003) & 1.18 (0.004) & 1.04 (0.003) & 1.15 (0.004)\\
\hspace{1em}WLS-P[(n,x)] & Design (\ref{eq:vspmk}) & 0.98 (0.003) & 1.16 (0.004) & 1.02 (0.003) & 1.13 (0.004)\\
\hspace{1em}\cellcolor{gray!10}{WLS-R[(n,x)]} & \cellcolor{gray!10}{Regression} & \cellcolor{gray!10}{0.88 (0.003)} & \cellcolor{gray!10}{1.01 (0.004)} & \cellcolor{gray!10}{0.89 (0.003)} & \cellcolor{gray!10}{0.98 (0.004)}\\
\hspace{1em}WLS-HT[(n,x)] & H-W Robust & 1.00 (0.005) & 1.01 (0.005) & 1.01 (0.005) & 1.02 (0.005)\\
\hspace{1em}\cellcolor{gray!10}{IDPD[LOO DR, (n,x)]} & \cellcolor{gray!10}{Design (\ref{eq:mavar})} & \cellcolor{gray!10}{1.93 (0.010)} & \cellcolor{gray!10}{2.40 (0.013)} & \cellcolor{gray!10}{1.95 (0.010)} & \cellcolor{gray!10}{2.41 (0.013)}\\
\hspace{1em}IDPD[LOO WLS, (n,x)] & Design (\ref{eq:vloop}) & 1.02 (0.003) & 1.02 (0.003) & 1.02 (0.003) & 1.02 (0.003)\\
\bottomrule
\end{tabular}}

\endgroup
\caption{Simulated \textbf{relative bias} of variance estimators associated with each point estimator when there are \textbf{200 pairs} and a \textbf{pair effect} with simulation standard errors in parentheses.}
\label{tab:var-200-pair}
\end{table}


\begin{table}[ht]
\centering
\begingroup \footnotesize
\centering
\resizebox{\ifdim\width>\linewidth\linewidth\else\width\fi}{!}{
\begin{tabular}{llcccc}
\toprule
\multicolumn{6}{c}{\textbf{Coverage Probability with Variance Estimators}} \\
\multicolumn{6}{c}{\textbf{M = 20}} \\
\cmidrule(l{3pt}r{3pt}){1-6}
\multicolumn{2}{c}{\textbf{ }} & \multicolumn{4}{c}{\textbf{Pair Effect}} \\
\multicolumn{2}{c}{\textbf{ }} & \multicolumn{2}{c}{\textbf{Matched Size}} & \multicolumn{2}{c}{\textbf{Varied Size}} \\
\multicolumn{1}{c}{\textbf{\textbf{Method}}} & \multicolumn{1}{c}{\textbf{Var Estimator}} & \multicolumn{1}{c}{\textbf{Tr Const}} & \multicolumn{1}{c}{\textbf{Tr Cor. Size}} & \multicolumn{1}{c}{\textbf{Tr Const}} & \multicolumn{1}{c}{\textbf{Tr Cor. Size}}\\
\midrule
\addlinespace[0.3em]
\multicolumn{6}{l}{\textbf{Baseline}}\\
\hspace{1em}\cellcolor{gray!10}{IDPD[0, $\emptyset$]} & \cellcolor{gray!10}{H-W Robust} & \cellcolor{gray!10}{0.952} & \cellcolor{gray!10}{0.961} & \cellcolor{gray!10}{0.950} & \cellcolor{gray!10}{0.955}\\
\hspace{1em}\cellcolor{gray!10}{IDPD[0, $\emptyset$]} & \cellcolor{gray!10}{Design (\ref{eq:mavar})} & \cellcolor{gray!10}{0.965} & \cellcolor{gray!10}{0.971} & \cellcolor{gray!10}{0.963} & \cellcolor{gray!10}{0.966}\\
\hspace{1em}WLS[$\emptyset$] & H-W Robust & 0.947 & 0.950 & 0.946 & 0.948\\
\hspace{1em}WLS[$\emptyset$] & Regression & 0.940 & 0.944 & 0.939 & 0.941\\
\hspace{1em}WLS[$\emptyset$] & Design (\ref{eq:crhjvar}) & 0.933 & 0.937 & 0.935 & 0.937\\
\hspace{1em}\cellcolor{gray!10}{WLS-P[$\emptyset$]} & \cellcolor{gray!10}{H-W Robust} & \cellcolor{gray!10}{0.947} & \cellcolor{gray!10}{0.951} & \cellcolor{gray!10}{0.952} & \cellcolor{gray!10}{0.952}\\
\hspace{1em}\cellcolor{gray!10}{WLS-P[$\emptyset$]} & \cellcolor{gray!10}{Regression} & \cellcolor{gray!10}{0.942} & \cellcolor{gray!10}{0.945} & \cellcolor{gray!10}{0.946} & \cellcolor{gray!10}{0.946}\\
\hspace{1em}\cellcolor{gray!10}{WLS-P[$\emptyset$]} & \cellcolor{gray!10}{Design (\ref{eq:crhjvar})} & \cellcolor{gray!10}{0.941} & \cellcolor{gray!10}{0.944} & \cellcolor{gray!10}{0.943} & \cellcolor{gray!10}{0.943}\\
\hspace{1em}\cellcolor{gray!10}{WLS-P[$\emptyset$]} & \cellcolor{gray!10}{Design (\ref{eq:vikn})} & \cellcolor{gray!10}{0.954} & \cellcolor{gray!10}{0.961} & \cellcolor{gray!10}{0.952} & \cellcolor{gray!10}{0.954}\\
\hspace{1em}\cellcolor{gray!10}{WLS-P[$\emptyset$]} & \cellcolor{gray!10}{Design (\ref{eq:vspmk})} & \cellcolor{gray!10}{0.947} & \cellcolor{gray!10}{0.951} & \cellcolor{gray!10}{0.952} & \cellcolor{gray!10}{0.952}\\
\hspace{1em}WLS-R[$\emptyset$] & Regression & 0.922 & 0.926 & 0.923 & 0.925\\
\hspace{1em}\cellcolor{gray!10}{IDPD[LOI MI, $\emptyset$]} & \cellcolor{gray!10}{H-W Robust} & \cellcolor{gray!10}{0.947} & \cellcolor{gray!10}{0.950} & \cellcolor{gray!10}{0.948} & \cellcolor{gray!10}{0.947}\\
\hspace{1em}\cellcolor{gray!10}{IDPD[LOI MI, $\emptyset$]} & \cellcolor{gray!10}{Regression} & \cellcolor{gray!10}{0.940} & \cellcolor{gray!10}{0.945} & \cellcolor{gray!10}{0.944} & \cellcolor{gray!10}{0.944}\\
\hspace{1em}\cellcolor{gray!10}{IDPD[LOI MI, $\emptyset$]} & \cellcolor{gray!10}{Design (\ref{eq:vikn})} & \cellcolor{gray!10}{0.954} & \cellcolor{gray!10}{0.960} & \cellcolor{gray!10}{0.952} & \cellcolor{gray!10}{0.954}\\
\hspace{1em}IDPD[LOO MI, $\emptyset$] & Design (\ref{eq:vloop}) & 0.939 & 0.943 & 0.941 & 0.943\\
\addlinespace[0.3em]
\multicolumn{6}{l}{\textbf{Cluster Size}}\\
\hspace{1em}\cellcolor{gray!10}{WLS[n]} & \cellcolor{gray!10}{H-W Robust} & \cellcolor{gray!10}{0.947} & \cellcolor{gray!10}{0.948} & \cellcolor{gray!10}{0.945} & \cellcolor{gray!10}{0.947}\\
\hspace{1em}WLS-P[n] & H-W Robust & 0.944 & 0.947 & 0.949 & 0.949\\
\hspace{1em}WLS-P[n] & Design (\ref{eq:vspmk}) & 0.933 & 0.937 & 0.939 & 0.939\\
\hspace{1em}\cellcolor{gray!10}{WLS-R[n]} & \cellcolor{gray!10}{Regression} & \cellcolor{gray!10}{0.923} & \cellcolor{gray!10}{0.926} & \cellcolor{gray!10}{0.923} & \cellcolor{gray!10}{0.925}\\
\hspace{1em}WLS-HT[n] & H-W Robust & 0.947 & 0.947 & 0.945 & 0.945\\
\hspace{1em}\cellcolor{gray!10}{IDPD[LOO DR, n]} & \cellcolor{gray!10}{Design (\ref{eq:mavar})} & \cellcolor{gray!10}{0.966} & \cellcolor{gray!10}{0.969} & \cellcolor{gray!10}{0.963} & \cellcolor{gray!10}{0.966}\\
\hspace{1em}IDPD[LOO WLS, n] & Design (\ref{eq:vloop}) & 0.942 & 0.942 & 0.943 & 0.943\\
\addlinespace[0.3em]
\multicolumn{6}{l}{\textbf{Covariate and Cluster Size}}\\
\hspace{1em}\cellcolor{gray!10}{WLS[(n,x)]} & \cellcolor{gray!10}{H-W Robust} & \cellcolor{gray!10}{0.947} & \cellcolor{gray!10}{0.956} & \cellcolor{gray!10}{0.946} & \cellcolor{gray!10}{0.955}\\
\hspace{1em}WLS-P[(n,x)] & H-W Robust & 0.940 & 0.955 & 0.945 & 0.946\\
\hspace{1em}WLS-P[(n,x)] & Design (\ref{eq:vspmk}) & 0.916 & 0.934 & 0.923 & 0.923\\
\hspace{1em}\cellcolor{gray!10}{WLS-R[(n,x)]} & \cellcolor{gray!10}{Regression} & \cellcolor{gray!10}{0.924} & \cellcolor{gray!10}{0.942} & \cellcolor{gray!10}{0.924} & \cellcolor{gray!10}{0.934}\\
\hspace{1em}WLS-HT[(n,x)] & H-W Robust & 0.946 & 0.947 & 0.944 & 0.945\\
\hspace{1em}\cellcolor{gray!10}{IDPD[LOO DR, (n,x)]} & \cellcolor{gray!10}{Design (\ref{eq:mavar})} & \cellcolor{gray!10}{0.985} & \cellcolor{gray!10}{0.989} & \cellcolor{gray!10}{0.982} & \cellcolor{gray!10}{0.986}\\
\hspace{1em}IDPD[LOO WLS, (n,x)] & Design (\ref{eq:vloop}) & 0.944 & 0.944 & 0.945 & 0.945\\
\bottomrule
\end{tabular}}

\endgroup
\caption{Simulated \textbf{coverage} of associated point and variance estimators when there are \textbf{20 pairs} and \textbf{no pair effect}.}
\label{tab:covg-20-no}
\end{table}

\begin{table}[ht]
\centering
\begingroup \footnotesize
\centering
\resizebox{\ifdim\width>\linewidth\linewidth\else\width\fi}{!}{
\begin{tabular}{llcccc}
\toprule
\multicolumn{6}{c}{\textbf{Coverage Probability with Variance Estimators}} \\
\multicolumn{6}{c}{\textbf{M = 20}} \\
\cmidrule(l{3pt}r{3pt}){1-6}
\multicolumn{2}{c}{\textbf{ }} & \multicolumn{4}{c}{\textbf{Pair Effect}} \\
\multicolumn{2}{c}{\textbf{ }} & \multicolumn{2}{c}{\textbf{Matched Size}} & \multicolumn{2}{c}{\textbf{Varied Size}} \\
\multicolumn{1}{c}{\textbf{\textbf{Method}}} & \multicolumn{1}{c}{\textbf{Var Estimator}} & \multicolumn{1}{c}{\textbf{Tr Const}} & \multicolumn{1}{c}{\textbf{Tr Cor. Size}} & \multicolumn{1}{c}{\textbf{Tr Const}} & \multicolumn{1}{c}{\textbf{Tr Cor. Size}}\\
\midrule
\addlinespace[0.3em]
\multicolumn{6}{l}{\textbf{Baseline}}\\
\hspace{1em}\cellcolor{gray!10}{IDPD[0, $\emptyset$]} & \cellcolor{gray!10}{H-W Robust} & \cellcolor{gray!10}{1.000} & \cellcolor{gray!10}{1.000} & \cellcolor{gray!10}{0.999} & \cellcolor{gray!10}{0.999}\\
\hspace{1em}\cellcolor{gray!10}{IDPD[0, $\emptyset$]} & \cellcolor{gray!10}{Design (\ref{eq:mavar})} & \cellcolor{gray!10}{1.000} & \cellcolor{gray!10}{1.000} & \cellcolor{gray!10}{1.000} & \cellcolor{gray!10}{0.999}\\
\hspace{1em}WLS[$\emptyset$] & H-W Robust & 1.000 & 1.000 & 0.999 & 0.999\\
\hspace{1em}WLS[$\emptyset$] & Regression & 1.000 & 1.000 & 0.999 & 0.999\\
\hspace{1em}WLS[$\emptyset$] & Design (\ref{eq:crhjvar}) & 0.933 & 0.948 & 0.934 & 0.940\\
\hspace{1em}\cellcolor{gray!10}{WLS-P[$\emptyset$]} & \cellcolor{gray!10}{H-W Robust} & \cellcolor{gray!10}{0.947} & \cellcolor{gray!10}{0.960} & \cellcolor{gray!10}{0.952} & \cellcolor{gray!10}{0.953}\\
\hspace{1em}\cellcolor{gray!10}{WLS-P[$\emptyset$]} & \cellcolor{gray!10}{Regression} & \cellcolor{gray!10}{0.942} & \cellcolor{gray!10}{0.957} & \cellcolor{gray!10}{0.946} & \cellcolor{gray!10}{0.947}\\
\hspace{1em}\cellcolor{gray!10}{WLS-P[$\emptyset$]} & \cellcolor{gray!10}{Design (\ref{eq:crhjvar})} & \cellcolor{gray!10}{0.941} & \cellcolor{gray!10}{0.955} & \cellcolor{gray!10}{0.943} & \cellcolor{gray!10}{0.943}\\
\hspace{1em}\cellcolor{gray!10}{WLS-P[$\emptyset$]} & \cellcolor{gray!10}{Design (\ref{eq:vikn})} & \cellcolor{gray!10}{0.960} & \cellcolor{gray!10}{0.979} & \cellcolor{gray!10}{0.957} & \cellcolor{gray!10}{0.962}\\
\hspace{1em}\cellcolor{gray!10}{WLS-P[$\emptyset$]} & \cellcolor{gray!10}{Design (\ref{eq:vspmk})} & \cellcolor{gray!10}{0.947} & \cellcolor{gray!10}{0.960} & \cellcolor{gray!10}{0.952} & \cellcolor{gray!10}{0.953}\\
\hspace{1em}WLS-R[$\emptyset$] & Regression & 0.934 & 0.950 & 0.937 & 0.940\\
\hspace{1em}\cellcolor{gray!10}{IDPD[LOI MI, $\emptyset$]} & \cellcolor{gray!10}{H-W Robust} & \cellcolor{gray!10}{1.000} & \cellcolor{gray!10}{1.000} & \cellcolor{gray!10}{1.000} & \cellcolor{gray!10}{0.999}\\
\hspace{1em}\cellcolor{gray!10}{IDPD[LOI MI, $\emptyset$]} & \cellcolor{gray!10}{Regression} & \cellcolor{gray!10}{1.000} & \cellcolor{gray!10}{1.000} & \cellcolor{gray!10}{1.000} & \cellcolor{gray!10}{0.999}\\
\hspace{1em}\cellcolor{gray!10}{IDPD[LOI MI, $\emptyset$]} & \cellcolor{gray!10}{Design (\ref{eq:vikn})} & \cellcolor{gray!10}{0.960} & \cellcolor{gray!10}{0.979} & \cellcolor{gray!10}{0.956} & \cellcolor{gray!10}{0.960}\\
\hspace{1em}IDPD[LOO MI, $\emptyset$] & Design (\ref{eq:vloop}) & 0.939 & 0.956 & 0.942 & 0.949\\
\addlinespace[0.3em]
\multicolumn{6}{l}{\textbf{Cluster Size}}\\
\hspace{1em}\cellcolor{gray!10}{WLS[n]} & \cellcolor{gray!10}{H-W Robust} & \cellcolor{gray!10}{1.000} & \cellcolor{gray!10}{1.000} & \cellcolor{gray!10}{0.999} & \cellcolor{gray!10}{0.999}\\
\hspace{1em}WLS-P[n] & H-W Robust & 0.943 & 0.955 & 0.950 & 0.950\\
\hspace{1em}WLS-P[n] & Design (\ref{eq:vspmk}) & 0.932 & 0.946 & 0.940 & 0.940\\
\hspace{1em}\cellcolor{gray!10}{WLS-R[n]} & \cellcolor{gray!10}{Regression} & \cellcolor{gray!10}{0.934} & \cellcolor{gray!10}{0.949} & \cellcolor{gray!10}{0.936} & \cellcolor{gray!10}{0.938}\\
\hspace{1em}WLS-HT[n] & H-W Robust & 1.000 & 1.000 & 0.999 & 0.998\\
\hspace{1em}\cellcolor{gray!10}{IDPD[LOO DR, n]} & \cellcolor{gray!10}{Design (\ref{eq:mavar})} & \cellcolor{gray!10}{1.000} & \cellcolor{gray!10}{1.000} & \cellcolor{gray!10}{0.999} & \cellcolor{gray!10}{0.999}\\
\hspace{1em}IDPD[LOO WLS, n] & Design (\ref{eq:vloop}) & 0.944 & 0.944 & 0.947 & 0.947\\
\addlinespace[0.3em]
\multicolumn{6}{l}{\textbf{Covariate and Cluster Size}}\\
\hspace{1em}\cellcolor{gray!10}{WLS[(n,x)]} & \cellcolor{gray!10}{H-W Robust} & \cellcolor{gray!10}{0.947} & \cellcolor{gray!10}{0.957} & \cellcolor{gray!10}{0.947} & \cellcolor{gray!10}{0.956}\\
\hspace{1em}WLS-P[(n,x)] & H-W Robust & 0.940 & 0.955 & 0.945 & 0.946\\
\hspace{1em}WLS-P[(n,x)] & Design (\ref{eq:vspmk}) & 0.916 & 0.934 & 0.923 & 0.923\\
\hspace{1em}\cellcolor{gray!10}{WLS-R[(n,x)]} & \cellcolor{gray!10}{Regression} & \cellcolor{gray!10}{0.924} & \cellcolor{gray!10}{0.943} & \cellcolor{gray!10}{0.925} & \cellcolor{gray!10}{0.935}\\
\hspace{1em}WLS-HT[(n,x)] & H-W Robust & 0.946 & 0.948 & 0.946 & 0.946\\
\hspace{1em}\cellcolor{gray!10}{IDPD[LOO DR, (n,x)]} & \cellcolor{gray!10}{Design (\ref{eq:mavar})} & \cellcolor{gray!10}{0.988} & \cellcolor{gray!10}{0.993} & \cellcolor{gray!10}{0.985} & \cellcolor{gray!10}{0.990}\\
\hspace{1em}IDPD[LOO WLS, (n,x)] & Design (\ref{eq:vloop}) & 0.945 & 0.945 & 0.946 & 0.946\\
\bottomrule
\end{tabular}}

\endgroup
\caption{Simulated \textbf{coverage} of associated point and variance estimators when there are \textbf{20 pairs} and a \textbf{pair effect}.}
\label{tab:covg-20-pair}
\end{table}

\begin{table}[ht]
\centering
\begingroup \footnotesize
\centering
\resizebox{\ifdim\width>\linewidth\linewidth\else\width\fi}{!}{
\begin{tabular}{llcccc}
\toprule
\multicolumn{6}{c}{\textbf{Coverage Probability with Variance Estimators}} \\
\multicolumn{6}{c}{\textbf{M = 200}} \\
\cmidrule(l{3pt}r{3pt}){1-6}
\multicolumn{2}{c}{\textbf{ }} & \multicolumn{4}{c}{\textbf{Pair Effect}} \\
\multicolumn{2}{c}{\textbf{ }} & \multicolumn{2}{c}{\textbf{Matched Size}} & \multicolumn{2}{c}{\textbf{Varied Size}} \\
\multicolumn{1}{c}{\textbf{\textbf{Method}}} & \multicolumn{1}{c}{\textbf{Var Estimator}} & \multicolumn{1}{c}{\textbf{Tr Const}} & \multicolumn{1}{c}{\textbf{Tr Cor. Size}} & \multicolumn{1}{c}{\textbf{Tr Const}} & \multicolumn{1}{c}{\textbf{Tr Cor. Size}}\\
\midrule
\addlinespace[0.3em]
\multicolumn{6}{l}{\textbf{Baseline}}\\
\hspace{1em}\cellcolor{gray!10}{IDPD[0, $\emptyset$]} & \cellcolor{gray!10}{H-W Robust} & \cellcolor{gray!10}{0.952} & \cellcolor{gray!10}{0.961} & \cellcolor{gray!10}{0.952} & \cellcolor{gray!10}{0.956}\\
\hspace{1em}\cellcolor{gray!10}{IDPD[0, $\emptyset$]} & \cellcolor{gray!10}{Design (\ref{eq:mavar})} & \cellcolor{gray!10}{0.965} & \cellcolor{gray!10}{0.973} & \cellcolor{gray!10}{0.965} & \cellcolor{gray!10}{0.969}\\
\hspace{1em}WLS[$\emptyset$] & H-W Robust & 0.949 & 0.952 & 0.950 & 0.952\\
\hspace{1em}WLS[$\emptyset$] & Regression & 0.940 & 0.944 & 0.941 & 0.943\\
\hspace{1em}WLS[$\emptyset$] & Design (\ref{eq:crhjvar}) & 0.948 & 0.952 & 0.949 & 0.951\\
\hspace{1em}\cellcolor{gray!10}{WLS-P[$\emptyset$]} & \cellcolor{gray!10}{H-W Robust} & \cellcolor{gray!10}{0.950} & \cellcolor{gray!10}{0.953} & \cellcolor{gray!10}{0.954} & \cellcolor{gray!10}{0.941}\\
\hspace{1em}\cellcolor{gray!10}{WLS-P[$\emptyset$]} & \cellcolor{gray!10}{Regression} & \cellcolor{gray!10}{0.941} & \cellcolor{gray!10}{0.944} & \cellcolor{gray!10}{0.945} & \cellcolor{gray!10}{0.931}\\
\hspace{1em}\cellcolor{gray!10}{WLS-P[$\emptyset$]} & \cellcolor{gray!10}{Design (\ref{eq:crhjvar})} & \cellcolor{gray!10}{0.949} & \cellcolor{gray!10}{0.952} & \cellcolor{gray!10}{0.950} & \cellcolor{gray!10}{0.936}\\
\hspace{1em}\cellcolor{gray!10}{WLS-P[$\emptyset$]} & \cellcolor{gray!10}{Design (\ref{eq:vikn})} & \cellcolor{gray!10}{0.953} & \cellcolor{gray!10}{0.961} & \cellcolor{gray!10}{0.952} & \cellcolor{gray!10}{0.942}\\
\hspace{1em}\cellcolor{gray!10}{WLS-P[$\emptyset$]} & \cellcolor{gray!10}{Design (\ref{eq:vspmk})} & \cellcolor{gray!10}{0.950} & \cellcolor{gray!10}{0.953} & \cellcolor{gray!10}{0.954} & \cellcolor{gray!10}{0.941}\\
\hspace{1em}WLS-R[$\emptyset$] & Regression & 0.931 & 0.935 & 0.933 & 0.935\\
\hspace{1em}\cellcolor{gray!10}{IDPD[LOI MI, $\emptyset$]} & \cellcolor{gray!10}{H-W Robust} & \cellcolor{gray!10}{0.949} & \cellcolor{gray!10}{0.952} & \cellcolor{gray!10}{0.950} & \cellcolor{gray!10}{0.934}\\
\hspace{1em}\cellcolor{gray!10}{IDPD[LOI MI, $\emptyset$]} & \cellcolor{gray!10}{Regression} & \cellcolor{gray!10}{0.940} & \cellcolor{gray!10}{0.944} & \cellcolor{gray!10}{0.945} & \cellcolor{gray!10}{0.929}\\
\hspace{1em}\cellcolor{gray!10}{IDPD[LOI MI, $\emptyset$]} & \cellcolor{gray!10}{Design (\ref{eq:vikn})} & \cellcolor{gray!10}{0.953} & \cellcolor{gray!10}{0.960} & \cellcolor{gray!10}{0.952} & \cellcolor{gray!10}{0.939}\\
\hspace{1em}IDPD[LOO MI, $\emptyset$] & Design (\ref{eq:vloop}) & 0.949 & 0.953 & 0.950 & 0.952\\
\addlinespace[0.3em]
\multicolumn{6}{l}{\textbf{Cluster Size}}\\
\hspace{1em}\cellcolor{gray!10}{WLS[n]} & \cellcolor{gray!10}{H-W Robust} & \cellcolor{gray!10}{0.949} & \cellcolor{gray!10}{0.951} & \cellcolor{gray!10}{0.950} & \cellcolor{gray!10}{0.952}\\
\hspace{1em}WLS-P[n] & H-W Robust & 0.950 & 0.953 & 0.954 & 0.941\\
\hspace{1em}WLS-P[n] & Design (\ref{eq:vspmk}) & 0.949 & 0.952 & 0.953 & 0.940\\
\hspace{1em}\cellcolor{gray!10}{WLS-R[n]} & \cellcolor{gray!10}{Regression} & \cellcolor{gray!10}{0.931} & \cellcolor{gray!10}{0.935} & \cellcolor{gray!10}{0.933} & \cellcolor{gray!10}{0.935}\\
\hspace{1em}WLS-HT[n] & H-W Robust & 0.949 & 0.949 & 0.950 & 0.950\\
\hspace{1em}\cellcolor{gray!10}{IDPD[LOO DR, n]} & \cellcolor{gray!10}{Design (\ref{eq:mavar})} & \cellcolor{gray!10}{0.964} & \cellcolor{gray!10}{0.970} & \cellcolor{gray!10}{0.965} & \cellcolor{gray!10}{0.970}\\
\hspace{1em}IDPD[LOO WLS, n] & Design (\ref{eq:vloop}) & 0.949 & 0.949 & 0.951 & 0.951\\
\addlinespace[0.3em]
\multicolumn{6}{l}{\textbf{Covariate and Cluster Size}}\\
\hspace{1em}\cellcolor{gray!10}{WLS[(n,x)]} & \cellcolor{gray!10}{H-W Robust} & \cellcolor{gray!10}{0.949} & \cellcolor{gray!10}{0.959} & \cellcolor{gray!10}{0.950} & \cellcolor{gray!10}{0.960}\\
\hspace{1em}WLS-P[(n,x)] & H-W Robust & 0.949 & 0.964 & 0.953 & 0.883\\
\hspace{1em}WLS-P[(n,x)] & Design (\ref{eq:vspmk}) & 0.947 & 0.963 & 0.951 & 0.879\\
\hspace{1em}\cellcolor{gray!10}{WLS-R[(n,x)]} & \cellcolor{gray!10}{Regression} & \cellcolor{gray!10}{0.933} & \cellcolor{gray!10}{0.950} & \cellcolor{gray!10}{0.934} & \cellcolor{gray!10}{0.944}\\
\hspace{1em}WLS-HT[(n,x)] & H-W Robust & 0.948 & 0.950 & 0.950 & 0.950\\
\hspace{1em}\cellcolor{gray!10}{IDPD[LOO DR, (n,x)]} & \cellcolor{gray!10}{Design (\ref{eq:mavar})} & \cellcolor{gray!10}{0.992} & \cellcolor{gray!10}{0.997} & \cellcolor{gray!10}{0.993} & \cellcolor{gray!10}{0.997}\\
\hspace{1em}IDPD[LOO WLS, (n,x)] & Design (\ref{eq:vloop}) & 0.950 & 0.950 & 0.950 & 0.950\\
\bottomrule
\end{tabular}}

\endgroup
\caption{Simulated \textbf{coverage} of associated point and variance estimators when there are \textbf{200 pairs} and \textbf{no pair effect}.}
\label{tab:covg-200-no}
\end{table}

\begin{table}[ht]
\centering
\begingroup \footnotesize
\centering
\resizebox{\ifdim\width>\linewidth\linewidth\else\width\fi}{!}{
\begin{tabular}{llcccc}
\toprule
\multicolumn{6}{c}{\textbf{Coverage Probability with Variance Estimators}} \\
\multicolumn{6}{c}{\textbf{M = 200}} \\
\cmidrule(l{3pt}r{3pt}){1-6}
\multicolumn{2}{c}{\textbf{ }} & \multicolumn{4}{c}{\textbf{Pair Effect}} \\
\multicolumn{2}{c}{\textbf{ }} & \multicolumn{2}{c}{\textbf{Matched Size}} & \multicolumn{2}{c}{\textbf{Varied Size}} \\
\multicolumn{1}{c}{\textbf{\textbf{Method}}} & \multicolumn{1}{c}{\textbf{Var Estimator}} & \multicolumn{1}{c}{\textbf{Tr Const}} & \multicolumn{1}{c}{\textbf{Tr Cor. Size}} & \multicolumn{1}{c}{\textbf{Tr Const}} & \multicolumn{1}{c}{\textbf{Tr Cor. Size}}\\
\midrule
\addlinespace[0.3em]
\multicolumn{6}{l}{\textbf{Baseline}}\\
\hspace{1em}\cellcolor{gray!10}{IDPD[0, $\emptyset$]} & \cellcolor{gray!10}{H-W Robust} & \cellcolor{gray!10}{1.000} & \cellcolor{gray!10}{1.000} & \cellcolor{gray!10}{1.000} & \cellcolor{gray!10}{1.000}\\
\hspace{1em}\cellcolor{gray!10}{IDPD[0, $\emptyset$]} & \cellcolor{gray!10}{Design (\ref{eq:mavar})} & \cellcolor{gray!10}{1.000} & \cellcolor{gray!10}{1.000} & \cellcolor{gray!10}{1.000} & \cellcolor{gray!10}{1.000}\\
\hspace{1em}WLS[$\emptyset$] & H-W Robust & 1.000 & 1.000 & 1.000 & 1.000\\
\hspace{1em}WLS[$\emptyset$] & Regression & 1.000 & 1.000 & 1.000 & 1.000\\
\hspace{1em}WLS[$\emptyset$] & Design (\ref{eq:crhjvar}) & 0.948 & 0.961 & 0.949 & 0.955\\
\hspace{1em}\cellcolor{gray!10}{WLS-P[$\emptyset$]} & \cellcolor{gray!10}{H-W Robust} & \cellcolor{gray!10}{0.950} & \cellcolor{gray!10}{0.962} & \cellcolor{gray!10}{0.954} & \cellcolor{gray!10}{0.903}\\
\hspace{1em}\cellcolor{gray!10}{WLS-P[$\emptyset$]} & \cellcolor{gray!10}{Regression} & \cellcolor{gray!10}{0.941} & \cellcolor{gray!10}{0.955} & \cellcolor{gray!10}{0.946} & \cellcolor{gray!10}{0.890}\\
\hspace{1em}\cellcolor{gray!10}{WLS-P[$\emptyset$]} & \cellcolor{gray!10}{Design (\ref{eq:crhjvar})} & \cellcolor{gray!10}{0.949} & \cellcolor{gray!10}{0.961} & \cellcolor{gray!10}{0.950} & \cellcolor{gray!10}{0.895}\\
\hspace{1em}\cellcolor{gray!10}{WLS-P[$\emptyset$]} & \cellcolor{gray!10}{Design (\ref{eq:vikn})} & \cellcolor{gray!10}{0.960} & \cellcolor{gray!10}{0.982} & \cellcolor{gray!10}{0.958} & \cellcolor{gray!10}{0.923}\\
\hspace{1em}\cellcolor{gray!10}{WLS-P[$\emptyset$]} & \cellcolor{gray!10}{Design (\ref{eq:vspmk})} & \cellcolor{gray!10}{0.950} & \cellcolor{gray!10}{0.962} & \cellcolor{gray!10}{0.954} & \cellcolor{gray!10}{0.903}\\
\hspace{1em}WLS-R[$\emptyset$] & Regression & 0.940 & 0.954 & 0.944 & 0.903\\
\hspace{1em}\cellcolor{gray!10}{IDPD[LOI MI, $\emptyset$]} & \cellcolor{gray!10}{H-W Robust} & \cellcolor{gray!10}{1.000} & \cellcolor{gray!10}{1.000} & \cellcolor{gray!10}{1.000} & \cellcolor{gray!10}{0.999}\\
\hspace{1em}\cellcolor{gray!10}{IDPD[LOI MI, $\emptyset$]} & \cellcolor{gray!10}{Regression} & \cellcolor{gray!10}{1.000} & \cellcolor{gray!10}{1.000} & \cellcolor{gray!10}{1.000} & \cellcolor{gray!10}{0.999}\\
\hspace{1em}\cellcolor{gray!10}{IDPD[LOI MI, $\emptyset$]} & \cellcolor{gray!10}{Design (\ref{eq:vikn})} & \cellcolor{gray!10}{0.960} & \cellcolor{gray!10}{0.982} & \cellcolor{gray!10}{0.956} & \cellcolor{gray!10}{0.913}\\
\hspace{1em}IDPD[LOO MI, $\emptyset$] & Design (\ref{eq:vloop}) & 0.949 & 0.964 & 0.950 & 0.957\\
\addlinespace[0.3em]
\multicolumn{6}{l}{\textbf{Cluster Size}}\\
\hspace{1em}\cellcolor{gray!10}{WLS[n]} & \cellcolor{gray!10}{H-W Robust} & \cellcolor{gray!10}{1.000} & \cellcolor{gray!10}{1.000} & \cellcolor{gray!10}{1.000} & \cellcolor{gray!10}{1.000}\\
\hspace{1em}WLS-P[n] & H-W Robust & 0.950 & 0.962 & 0.954 & 0.898\\
\hspace{1em}WLS-P[n] & Design (\ref{eq:vspmk}) & 0.949 & 0.961 & 0.953 & 0.896\\
\hspace{1em}\cellcolor{gray!10}{WLS-R[n]} & \cellcolor{gray!10}{Regression} & \cellcolor{gray!10}{0.940} & \cellcolor{gray!10}{0.955} & \cellcolor{gray!10}{0.944} & \cellcolor{gray!10}{0.898}\\
\hspace{1em}WLS-HT[n] & H-W Robust & 1.000 & 1.000 & 1.000 & 1.000\\
\hspace{1em}\cellcolor{gray!10}{IDPD[LOO DR, n]} & \cellcolor{gray!10}{Design (\ref{eq:mavar})} & \cellcolor{gray!10}{1.000} & \cellcolor{gray!10}{1.000} & \cellcolor{gray!10}{1.000} & \cellcolor{gray!10}{1.000}\\
\hspace{1em}IDPD[LOO WLS, n] & Design (\ref{eq:vloop}) & 0.950 & 0.950 & 0.951 & 0.951\\
\addlinespace[0.3em]
\multicolumn{6}{l}{\textbf{Covariate and Cluster Size}}\\
\hspace{1em}\cellcolor{gray!10}{WLS[(n,x)]} & \cellcolor{gray!10}{H-W Robust} & \cellcolor{gray!10}{0.949} & \cellcolor{gray!10}{0.959} & \cellcolor{gray!10}{0.950} & \cellcolor{gray!10}{0.960}\\
\hspace{1em}WLS-P[(n,x)] & H-W Robust & 0.949 & 0.964 & 0.953 & 0.883\\
\hspace{1em}WLS-P[(n,x)] & Design (\ref{eq:vspmk}) & 0.947 & 0.963 & 0.951 & 0.879\\
\hspace{1em}\cellcolor{gray!10}{WLS-R[(n,x)]} & \cellcolor{gray!10}{Regression} & \cellcolor{gray!10}{0.933} & \cellcolor{gray!10}{0.950} & \cellcolor{gray!10}{0.935} & \cellcolor{gray!10}{0.945}\\
\hspace{1em}WLS-HT[(n,x)] & H-W Robust & 0.949 & 0.950 & 0.950 & 0.951\\
\hspace{1em}\cellcolor{gray!10}{IDPD[LOO DR, (n,x)]} & \cellcolor{gray!10}{Design (\ref{eq:mavar})} & \cellcolor{gray!10}{0.992} & \cellcolor{gray!10}{0.997} & \cellcolor{gray!10}{0.993} & \cellcolor{gray!10}{0.997}\\
\hspace{1em}IDPD[LOO WLS, (n,x)] & Design (\ref{eq:vloop}) & 0.950 & 0.950 & 0.950 & 0.950\\
\bottomrule
\end{tabular}}

\endgroup
\caption{Simulated \textbf{coverage} of associated point and variance estimators when there are \textbf{200 pairs} and a \textbf{pair effect}.}
\label{tab:covg-200-pair}
\end{table}

\FloatBarrier

\setcounter{table}{0}
\setcounter{figure}{0}
\section{Additional Figures: Simulations with Outcomes Correlated with Cluster Size}\label{chpt:appn_pcrt10}

This supplement contains the full simulation results all settings and point and variance estimators, for a different data generating scheme -- where the control potential outcomes are correlated with cluster size.

The simulation design is the same as that described in Supplement~\ref{chpt:appn_pcrt8}, except that we generate the individual potential outcomes as:

$$y_{ik\ell}^c = \alpha_0 + x_{ik} + \gamma_{ik} + \beta_n \nk + \epsilon_{ik\ell} \hspace{2cm} y_{ik\ell}^t = y_{ik\ell}^c + \tau_{ik}$$
with $\beta_n = \sqrt{.0125/\V[\nk]}$. Therefore, the cluster size explains $\approx5\%$ of the variance in the control potential outcomes. We still let $\approx15\%$ of the variance remain unexplained, so we distribute the variance between the errors as follows to create the pair effect settings:

\begin{enumerate}
    \item \textbf{No Pair Effect:} $\sigma^2_{\alpha} = 0$,  $\sigma^2_{z} = .2$,  $\sigma^2_{\gamma} = .0375$
    \item \textbf{Pair Effect:} $\sigma^2_{\alpha} = .1875$,  $\sigma^2_{z} = .0125$,  $\sigma^2_{\gamma} = .0375$
\end{enumerate}

As in Supplement~\ref{chpt:appn_pcrt9}, we include additional estimators discussed in the literature that we did not include in the main paper results for brevity and focus. The additional point estimators include WLS with a pair random effect instead of a pair fixed effect (WLS-R[$\cdot$]), the Des Raj difference estimator (IDPD[LOO DR, $\cdot$]) as implemented in \cite{middleton_unbiased_2015} and described in Supplement~\ref{chpt:appn_pcrt1}, and the weighted regression estimator of \cite{su_model-assisted_2021} described in Supplement~\ref{chpt:appn_pcrt1} (WLS-HT[$\cdot$]). For each point estimator, we also show full results for available, associated variance estimators. For design-based variance estimators, the relevant equation is included. In addition to the Huber-White robust variance estimators, we also show results for the typical parametric variance estimator for OLS or WLS (``Regression'').


\begin{figure}[ht]
    \centering
    \subfloat[M=20]{%
    \includegraphics[width = .9\textwidth]{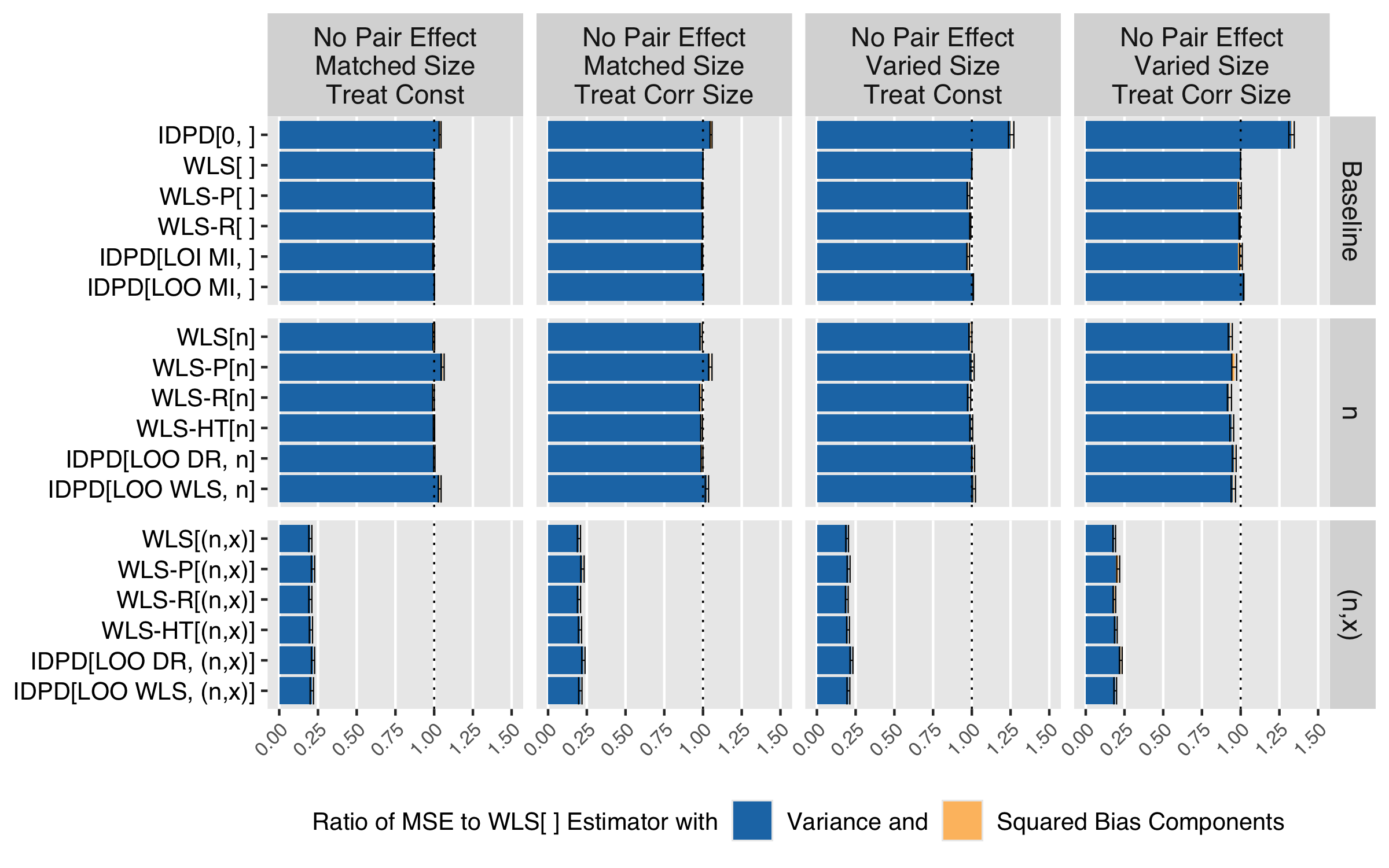}%
    }
    
    \subfloat[M=200]{%
     \includegraphics[width = .9\textwidth]{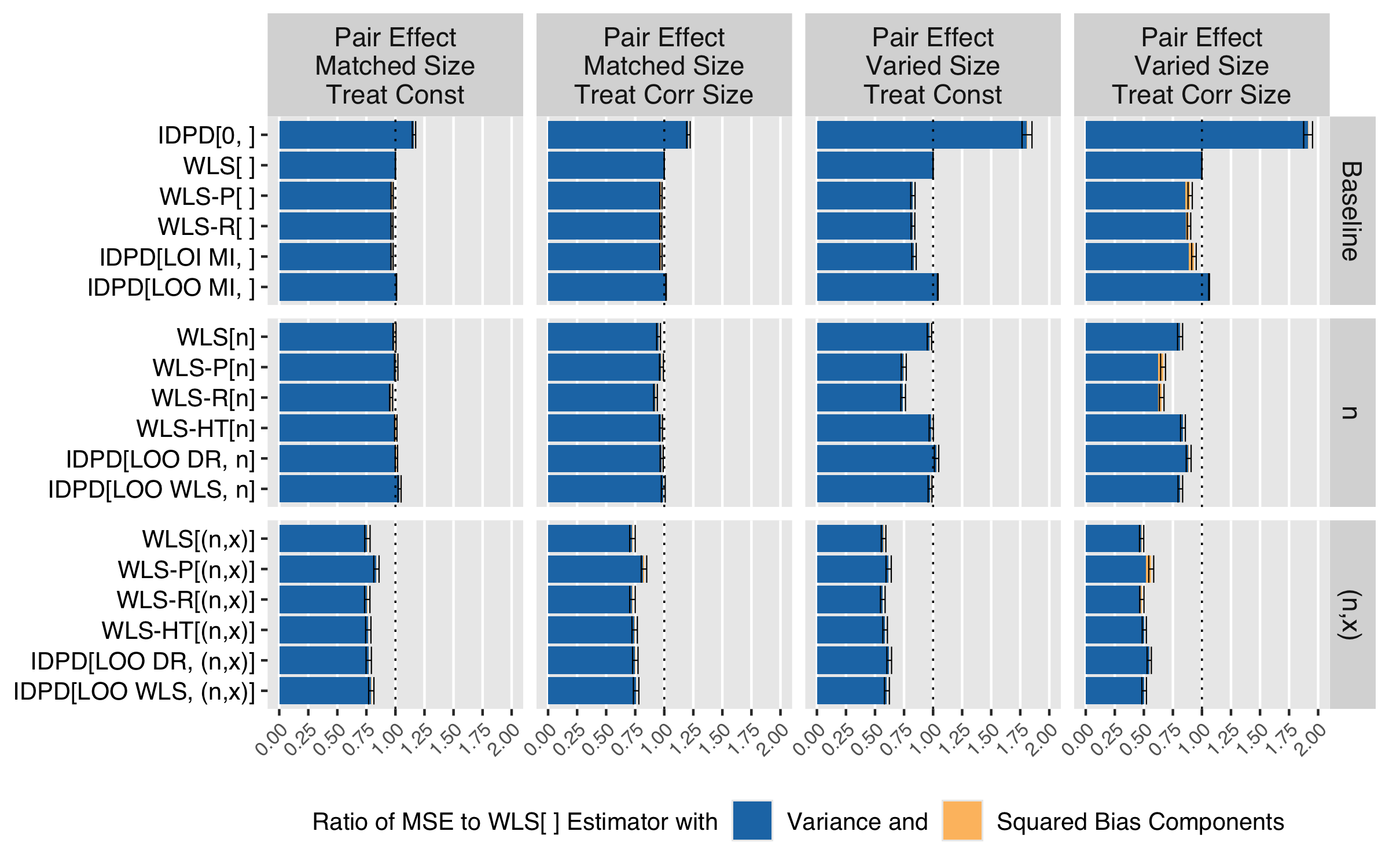}%
    }
    \caption{Ratio of the MSE of treatment effect estimators, compared with the H\'ajek estimator ($\hajek$) when there is no pair effect, outcomes are correlated with cluster size, and there are $M = 20$ or $M= 200$ pairs.}
    \label{fig:mse-200-apdx-neff}
\end{figure}

\begin{figure}[ht]
    \centering
    \subfloat[M=20]{%
    \includegraphics[width = .9\textwidth]{figures/mse-appendix-neff-paireffect-m=20.png}%
    }
    
    \subfloat[M=200]{%
     \includegraphics[width = .9\textwidth]{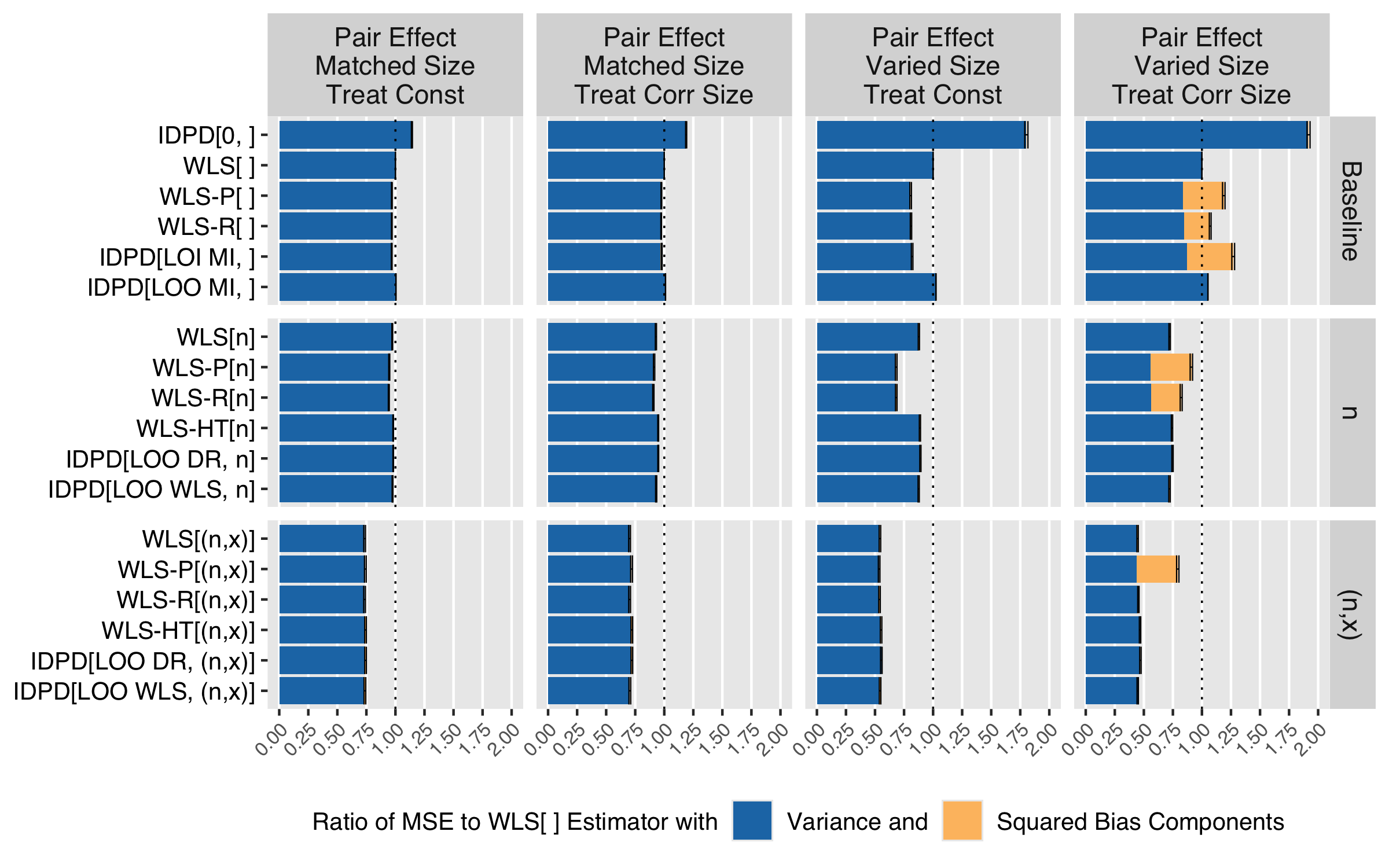}%
    }
    \caption{Ratio of the MSE of treatment effect estimators, compared with the H\'ajek estimator ($\hajek$) when there is a pair effect, outcomes are correlated with cluster size, and there are $M = 20$ or $M= 200$ pairs.}
    \label{fig:mse-200-appdx-neff}
\end{figure}


\begin{table}[ht]
\centering
\begingroup \footnotesize

\endgroup
\caption{Simulated \textbf{relative bias} of variance estimators associated with each point estimator when there are \textbf{20 pairs}, \textbf{no pair effect}, and the outcomes are correlated with cluster size, with simulation standard errors in parentheses. Relative bias is calculated as the empirical mean of the variance estimator divided by the empirical variance of the point estimator.}
\label{tab:var-20-no-neff}
\end{table}

\begin{table}[ht]
\centering
\begingroup \footnotesize

\endgroup
\caption{Simulated \textbf{relative bias} of variance estimators associated with each point estimator when there are \textbf{20 pairs}, a \textbf{pair effect}, and the outcomes are correlated with cluster size, with simulation standard errors in parentheses.}
\label{tab:var-20-pair-neff}
\end{table}

\begin{table}[ht]
\centering
\begingroup \footnotesize

\endgroup
\caption{Simulated \textbf{relative bias} of variance estimators associated with each point estimator when there are \textbf{200 pairs}, \textbf{no pair effect}, and the outcomes are correlated with cluster size, with simulation standard errors in parentheses.}
\label{tab:var-200-no-neff}
\end{table}

\begin{table}[ht]
\centering
\begingroup \footnotesize

\endgroup
\caption{Simulated \textbf{relative bias} of variance estimators associated with each point estimator when there are \textbf{200 pairs} and a \textbf{pair effect} with simulation standard errors in parentheses.}
\label{tab:var-200-pair-neff}
\end{table}


\begin{table}[ht]
\centering
\begingroup \footnotesize

\endgroup
\caption{Simulated \textbf{coverage} of associated point and variance estimators when there are \textbf{20 pairs}, \textbf{no pair effect}, and the outcomes are correlated with cluster size.}
\label{tab:covg-20-no-neff}
\end{table}

\begin{table}[ht]
\centering
\begingroup \footnotesize

\endgroup
\caption{Simulated \textbf{coverage} of associated point and variance estimators when there are \textbf{20 pairs}, a \textbf{pair effect}, and the outcomes are correlated with cluster size.}
\label{tab:covg-20-pair-neff}
\end{table}

\begin{table}[ht]
\centering
\begingroup \footnotesize

\endgroup
\caption{Simulated \textbf{coverage} of associated point and variance estimators when there are \textbf{200 pairs}, \textbf{no pair effect}, and the outcomes are correlated with cluster size.}
\label{tab:covg-200-no-neff}
\end{table}

\begin{table}[ht]
\centering
\begingroup \footnotesize

\endgroup
\caption{Simulated \textbf{coverage} of associated point and variance estimators when there are \textbf{200 pairs}, a \textbf{pair effect}, and the outcomes are correlated with cluster size.}
\label{tab:covg-200-pair-neff}
\end{table}

\FloatBarrier

\setcounter{table}{0}
\setcounter{figure}{0}
\section{Additional Results: Real Data Simulations}\label{chpt:appn_pcrt11}

\begin{figure}[ht]
    \centering
    \subfloat[All Estimators]{%
    \includegraphics[width = .9\textwidth]{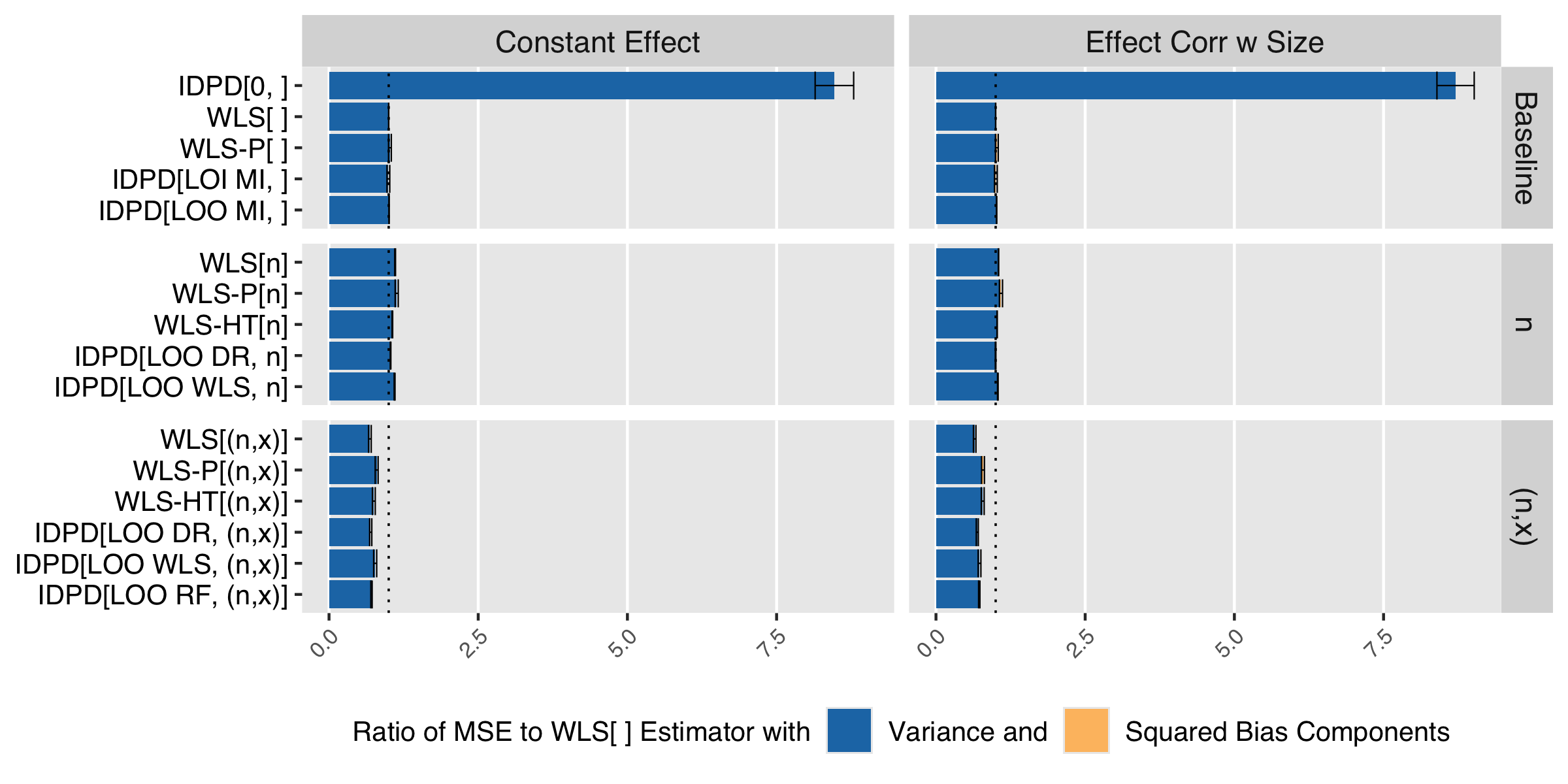}%
    }
    
    \subfloat[Excluding Horvitz-Thompson Estimator]{%
     \includegraphics[width = .9\textwidth]{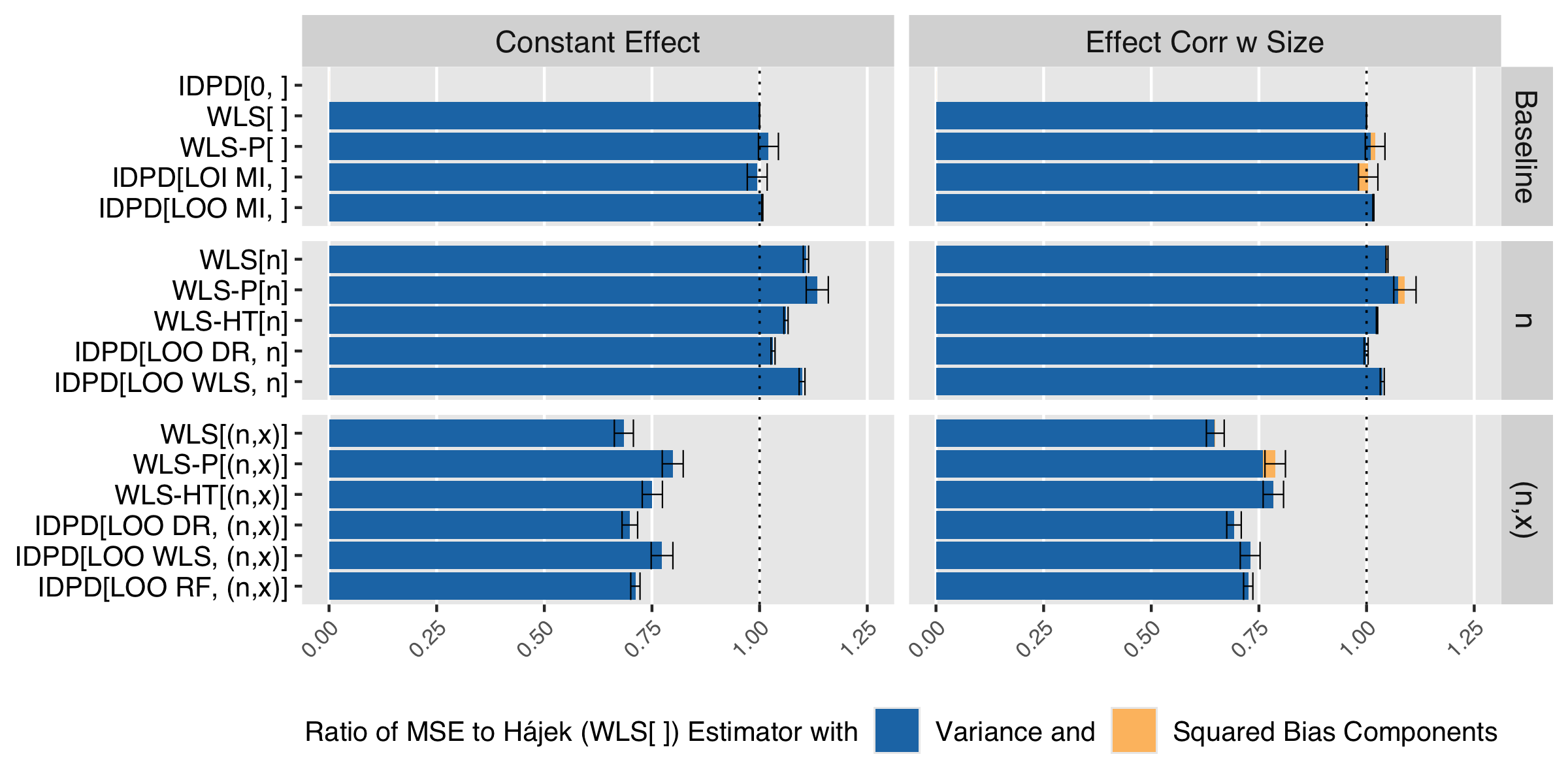}%
    }
    \caption{Ratio of the MSE of treatment effect estimators, compared with the H\'ajek estimator ($\hajek$) under CTAI real data simulations. (a) and (b) show the same data, with (b) just excluding the Horvitz-Thompson estimator in order to compare the other estimators.}
    \label{fig:mse-cta-apx}
\end{figure}

\begin{table}[ht]
\centering
\begingroup \footnotesize
\centering
\resizebox{\ifdim\width>\linewidth\linewidth\else\width\fi}{!}{
\begin{tabular}{llcccc}
\toprule
\multicolumn{2}{c}{\textbf{Estimator}} & \multicolumn{2}{c}{\textbf{Constant Effect}} & \multicolumn{2}{c}{\textbf{Effect Corr w Size}} \\
\cmidrule(l{3pt}r{3pt}){1-2} \cmidrule(l{3pt}r{3pt}){3-4} \cmidrule(l{3pt}r{3pt}){5-6}
\multicolumn{1}{c}{\textbf{Point}} & \multicolumn{1}{c}{\textbf{Variance}} & \multicolumn{1}{c}{\textbf{Var Rel Bias}} & \multicolumn{1}{c}{\textbf{Covg.}} & \multicolumn{1}{c}{\textbf{Var Rel Bias}} & \multicolumn{1}{c}{\textbf{Covg.}}\\
\midrule
\addlinespace[0.3em]
\multicolumn{6}{l}{\textbf{Baseline}}\\
\hspace{1em}IDPD[0, $\emptyset$] & Design (\ref{eq:mavar}) & 1.56 (0.02) & 0.99 & 1.57 (0.02) & 0.99\\
\hspace{1em}WLS[$\emptyset$] & Design (\ref{eq:crhjvar}) & 0.94 (0.01) & 0.94 & 0.98 (0.01) & 0.94\\
\hspace{1em}WLS[$\emptyset$] & H-W Robust & 2.60 (0.04) & 1.00 & 2.42 (0.03) & 1.00\\
\hspace{1em}WLS-P[$\emptyset$] & H-W Robust & 1.01 (0.01) & 0.95 & 1.05 (0.01) & 0.95\\
\hspace{1em}IDPD[LOI MI, $\emptyset$] & Design (\ref{eq:vikn}) & 1.05 (0.01) & 0.95 & 1.21 (0.02) & 0.95\\
\hspace{1em}IDPD[LOO MI, $\emptyset$] & Design (\ref{eq:vloop}) & 1.04 (0.01) & 0.95 & 1.10 (0.02) & 0.95\\
\addlinespace[0.3em]
\multicolumn{6}{l}{\textbf{Cluster Size}}\\
\hspace{1em}WLS[n] & H-W Robust & 2.42 (0.03) & 1.00 & 2.44 (0.03) & 1.00\\
\hspace{1em}WLS-P[n] & H-W Robust & 0.98 (0.01) & 0.94 & 1.02 (0.01) & 0.94\\
\hspace{1em}WLS-HT[n] & H-W Robust & 2.59 (0.04) & 1.00 & 2.49 (0.04) & 1.00\\
\hspace{1em}IDPD[LOO DR, n] & Design (\ref{eq:mavar}) & 3.00 (0.04) & 1.00 & 3.06 (0.04) & 1.00\\
\hspace{1em}IDPD[LOO WLS, n] & Design (\ref{eq:vloop}) & 1.12 (0.02) & 0.95 & 1.12 (0.02) & 0.95\\
\addlinespace[0.3em]
\multicolumn{6}{l}{\textbf{Cluster Size and Pretest}}\\
\hspace{1em}WLS[(n,x)] & H-W Robust & 0.67 (0.01) & 0.88 & 0.71 (0.01) & 0.89\\
\hspace{1em}WLS-P[(n,x)] & H-W Robust & 0.92 (0.01) & 0.92 & 0.95 (0.01) & 0.92\\
\hspace{1em}WLS-HT[(n,x)] & H-W Robust & 0.65 (0.01) & 0.86 & 0.66 (0.01) & 0.86\\
\hspace{1em}IDPD[LOO DR, (n,x)] & Design (\ref{eq:mavar}) & 1.96 (0.03) & 0.99 & 2.18 (0.03) & 0.99\\
\hspace{1em}IDPD[LOO WLS, (n,x)] & Design (\ref{eq:vloop}) & 1.22 (0.02) & 0.90 & 1.22 (0.02) & 0.90\\
\hspace{1em}IDPD[LOO RF, (n,x)] & Design (\ref{eq:vloop}) & 1.13 (0.02) & 0.95 & 1.17 (0.02) & 0.96\\
\bottomrule
\end{tabular}}

\endgroup
\caption{Simulated relative bias (``Var Rel Bias'') and coverage (``Covg.'') of treatment effect estimators under CTAI real data simulations. Simulation standard errors are shown in parentheses. The simulation standard errors for coverage probabilities are all less than 0.0035.  Design-based variance estimators include a reference to the relevant equation in this chapter. ``H-W Robust'' refers to the Huber-White heteroskedasticity robust variance estimator with HC1 structure.}
\label{tab:var-cta-apx}
\end{table}

\end{document}